\documentclass[floats,floatfix,showpacs,amssymb,prd,twocolumn,superscriptaddress,nofootinbib,longbibliography,reprint]{revtex4-1}

\usepackage{amssymb,amsmath,verbatim,mathtools,needspace,enumitem,etoolbox,graphicx,physics,microtype,afterpage,xspace,tabularx,lmodern,multirow}
\usepackage{gensymb}
\usepackage[normalem]{ulem}
\usepackage[dvipsnames, usenames]{xcolor}
\definecolor{linkcolor}{rgb}{0.0,0.3,0.5}
\usepackage[unicode, colorlinks=true, linkcolor=linkcolor, citecolor=linkcolor, filecolor=linkcolor, urlcolor=linkcolor, linktocpage, breaklinks]{hyperref}
\usepackage[all]{hypcap}
\usepackage[T1]{fontenc}
\usepackage[utf8]{inputenc}
\usepackage[usenames,dvipsnames]{xcolor}
\hypersetup{colorlinks=true,citecolor=romared,linkcolor=romared,urlcolor=romared}
\definecolor{romared}{RGB}{142,0,28}
\usepackage{aas_macros}
\usepackage{makecell}
\usepackage{soul}
\usepackage{booktabs}
\usepackage{longtable}

\newcommand{\jhu}{\affiliation{William H. Miller III Department of Physics and Astronomy, Johns Hopkins University, 3400 North Charles
Street, Baltimore, Maryland 21218, USA}}

\begin{document}
	
\title{Predicting intermediate-mass black hole formation in star clusters \texorpdfstring{\\}{} with machine learning} 

\begin{abstract}
    Whether intermediate-mass black holes reside in nearby star clusters has remained contested for decades. We address this question by training neural network and random forest regressors on synthetic catalogs generated with the {\sc Rapster} cluster evolution code, mapping observable cluster properties such as total mass and half-mass radius onto the mass of the heaviest black hole built up through repeated mergers. Applying these models to nearby globular and nuclear star clusters, we forecast the intermediate-mass black hole population that each system may host. Globular clusters are unlikely to contain black holes more massive than $\sim 100\,M_\odot$, with an occupation fraction near 0.02, although they can produce remnants within the upper mass gap with masses approaching $100\,M_\odot$. Among nuclear star clusters, a handful of cases, including NGC~5102 and NGC~5206, yield predicted central black hole masses above $100\,M_\odot$, which we contrast with kinematically inferred estimates. Where the observationally claimed masses exceed our predictions, the implication is that the assembly history involved processes beyond hierarchical mergers, most plausibly accretion of gas and stars. Finally, we employ a normalizing flow to quantify, for individual globular clusters, the likelihood that their initial conditions were favorable to a collisional runaway during the first few million years after formation.
\end{abstract}

\author{Konstantinos Kritos}
\email{kkritos1@jhu.edu}
\jhu

\author{Digvijay Wadekar}
\jhu
\affiliation{\mbox{Center for gravitational physics, University of Texas at Austin, Austin, TX 78712, USA}}

\author{Emanuele Berti}
\jhu

\date{\today}
\maketitle

\section{Introduction}
\label{sec:Introduction}

Intermediate-mass black holes (IMBHs), typically defined to have masses in the range $10^2$–$10^6\,M_\odot$~\cite{2020ARA&A..58..257G}, are a long-sought but still elusive population that bridges the gap between stellar-mass and supermassive black holes (BHs). Dense stellar systems such as globular clusters (GCs) and nuclear star clusters (NSCs), with their high stellar densities and frequent dynamical interactions, provide natural environments for their formation and growth. These systems are ubiquitous across cosmic time, following a near-universal $-2$ power-law mass function~\cite{1996ApJ...466..802E}, and are therefore expected to play a fundamental role in seeding and assembling BHs across a wide mass spectrum. Recent observations with the James Webb Space Telescope have further highlighted massive star clusters at high redshift as potential birthplaces of IMBHs~\cite{Schleicher:2026rcc}.

Despite strong theoretical motivation, no IMBH has yet been unambiguously confirmed. A definitive detection requires a robust dynamical measurement of a central point mass, which remains observationally challenging given current spatial resolution limits, even for nearby systems in the Local Group~\cite{2018ApJ...858..118N}. Nevertheless, a number of candidates have been identified across different mass scales and environments. At the high-mass end ($\sim10^5$–$10^6\,M_\odot$), IMBH candidates have been inferred from optically selected active galactic nuclei in dwarf galaxies~\cite{Reines:2013pia}, including well-studied systems such as NGC~4395~\cite{Peterson:2005yp,2015ApJ...809..101D,2019NatAs...3..755W}. However, population studies suggest that IMBH occupation fractions in dwarf galaxies are significantly below unity~\cite{Greene:2021xbx,2021ApJ...917...17G}. Additional candidates arise in ultracompact dwarf galaxies and stripped galactic nuclei, where dynamical measurements reveal central massive objects~\cite{2025ApJ...991L..24T}.

At lower masses, gravitational-wave observations have opened a new window into the IMBH regime. The detection of confident binary BH mergers with remnant masses of $\sim140\,M_\odot$~\cite{LIGOScientific:2020iuh} and $\sim200\,M_\odot$~\cite{LIGOScientific:2025rsn} (and a number of additional near-threshold high-mass mergers~\cite{Wadekar:2023gea}) provides the first direct evidence for BHs entering the IMBH range ``from below.'' Interpreting these events requires understanding the relative contributions of isolated binary evolution and dynamical assembly channels~\cite{Sedda:2026xqr}, as well as hierarchical merger processes in dense stellar systems~\cite{Mai:2025jmk,2026arXiv260204176N,Tiwari:2026fta}. This hierarchical growth is expected to be particularly efficient in GCs and NSCs, where repeated mergers can build up IMBHs over time~\cite{OLeary:2005vqo,Leigh:2014oda}.

In Galactic GCs, extensive observational campaigns have placed stringent constraints on the presence of IMBHs. Deep radio surveys exploiting the fundamental plane of BH activity find no compelling evidence for accreting IMBHs, placing upper limits of $\lesssim10^3\,M_\odot$ in several clusters~\cite{Tremou:2018rvq,DeRijcke:2006tu}. X-ray observations similarly constrain central accreting sources~\cite{Grindlay:2001xh}. However, dynamical studies remain inconclusive. While some analyses report rising central velocity dispersions consistent with IMBHs, including early claims in $\omega$ Centauri~\cite{2008ApJ...676.1008N,2010ApJ...719L..60N} and 47~Tucanae~\cite{2017Natur.542..203K}, subsequent modeling often finds that these signatures can be explained without invoking a central BH~\cite{2010ApJ...710.1063V,2019MNRAS.488.5340B,2025A&A...693A.104B}.

This ambiguity reflects a fundamental degeneracy in the interpretation of cluster dynamics. Observed increases in central mass-to-light ratios and velocity dispersion profiles can arise not only from a single massive BH, but also from a centrally concentrated population of stellar remnants (white dwarfs, neutron stars, and stellar-mass BHs), orbital anisotropy, or an extended dark mass distribution~\cite{2016A&A...588A.149K,2021A&A...646A..63V,Vitral:2022apu,Vitral:2023zhl}. Evidence for such diffuse dark components has been reported in clusters such as NGC~6397~\cite{2021A&A...646A..63V}, M4~\cite{Vitral:2023zhl}, and M62~\cite{2019ApJ...884L...9A}. Even in systems often cited as strong IMBH candidates, such as $\omega$ Centauri and M31/G1 (Mayall II), both IMBH and extended-mass interpretations remain viable~\cite{Baumgardt:2003an,Gebhardt:2005cy,2022ApJ...924...48P,2017MNRAS.464.2174B}.

Recent high-resolution observations have further complicated the picture. In $\omega$ Centauri, the discovery of fast-moving stars whose velocities exceed the nominal escape speed suggests the presence of a central dark mass of at least $\sim8.2\times10^3\,M_\odot$~\cite{2024Natur.631..285H}, potentially consistent with an IMBH. Similarly, hypervelocity stars ejected via the Hills mechanism provide indirect evidence for IMBHs in clusters such as M15~\cite{Huang:2024gpv}. However, these interpretations are not unique, and alternative scenarios involving dense subsystems of stellar remnants remain plausible.

Transient phenomena offer an additional probe of IMBHs. Highly luminous X-ray flares in the local Universe ($z<0.1$) with peak luminosities $\gtrsim10^{41}\,\mathrm{erg\,s^{-1}}$ have been interpreted as tidal disruption events (TDEs) involving IMBHs~\cite{Soria:2017ght,Lin:2018dev,Lin:2020exl,Jin:2025izu,Grotova:2025tdm,Chang:2025ucz}. Some of these events are associated with compact stellar systems, including hyperluminous X-ray sources such as HLX-1 and 3XMM~J215022.4-055108, whose optical counterparts suggest star cluster hosts. However, the inferred rate of such off-nuclear TDEs is low, $\lesssim10^{-7}\,\mathrm{yr^{-1}}$ per cluster~\cite{2024MNRAS.530.3043P}, significantly below rates in galactic nuclei.

From a theoretical standpoint, multiple formation pathways have been proposed for IMBHs. These include runaway stellar collisions in young dense clusters~\cite{Rasio:2003sz}, gas accretion, repeated tidal interactions, and hierarchical mergers of stellar-mass BHs. The growth and retention of IMBHs depend sensitively on cluster dynamics, including processes such as the Miller–Davies instability~\cite{Miller:2012ys} and the interplay between stellar-mass BH and IMBH populations~\cite{2014MNRAS.444...29L}. Cluster Monte Carlo calculations have also shown that dense GCs can dynamically assemble merging BH binaries and populate masses above those expected from isolated binary evolution~\cite{Rodriguez:2015oxa,Rodriguez:2016kxx,Rodriguez:2021qhl}. Recent work has emphasized the importance of NSCs as sites of in-situ BH growth~\cite{2026arXiv260310581S,Partmann:2024ees}, as well as the survival and detectability of IMBHs over cosmic time~\cite{Martinez:2026jnc}. Numerical studies further demonstrate that IMBH formation can occur during the evolution from young massive clusters to present-day GCs~\cite{Sharma:2024ndf}, and may proceed efficiently in systems such as $\omega$ Centauri~\cite{Prieto:2025kqv}.

Empirical scaling relations provide another avenue for constraining IMBH populations. While supermassive BHs obey tight correlations, such as the $M_\mathrm{BH}$–$\sigma_\star$ relation, extensions of these correlations to the IMBH regime remain uncertain. Observations suggest a shallower slope for GCs~\cite{Lutzgendorf:2013csa}, with significant scatter that may reflect the diversity of formation pathways. Recent efforts using machine learning and symbolic regression have demonstrated the potential to uncover more complex, multivariate relations and reduce intrinsic scatter~\cite{2023arXiv231019406J,Wadekar:2022cyw}. Moreover, machine-learning models trained on GC simulations have also been used to identify IMBH host candidates among observed clusters, combining predictive accuracy with physical explainability~\cite{2024ApJ...965...89P}.

In this work, we adopt a data-driven approach to the IMBH inference problem. We train supervised machine-learning models on {\sc Rapster} simulations to approximate the conditional distribution $p(M_\mathrm{BH} \mid M_\mathrm{cl}, r_\mathrm{h}, R_\mathrm{g}, \dots)$ of the heaviest retained BH mass $M_\mathrm{BH}$ given observable cluster properties such as their mass $M_\mathrm{cl}$, half-mass radius $r_\mathrm{h}$, and galactocentric radius $R_\mathrm{g}$. This approach is motivated by the high dimensionality and intrinsic nonlinearity of the parameter space, which are not adequately captured by traditional one-parameter scaling relations. By training on large suites of cluster simulations, we can marginalize over uncertain initial conditions and stochastic merger histories. The rapid cluster evolution code {\sc Rapster} enables efficient exploration of clusters with $M_\mathrm{cl}\lesssim10^8\, M_\odot$. Such techniques are particularly well suited to disentangling the complex interplay between cluster structure, dynamical evolution, and BH growth.

Throughout this work, we adopt astrophysical units in which distances are measured in $\mathrm{pc}$, masses in $M_\odot$, times in $\mathrm{Myr}$, and velocities in $\mathrm{km\,s^{-1}} \approx \mathrm{pc\,Myr^{-1}}$. In these units, the gravitational constant is $G=(232)^{-1}$ and the speed of light is $c=3\times10^5$.

\section{Cluster simulation methods}
\label{sec:Cluster-simulations}

We simulate star clusters using {\sc Rapster},\footnote{The code is publicly available at the URL: \url{https://github.com/Kkritos/Rapster}} a population synthesis code for the formation of binary BHs in star clusters. The code relies on the Breen \& Heggie theory~\cite{Breen:2013vla} to evolve star clusters containing stars and BHs, with the latter segregating toward the center in a more compact configuration than the stellar population. It assumes a balance between the energy produced in the BH core and the heat flux through the half-mass radii of the BH subsystem and of the bulk stellar population (H\'enon's principle).
Moreover, BH masses are computed directly from initial star masses drawn from the Kroupa initial mass function in the range $[0.08,150]\,M_\odot$ with the {\tt SEVN} code~\cite{Spera:2017fyx}.

The {\sc Rapster} code accounts for the relevant dynamical processes of binary BH formation and evolution in star clusters, including three-body binary formation, dynamical BH-BH capture, binary-single exchanges, binary hardening, and prescriptions for binary-binary interactions, BH triple formation, and ejections (both relativistic and Newtonian). {\sc Rapster} is computationally efficient because it computes the probability of a given process occurring in the cluster, rather than directly integrating the dynamics of few-body systems, and treats macroscopic cluster properties statistically rather than evolving the full microphysical state. This approach is complementary to star-by-star Cluster Monte Carlo simulations, which use H\'enon's method to follow relaxation-driven cluster evolution and strong few-body encounters over many gigayears~\cite{Rodriguez:2015oxa,Rodriguez:2016kxx,Rodriguez:2021qhl}. Given the properties of the BH core, such as density $n_{\rm core}$ and velocity dispersion $v_{\rm core}$, which are determined from the balanced evolution condition, we compute the volumetric rate density of these processes based on an $n_{\rm core}^p\sigma_{_{\rm X}} v_{\rm core}$ computation, where $\sigma_{_{\rm X}}$ is the cross section of a particular process $\rm X$ and $p=2$ ($p=3$) for a two-body (three-body) interaction.

Binary BH merger channels include in-cluster mergers resulting from the emission of gravitational waves following the dynamical decoupling of the binary, single-single and binary-single BH-BH captures, mergers from the von Zaipel-Lidov-Kozai mechanism in a BH triple, as well as binaries that merge after being ejected from the cluster. The properties of the merger remnant (final mass, final spin, and gravitational-wave recoil velocity) are computed using analytic formulas calibrated with numerical relativity simulations for mass ratios up to $\sim$1:8 and interpolated to match the test-particle limit for extreme-mass-ratio inspirals (see, e.g. Sec.~V in Ref.~\cite{Gerosa:2016sys}). For more recent kick models, see also Refs.~\cite{Islam:2025drw,Islam:2026yxx}.

\begin{figure*}
    \centering
    \includegraphics[width=0.49\textwidth]{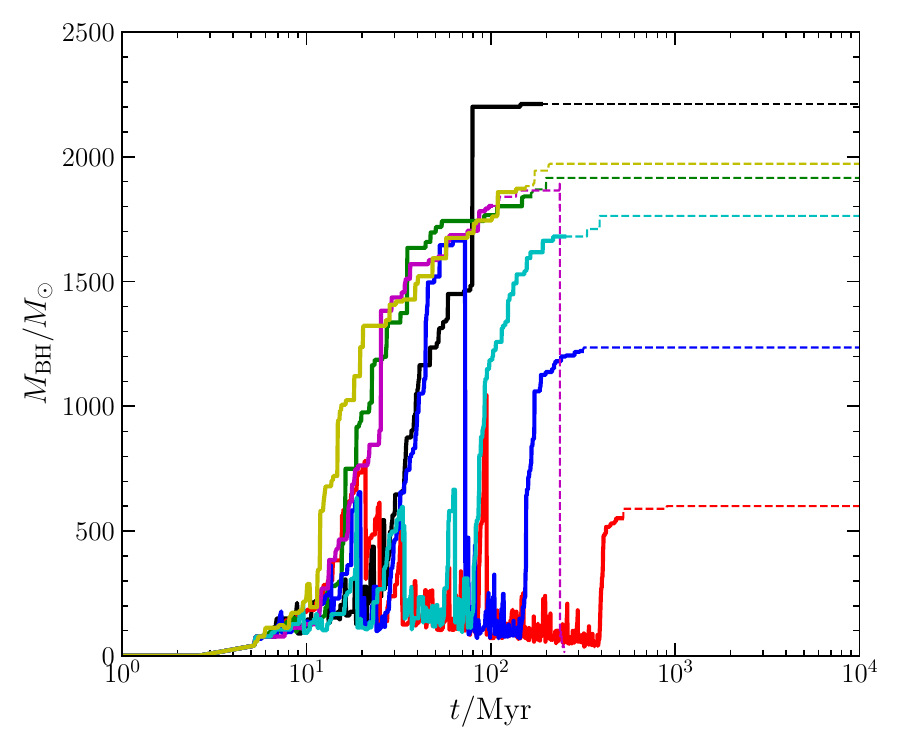}
    \includegraphics[width=0.49\textwidth]{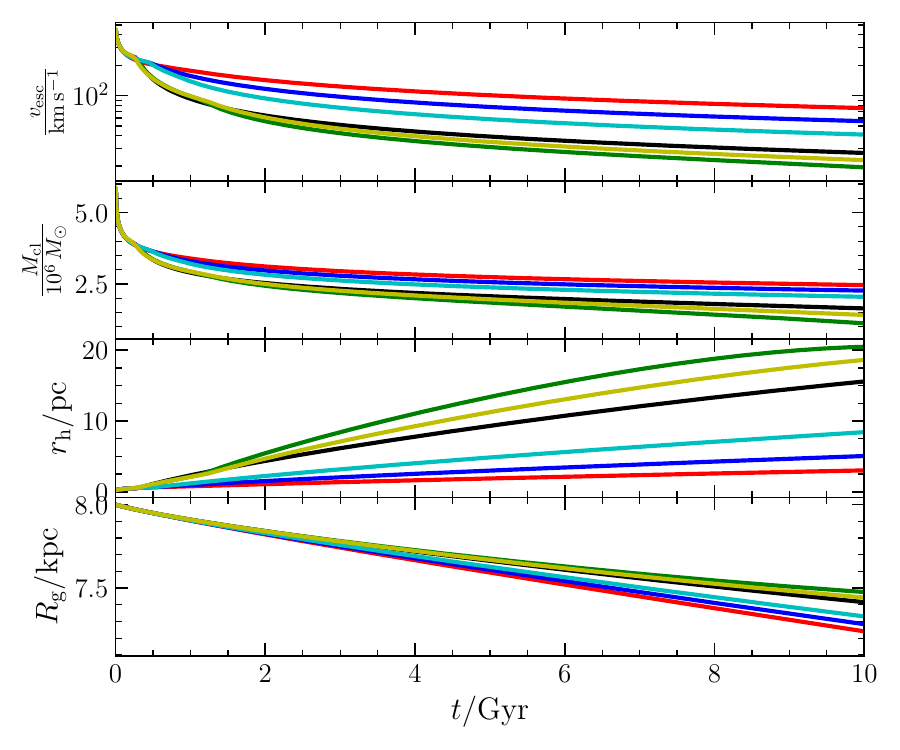}
    \includegraphics[width=0.49\textwidth]{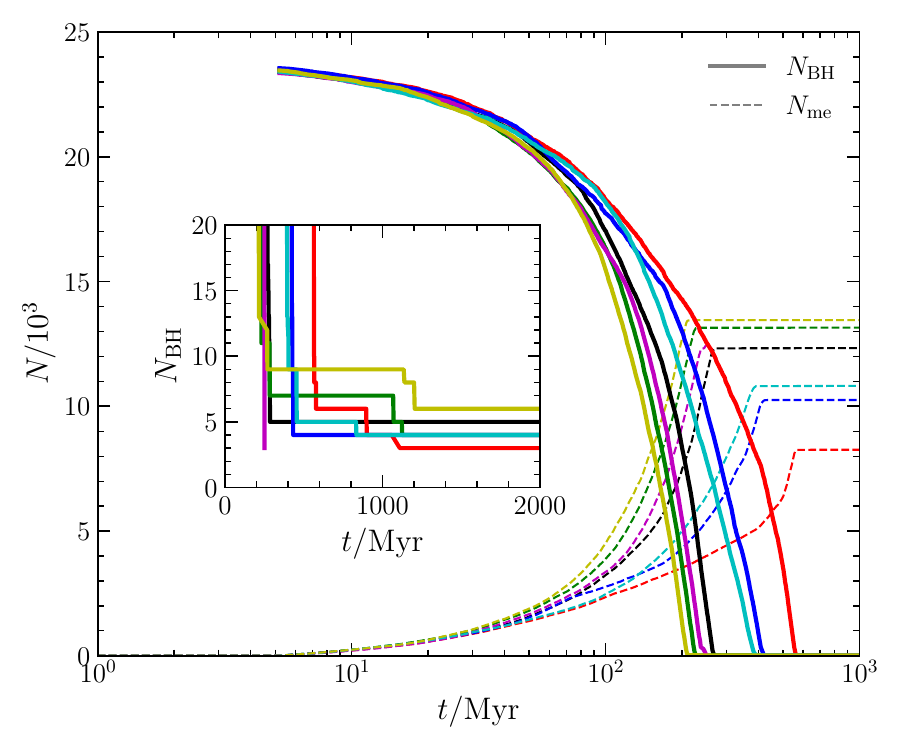}
    \includegraphics[width=0.49\textwidth]{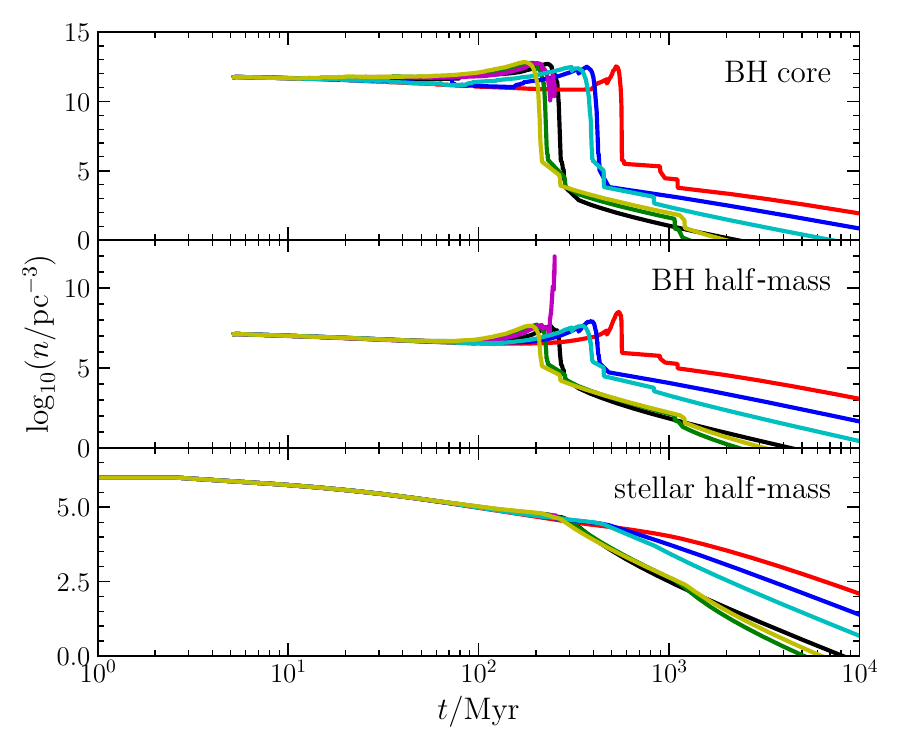}
    \caption{Seven realizations of star cluster simulations with $N_0=10^7$, $r_{\rm h,0}=0.2\,\rm pc$, $Z=0.001$, and $R_{\rm g,0}=8\,\rm kpc$. The time evolution of the heaviest BH mass in the cluster is shown in the upper left panel. Transition of the solid lines to dashed in the upper left panel designates the point at which the wandering radius of the heaviest BH in the system becomes smaller than its influence radius. The evolution of the escape velocity $v_{\rm esc}$, cluster mass $M_{\rm cl}$, half-mass radius $r_{\rm h}$, and galactocentric radius $R_{\rm g}$ are shown in the top right panel (from top to bottom). The total number of BHs in the cluster core (solid) and the cumulative number of BH-BH mergers (dashed) are shown in the bottom left panel, while the inset shows the number of BHs near the evaporation of the BH subsystem. Finally, the bottom right panels show the number density of BHs in the core (top), the number density of BHs at the half-mass radius of the BH subsystem (middle), and the half-mass stellar number density (bottom), respectively.}
    \label{fig:cluster-evolution-realizations}
\end{figure*}

\section{IMBH formation}
\label{sec:IMBH-formation}

If the remnant receives a recoil smaller than the escape speed of the cluster, then it is retained in the BH subsystem where it may form a binary BH and, if not ejected, merge again within the cluster. This process can continue, and then the mass of the BH can grow via repeated BH binary mergers. If the escape velocity is high enough to retain the successive merger products, an IMBH emerges from such a process of repeated merging. There is typically one IMBH that forms in a cluster; if two IMBHs were to form, they would preferentially merge into a single massive BH. A detailed description of the assumptions and of the code can be found in Ref.~\cite{Kritos:2022ggc}.

\subsection{Variation with cluster realization}

Before discussing the full star cluster simulation suite in Sec.~\ref{sec:cluster_suite}, we use a few isolated simulations to illustrate the variation in central BH properties across Monte Carlo realizations. 
We simulate seven different realizations of a star cluster with initial stellar number $N_{0}=10^7$ (corresponding to an initial total mass of $M_{\rm cl,0}=6\times10^{6}\,M_\odot$), initial half-mass radius $r_{\rm h,0}=0.2\,\rm pc$, absolute metallicity $Z=0.001$, and initial galactocentric radius $R_{\rm g,0}=8\,\rm kpc$. The simulations assume the same distribution of initial conditions, and each time, only the seed number (which controls all Monte Carlo draws in the simulation) is changed. The initial conditions have been chosen so that the initial escape speed of the cluster is around $450\,\rm km\, s^{-1}$, high enough to allow for BH growth through repeated mergers. In Fig.~\ref{fig:cluster-evolution-realizations} we show the mass growth of the BH (upper left panel) and the evolution of the cluster with time (upper right panel). Also shown are the evolution of the number of BHs in the subsystem and the cumulative number of mergers (lower left panel), as well as the time evolution of the BH and stellar number densities in the BH core and at the half-mass radii of the BH subsystem and of the stellar population (lower right panels).

Despite choosing the same initial distribution, we observe that the growth of the BH is stochastic, and the terminal mass depends on the Monte Carlo realization. In particular, the terminal mass of the heaviest BH in the cluster varies from $\simeq600\,M_\odot$ (the least massive) up to $\simeq2200\,M_\odot$ (the heaviest). In one realization (the magenta line of Fig.~\ref{fig:cluster-evolution-realizations}), no IMBH remains in the system, as it is ejected at $t\simeq250\,\rm Myr$ due to a Newtonian kick. In all cases, IMBH growth begins immediately after the formation of the BH subsystem in the core at $\simeq10\,\rm Myr$ and ends within a few $100\,\rm Myr$. The growth of the IMBH stops and the maximum BH mass plateaus, because only a few BHs remain in the cluster as a consequence of the evaporation of the BH subsystem. In particular, $N_{\rm BH}(t\to\infty)\simeq5\pm2$ including the IMBH, as seen in the inset of the lower left panel of Fig.~\ref{fig:cluster-evolution-realizations}.
The evaporation is fast and completes within only a few hundred $\rm Myr$ because the cluster has a small initial half-mass relaxation time between $10\,\rm Myr$ and $30\,\rm Myr$ during the runaway BH growth episode.

We also test the validity of the assumptions regarding energy production in the core of the BH subsystem in the presence of a massive BH. We compute the point in time at which this microphysical description should break down, indicated in the upper left panel of Fig.~\ref{fig:cluster-evolution-realizations} by the thin dashed lines. However, the BH mass plateaus at that point, and the BH subsystem rapidly evaporates. Thus, we are confident that the runaway is adequately described as in Ref.~\cite{Breen:2013vla}. The global macrophysical evolution of the cluster after that point is not dependent on the microphysical details of the core, and the long-term evolution of the cluster is correct (see Appendix~\ref{app:Validity-of-the-Breen-&-Heggie-theory} for details). In particular, we find that the merging chain runaway terminates roughly at the point where the wandering radius of the IMBH matches its radius of influence.

In specific simulations, such as those in the blue and cyan lines, the heaviest BH mass in the system is not a monotonic function of time. This occurs because the growing seeds receive a kick larger than the escape velocity at that point in time, the IMBH is ejected from the cluster, and the heaviest BH mass in the system becomes $<50\,M_\odot$. If a population of stellar-mass BHs remains in the core, a new growth chain may occur, leading to the formation of a new IMBH. Finally, the growth of IMBHs through repeated mergers predominantly occurs through a chain of intermediate-mass ratio inspirals~\cite{Mandel:2007hi}, which consist of the merger between the growing primary BH and one of the first-generation stellar-mass BHs in the BH subsystem. However, in one realization (the one in the black line), two heavy BHs with masses $\simeq1500\,M_\odot$ and $\simeq700\,M_\odot$ form simultaneously and rapidly in the cluster, and eventually merge at $t\simeq80\,\rm Myr$ to give the characteristic vertical step in the black line for $M_{\rm BH}(t)$. Nevertheless, such events are rarer than intermediate-mass ratio inspirals, and the growth proceeds in an oligarchic fashion, where a single IMBH emerges near the center within a more diffuse stellar-mass BH subcluster.

The global evolution of the system is dynamically coupled to the growth of the BH, in the sense that systems that happen to form a heavier IMBH tend to expand by a larger factor and lose more mass over time. The simulated models have an initial escape velocity of $\simeq450\,\rm km\,s^{-1}$, are massive, compact, and thus do not resemble the population of GCs observed in the Galactic halo. Nevertheless, all systems expand with time as a consequence of energy production in the BH subsystem, and lose mass as a consequence of escaping stars with velocities that exceed the escape speed. Systems that form heavier BHs produce more energy, and thus expand by larger factors. 
All clusters start expanding at $t=3\,\rm Myr$ due to adiabatic stellar mass loss in winds, and the core collapses at $t\approx250\,\rm Myr$, at which point balanced evolution commences. Thus, the rapid growth of the IMBH in these simulation examples occurs before core collapse.
Notice also that BHs segregate and form the BH subsystem on a timescale that is reduced by a factor of $\approx\overline{m}_\star/\overline{m}_{\rm BH}\approx24$ relative to the initial half-mass relaxation time, i.e., in $\approx3\,\rm Myr$. Thus, by the time the runaway process begins, we can assume that the BHs have already segregated in the core of the system.

In this simulation example, the half-mass stellar number density is of the order of $\sim10^6\,\rm pc^{-3}$, and drops as the cluster expands to below $300\,\rm pc^{-3}$ within $10\,\rm Gyr$ (cf.~Fig.~\ref{fig:cluster-evolution-realizations}, lower right panel). On the other hand, the corresponding BH half-mass density is $\sim\,\rm pc^{-3}$.
Moreover, the central number density of BHs in the BH core is of the order of $\sim10^{13}\,\rm pc^{-3}$.
As the BH subsystem evaporates and $N_{\rm BH}$ decreases, the BH subsystem contracts to meet the energy demands of the cluster according to le Chatelier's principle.
The lighter the mass of the final IMBH produced in the core, the higher the density drop in the long term.
The dramatic drop in the BH density is associated with the runaway evaporation of the BH subsystem.

Since the escape velocity $v_{\rm esc}$ is related to the half-mass radius $r_{\rm h}$ through $v_{\rm esc}\propto r_{\rm h}^{-1/2}$, these systems have a lower escape velocity and therefore lose more mass over $10\,\rm Gyr$. The range of final escape velocities, masses, and half-mass radii is $\sim20$--$75\,\rm km\,s^{-1}$, $\sim(0.8$--$2.4)\times 10^6\,M_\odot$, and $\sim3$--$20\,\rm pc$, respectively, which is compatible with the population of the heaviest observed Galactic GCs. Finally, the galactocentric radius decreases from $8\,\rm kpc$ to $\sim7.2$--$7.5\,\rm kpc$, with systems that host heavier IMBHs sinking by a smaller amount because of their smaller masses and weaker dynamical friction.

\subsection{Probability of IMBH growth}

\begin{figure}
    \centering
    \includegraphics[width=\linewidth]{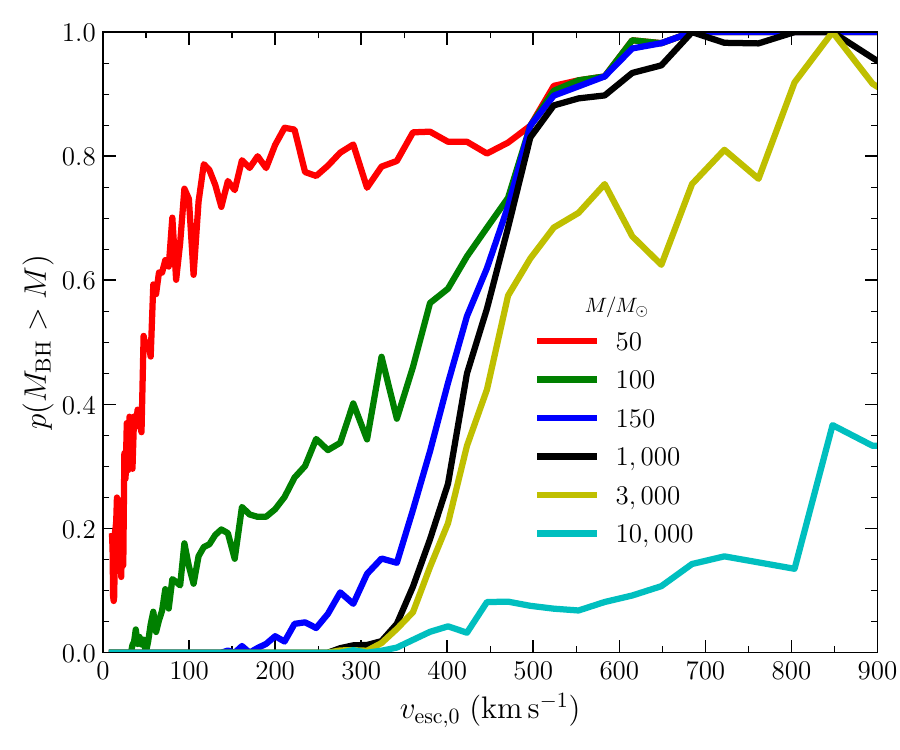}
    \caption{Probability $p(M_{\rm BH}>M)$ of forming a BH in a simulated cluster with a mass of at least $M=50\,M_\odot$ (red), $M=100\,M_\odot$ (green), $M=150\,M_\odot$ (blue), $M=1,000\,M_\odot$ (black), $M=3,000\,M_\odot$ (yellow), and $M=10,000\,M_\odot$ (cyan) as a function of the initial escape velocity in the system, $v_{\rm esc,0}$.}
    \label{fig:imbh-formation-probability}
\end{figure}

The production of an IMBH depends strongly on the initial conditions of the cluster. Larger initial escape velocities increase the probability of forming an IMBH that is retained in the system for several $\rm Gyr$. In Fig.~\ref{fig:imbh-formation-probability} we show the probability $p(M_{\rm BH}>M)$ of forming a BH with mass $M_{\rm BH}$ of at least $M$, for different values of $M$. Based on our simulations, we show that no matter how high the escape velocity is in the range up to $1000\,\rm km\,s^{-1}$ the formation of IMBHs with mass $>1000\,M_\odot$ is never certain, while efficient production of IMBHs with mass $>150\,M_\odot$ occurs above $\simeq400\,\rm km\,s^{-1}$ with probability $>50\%$. Nevertheless, the likelihood of forming an IMBH of a few hundred $M_\odot$ becomes unity only above $\simeq600\,\rm km\,s^{-1}$. Moreover, $80\%$ of simulations with $v_{\rm esc,0}>100\,\rm km\,s^{-1}$ result in a cluster with a BH with a mass of at least $50\,M_\odot$.

The escape velocity evolves with time, and may decrease by a factor of $\sim10$ over several $\rm Gyr$ as the cluster expands and loses mass (e.g., see the top left panel of Fig.~\ref{fig:cluster-evolution-realizations}). However, since IMBH production is a stochastic process and the evolution of the system is coupled nonlinearly with the mass of the IMBH, we cannot use the results of Fig.~\ref{fig:imbh-formation-probability} to compute the probability of a cluster hosting an IMBH based only on its final escape velocity. Instead, below we train supervised machine-learning models to connect the mass of the heaviest BH to the final properties of the star cluster based on a large set of simulations.

\begin{figure*}
    \centering
    \includegraphics[width=0.49\textwidth]{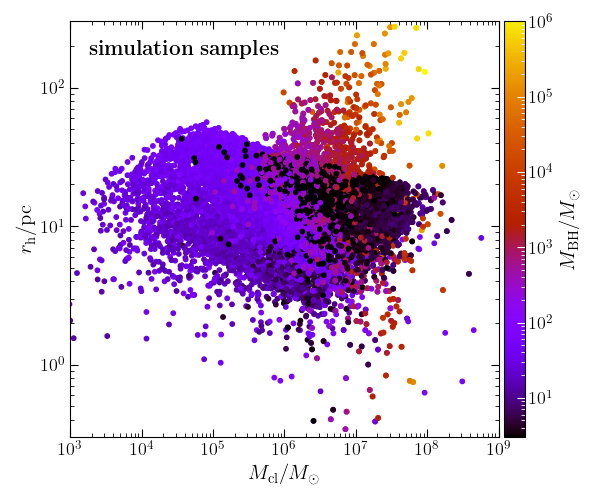}
    \includegraphics[width=0.49\textwidth]{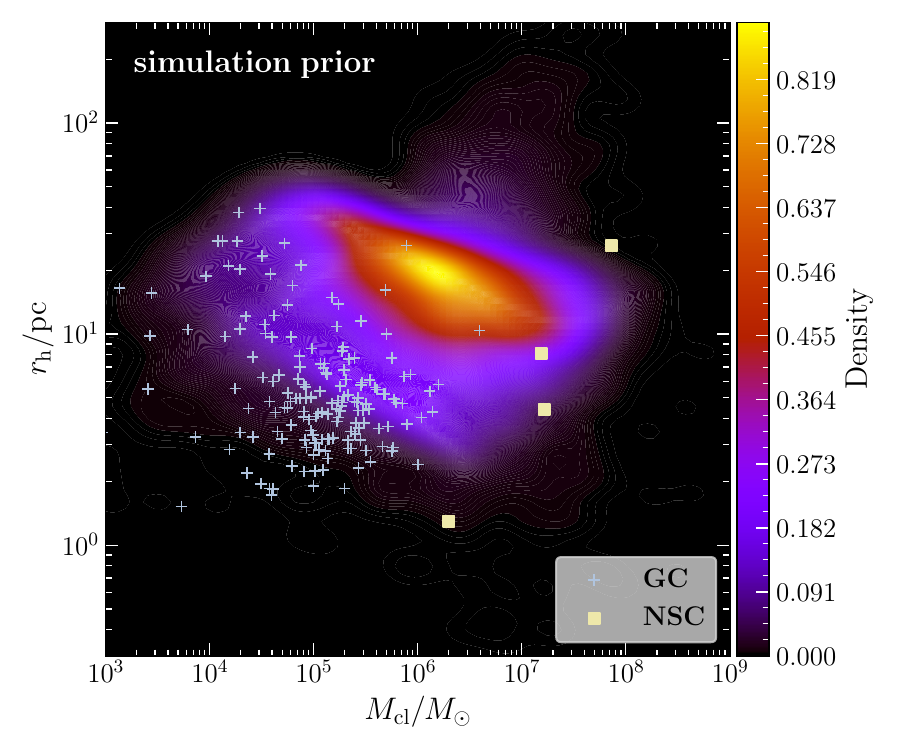}
    \includegraphics[width=0.49\textwidth]{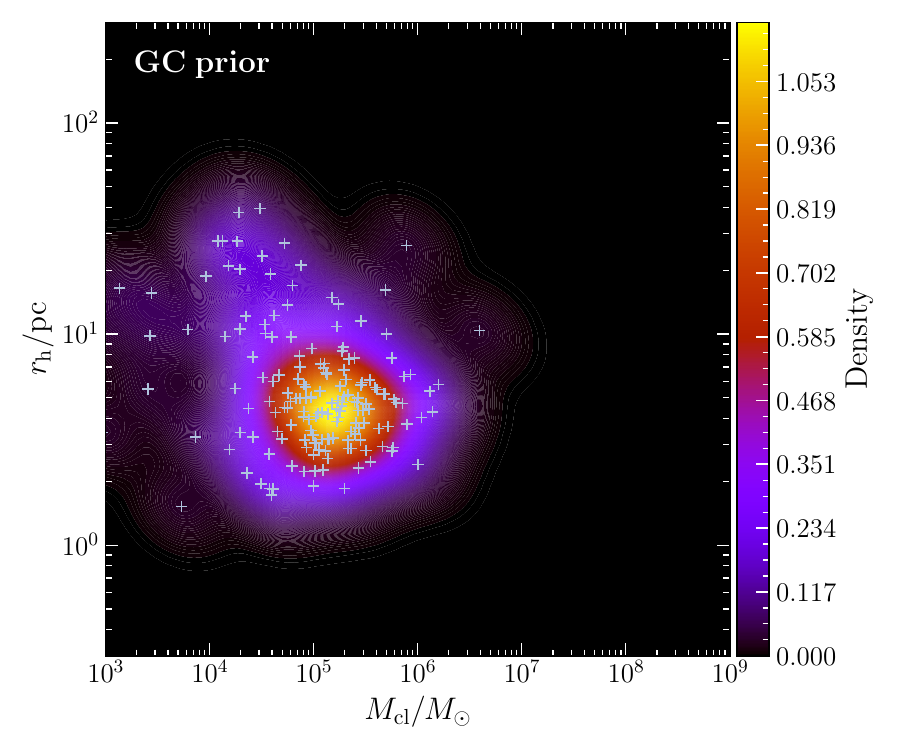}
    \includegraphics[width=0.49\textwidth]{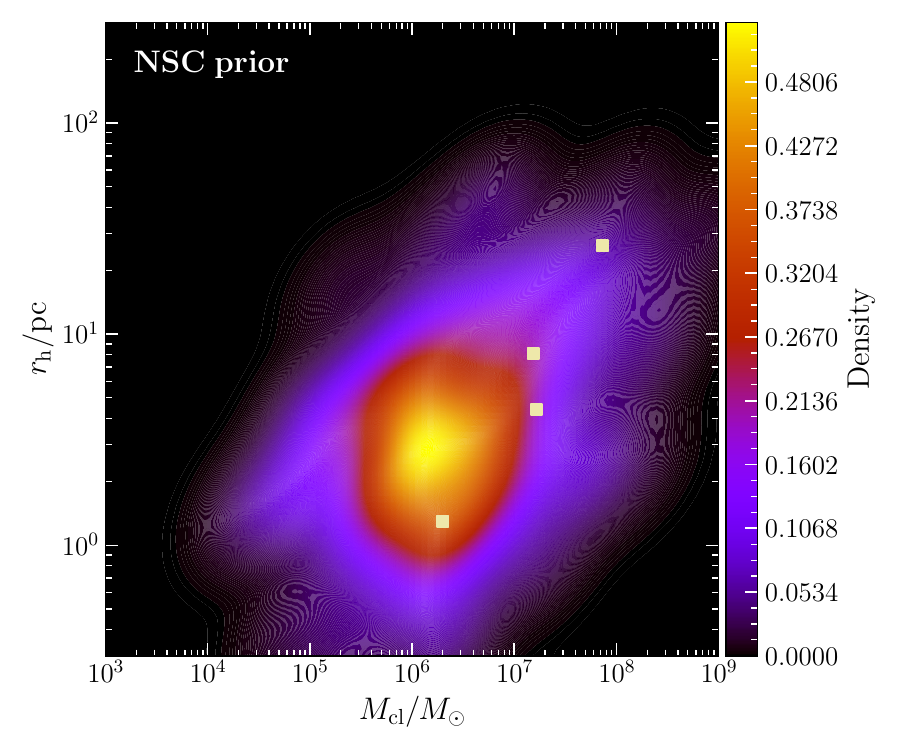}
    \caption[Simulation, GC, and NSC priors]{Suite of {\sc Rapster} simulations and priors used to train our machine learning models. Top left: final half-mass radius versus final cluster mass of our simulated star clusters that survived until $z=0$, along with the final mass of the heaviest BH in the system (shown in the colorbar). Top right: kernel density estimation of the simulated data from the upper left panel. The resulting kernel density estimation serves as our simulation prior in the $(M_{\rm cl},\,r_{\rm h})$ space. The white points show observational datasets for Milky Way GCs~\cite{Baumgardt:2018pyl} and NSCs~\cite{2018ApJ...858..118N}. Bottom left: kernel density estimation of the GC prior, taking all Milky Way GCs from the catalog of Ref.~\cite{Baumgardt:2018pyl}. Bottom right: kernel density estimation of the NSC prior, using observational data from NSCs in the local Universe from Ref.~\cite{2016MNRAS.457.2122G} along with the four Local Group NSCs from Ref.~\cite{2018ApJ...858..118N}.
    In simulation-based inference, the simulation set encodes prior information. We reweight the simulation samples separately for the GC and NSC target populations before training the population-specific models.}
    \label{fig:priors}
\end{figure*}

\section{Cluster simulation suite}
\label{sec:cluster_suite}

We perform a large number of star cluster simulations with {\sc Rapster}, varying the initial conditions, initial cluster mass $M_{\rm cl,0}$, initial half-mass radius $r_{\rm h,0}$, metallicity $Z$, galactocentric radius $R_{\rm g,0}$, and redshift of cluster formation $z_{\rm cl,0}$ using different seeds. We uniformly sample $\log_{10}(M_{\rm cl,0}/M_\odot)\in[4, 9]$, $\log_{10}(r_{\rm h,0}/{\rm pc})\in[-1, 1]$, $\log_{10}(Z)\in[-4, -1.7]$, $R_{\rm g,0}\in[0, 200]\,\rm kpc$, and $z_{\rm cl,0}\in[0, 10]$. The initial number of stars in the cluster is $N_0=M_{\rm cl,0}/\overline{m}$, where $\overline{m}\simeq0.6\,M_\odot$ is the average stellar mass assuming a Kroupa initial mass function in the range $0.08$--$150\,M_\odot$. The central stellar density is computed, assuming the Plummer profile, as $n\simeq0.5N/r_{\rm h}^3$. Clusters are assumed to evolve in the tidal field of an isothermal galaxy with circular velocity $220\,\rm km\,s^{-1}$.
We adopt the SEVN delayed model~\cite{Spera:2017fyx} to compute remnant masses from initial stellar masses, and we assume the spins of all first-generation BHs to be zero. However, our BHs may spin up to $\sim0.7$ after merging, and spin and mass-ratio effects are taken into consideration when computing gravitational-wave recoil during each merger. In addition, we assume natal kicks for the BHs following a Maxwellian with parameter $265\,\rm km\,s^{-1}$, reduced by mass fallback during the supernova. The termination criteria for the simulation depend on one of the following fates for the cluster:
\begin{itemize}
    \item the BH subsystem has evaporated ($N_{\rm BH}=0$);
    \item the cluster has dissolved ($M_{\rm cl}=0$);
    \item the cluster has reached the center of the galaxy ($R_{\rm g}=0$);
    \item redshift $z_{\rm cl}=0$ has been reached.
\end{itemize}
In total, we perform 29286 simulations, of which 18094 satisfy the last termination criterion above, i.e., survive until today. In our analysis, we discard clusters that have either evaporated, ejected all of their BHs (including the potential IMBH), or reached the center of the galaxy during their evolution.

We choose a broad sampling distribution $p_{\rm simul}(X_{0})$ for the initial conditions $X_{0}=\{M_{\rm cl,0}, r_{\rm h,0},...\}$, as described above. Thus, the raw simulation set is not a draw from any particular astrophysical star cluster population. When training models for a specific astrophysical population, either GCs or NSCs, we therefore apply importance weights that account for the mismatch between the simulation distribution and the target astrophysical population. In particular, the population of BH masses that result from our simulations does not correspond to the intrinsic population of BHs in the Universe, because the simulation prior was chosen for coverage of the parameter space of initial conditions rather than realism. Moreover, our nonlinear simulations map initial conditions $X_{0}$ to final conditions $X_{\rm f}$, and this mapping induces a simulation prior on the final conditions, $p_{\rm simul}(X_{\rm f})$.
In the upper left panel of Fig.~\ref{fig:priors} we show the final mass, half-mass radius, and heaviest BH mass for the simulated clusters. The heatmap in the upper right panel shows a kernel density estimation of the simulation distribution of the final cluster properties.

To obtain the mass distribution of BHs in GCs or NSCs, we reweight each simulation by the corresponding importance ratio, $p_{\rm GC}(X_{\rm f})/p_{\rm simul}(X_{\rm f})$ or $p_{\rm NSC}(X_{\rm f})/p_{\rm simul}(X_{\rm f})$. Here, $p_{\rm GC}$ is estimated with a kernel density estimator from the GC catalog of Ref.~\cite{Baumgardt:2018pyl} and shown in the lower left panel of Fig.~\ref{fig:priors}. Similarly, we estimate $p_{\rm NSC}$ from the local observational data of Ref.~\cite{2016MNRAS.457.2122G}; the result is shown in the lower right panel of Fig.~\ref{fig:priors}. The dependence of BH-mass inference on the simulation prior is discussed in Sec.~\ref{sec:Random-forest-regressors}, where we show that it is most important when the final-property inference is multimodal.

We use {\tt SciPy}'s default Gaussian kernel density estimator~\cite{2020SciPy-NMeth}, which implements Scott's algorithm~\cite{D.W.Scott}. We have also verified that our results are not sensitive to the choice of kernel density estimation by checking that different bandwidth factors of 0.2, 0.5, and 0.8 do not alter our findings.
We use the final cluster conditions to compute these weights, because GCs and NSCs are systems that formed at high redshift and have evolved to the present epoch. The fitted astrophysical priors do not correspond directly to the intrinsic underlying cluster populations, because they depend on observational data and are prone to selection effects. For GCs, the Milky Way census is largely complete at the masses relevant to IMBH formation ($M_\mathrm{cl} \gtrsim 10^5 M_\odot$), although a population of obscured bulge clusters remains under-sampled. For NSCs the situation is more delicate: the occupation fraction is a steep function of host galaxy mass, NSC detection is resolution and surface-brightness-limited, and the catalog is biased toward nearby, massive, gas-poor galaxies. As a result, our NSC prior should be understood as representative of currently observed NSCs rather than of the intrinsic NSC population in the local Universe.

The upper left panel of Fig.~\ref{fig:priors} shows that clusters that have formed an IMBH in their center tend to produce more energy and expand by a larger amount than clusters without an IMBH. This is consistent with the top-right panel of Fig.~\ref{fig:cluster-evolution-realizations}, where the expansion rate correlates weakly with the mass of the IMBH produced.
This correlation gives the final structural properties of the cluster predictive information about the BH mass, although the relation remains stochastic.

\begin{figure}
    \centering
    \includegraphics[width=\linewidth]{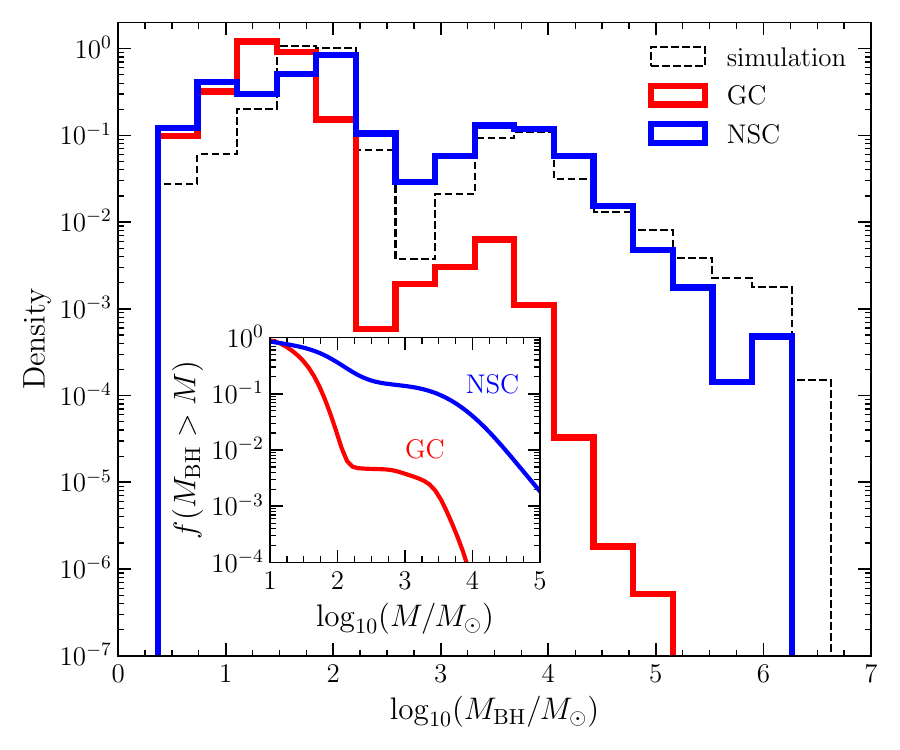}
    \caption[Predicted BH mass distributions for GCs and NSCs]{The predicted local population of the heaviest BH mass hosted in the centers of globular clusters (GC, red) and nuclear star clusters (NSC, blue). The population derived from our simulations is shown in the dotted black line and is reweighted by the GC or NSC prior to obtain the red and blue histograms, respectively. The inset panel is the cumulative version of the density plot showing the fractional density of clusters that host a BH with mass $M_{\rm BH}>M$ in the GC (red) and NSC (blue) populations, respectively.}
    \label{fig:MBH-histograms}
\end{figure}

Given the astrophysical priors for GCs and NSCs, we reweight the simulated BH mass distribution and obtain the normalized mass distributions for BHs in the two cluster classes, $df_{\rm GC}/d\log_{10}(M_{\rm BH}/M_\odot)$ and $df_{\rm NSC}/d\log_{10}(M_{\rm BH}/M_\odot)$, respectively. We show the results in Fig.~\ref{fig:MBH-histograms}. Note that the predicted distributions of IMBHs in GCs and NSCs are not merely power laws, but instead show population-dependent peaks. Moreover, the IMBH distribution has a small plateau just below the peak, while the high-mass tail in NSCs is longer.
This trend is qualitatively similar to Fig.~1 of Ref.~\cite{Sedda:2026xqr}, where an analogous peak is found at a somewhat larger mass value.

The occupation fraction of IMBHs in GCs and NSCs is found by numerical integration to be $f_{\rm GC}^{\rm IMBH}\simeq2\%$ and $f_{\rm NSC}^{\rm IMBH}\simeq36\%$, respectively. This is expected, because NSCs are heavier than GCs on average by a factor of $\sim10$ and retain more merger products due to their higher escape velocity. Here, IMBH denotes any BH with a mass above $100\,M_\odot$. 
Our estimated occupation fractions are likely to be lower limits due to contributions from other BH growth channels, such as gas accretion, tidal disruption of stars, and stellar collisions. Moreover, the occupation fraction of IMBHs in dwarf galaxies can be higher than our predicted $36\%$, but a $100\%$ occupation fraction has been excluded~\cite{2021ApJ...917...17G}.

To estimate the number density of IMBHs in GCs ($n_{\rm GC}^{\rm IMBH}$) and NSCs ($n_{\rm NSC}^{\rm IMBH}$) in the nearby Universe, we multiply these fractions by the number density of GCs ($n_{\rm GC}$) and NSCs ($n_{\rm NSC}$), respectively. The number density of GCs in the local Universe varies in the range $[0.33,\,2.31]\,\rm Mpc^{-3}$~\cite{Rodriguez:2016kxx}, while for the density of NSCs we use $0.15\,\rm Mpc^{-3}$, using predictions from cosmological simulations (see the right inset of Fig.~12 in Ref.~\cite{Kritos:2024sgd}). Using these numbers, we predict $n_{\rm GC}^{\rm IMBH}=f_{\rm GC}^{\rm IMBH}n_{\rm GC}\simeq0.0066$--$0.046\,\rm Mpc^{-3}$ and $n_{\rm NSC}^{\rm IMBH}=f_{\rm NSC}^{\rm IMBH}n_{\rm NSC}\simeq0.054\,\rm Mpc^{-3}$. We interpret the number density of IMBHs in NSCs as the number density of NSCs in dwarf galaxies that host an IMBH. Moreover, $n_{\rm GC}^{\rm IMBH}$ is a lower bound estimate on the number density of wandering IMBHs. Despite the fact that $n_{\rm NSC}<n_{\rm GC}$, we predict that $n_{\rm NSC}^{\rm IMBH}>n_{\rm GC}^{\rm IMBH}$, as a consequence of the higher efficiency of NSCs in producing IMBHs. This is also consistent with the higher observed rate of nuclear tidal disruption events compared with off-nuclear events~\cite{2024MNRAS.530.3043P}. 
The uncertainty in these estimates depends on our knowledge of the precise number density of GCs and NSCs in the local Universe, as well as on the accuracy of our simulations in estimating the fraction of systems that contain a BH with mass $>100\,M_\odot$.

\begin{figure*}
    \centering
    \includegraphics[width=0.24\linewidth]{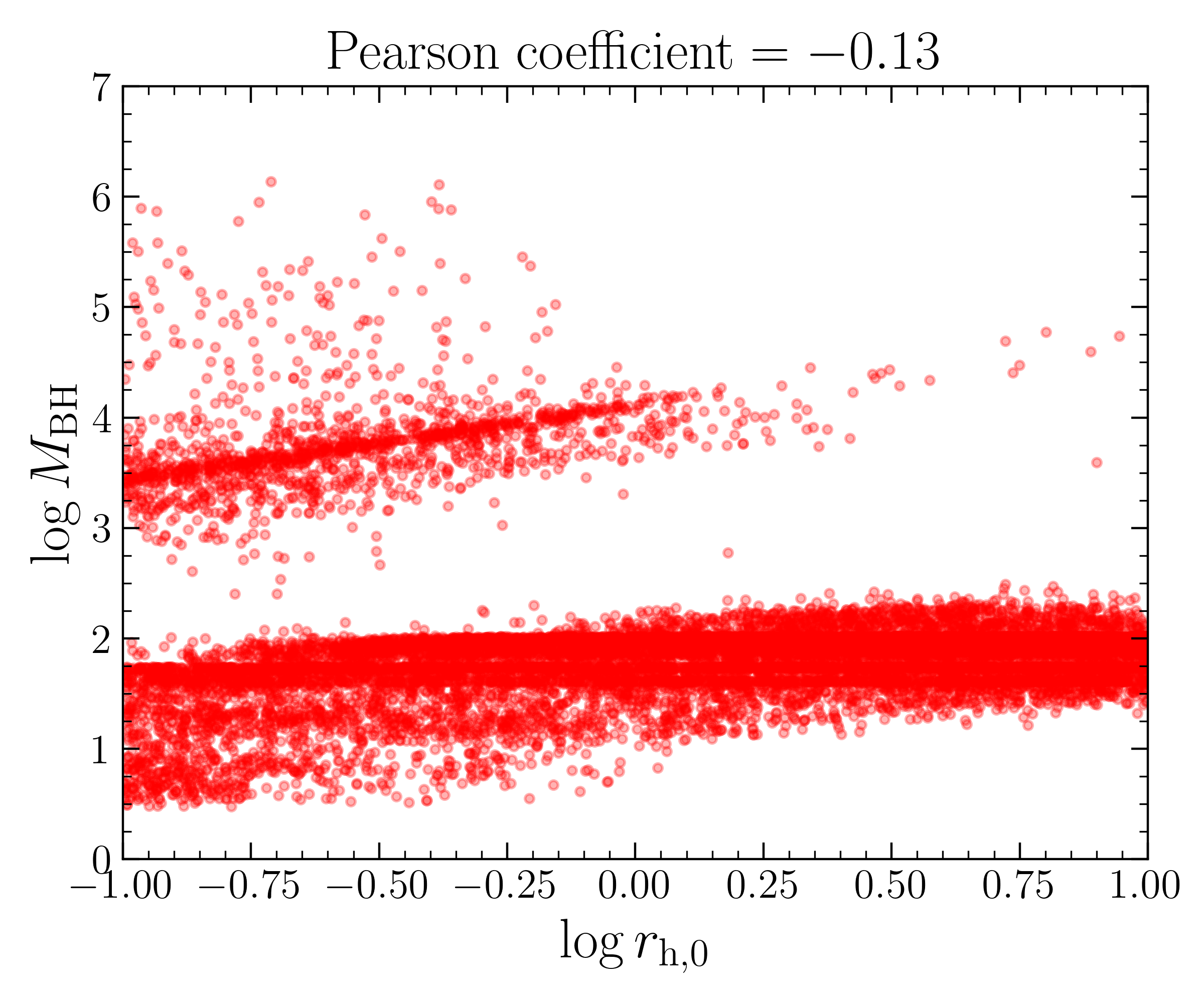}
    \includegraphics[width=0.24\linewidth]{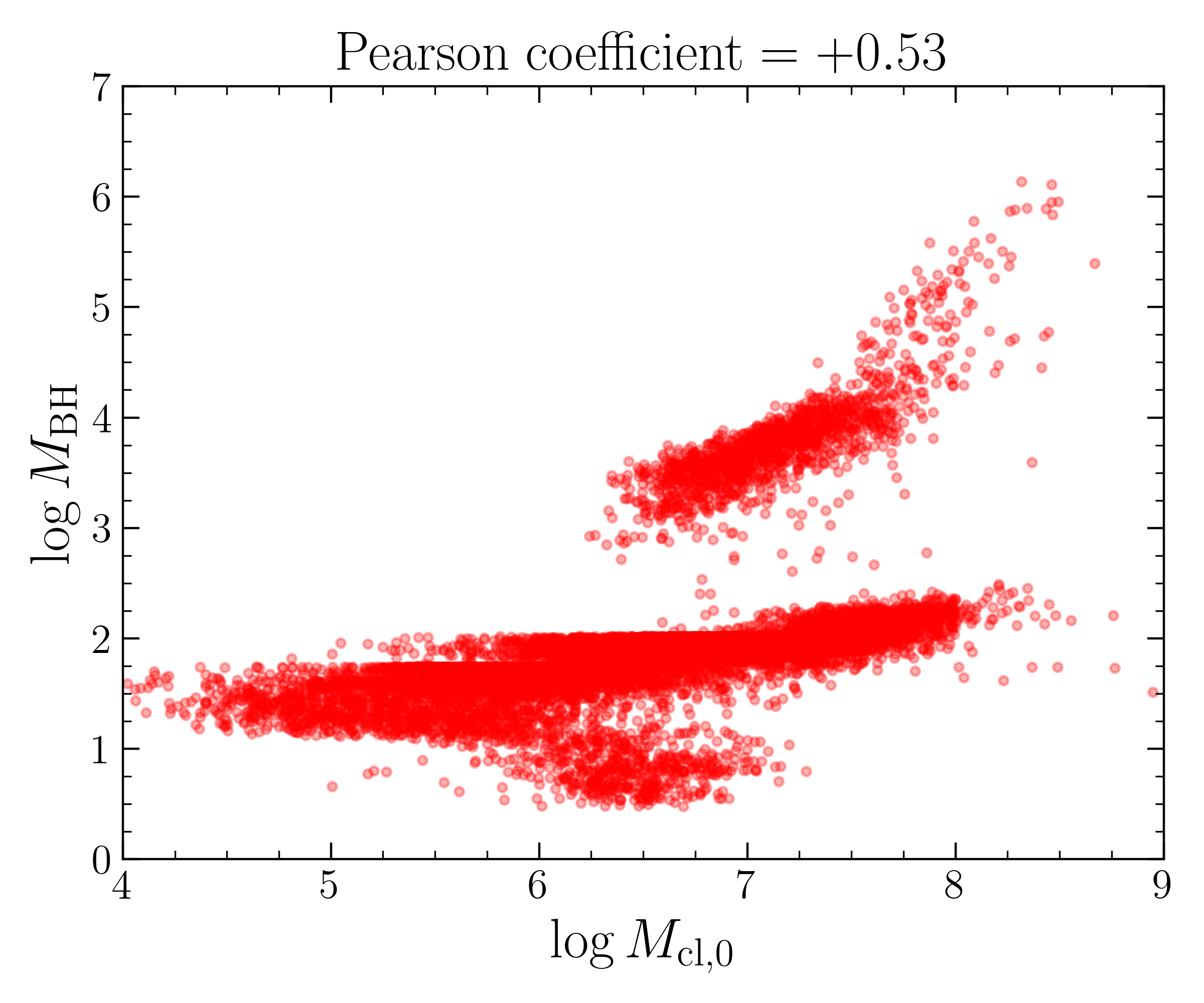}
    \includegraphics[width=0.24\linewidth]{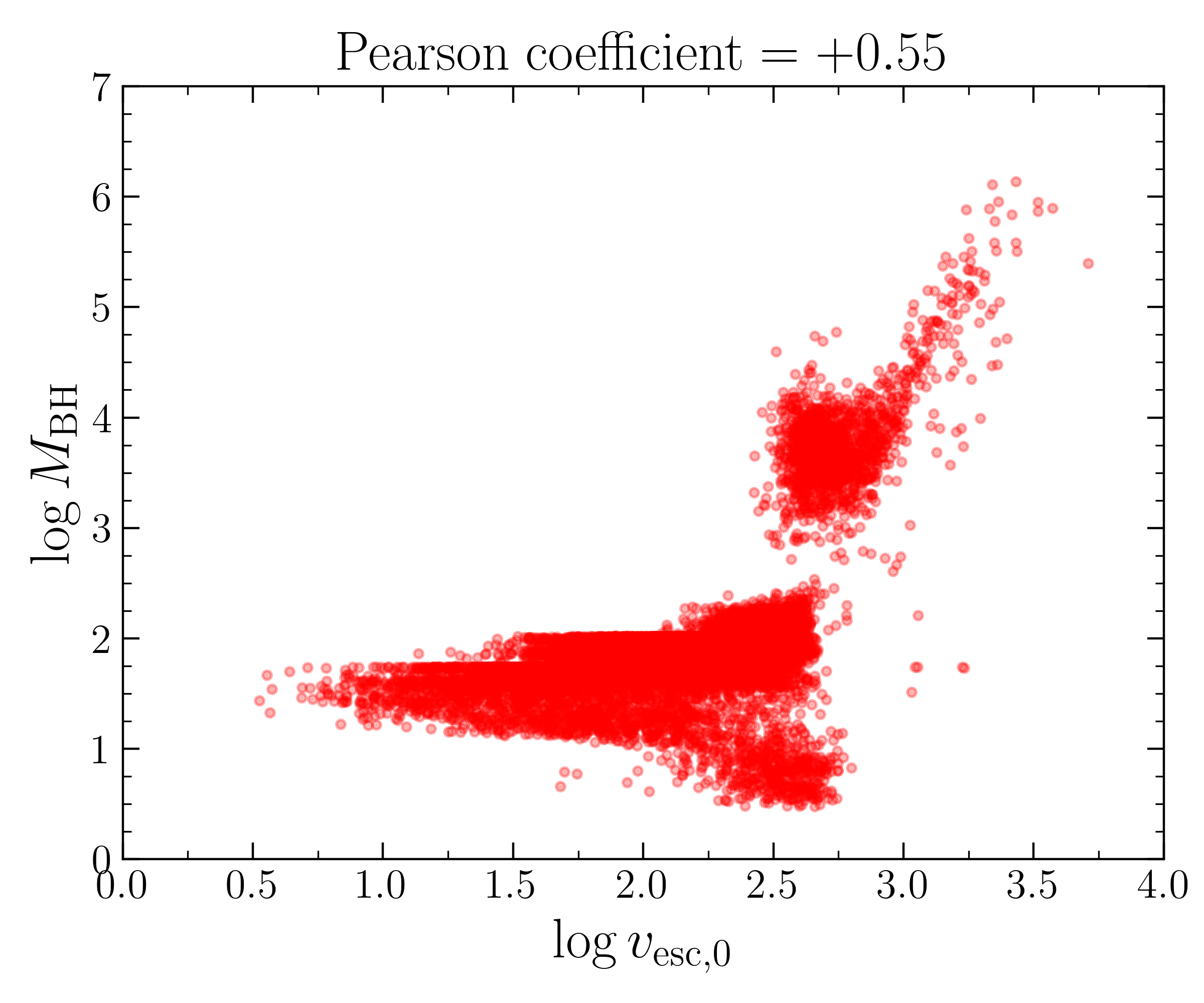}
    \includegraphics[width=0.24\linewidth]{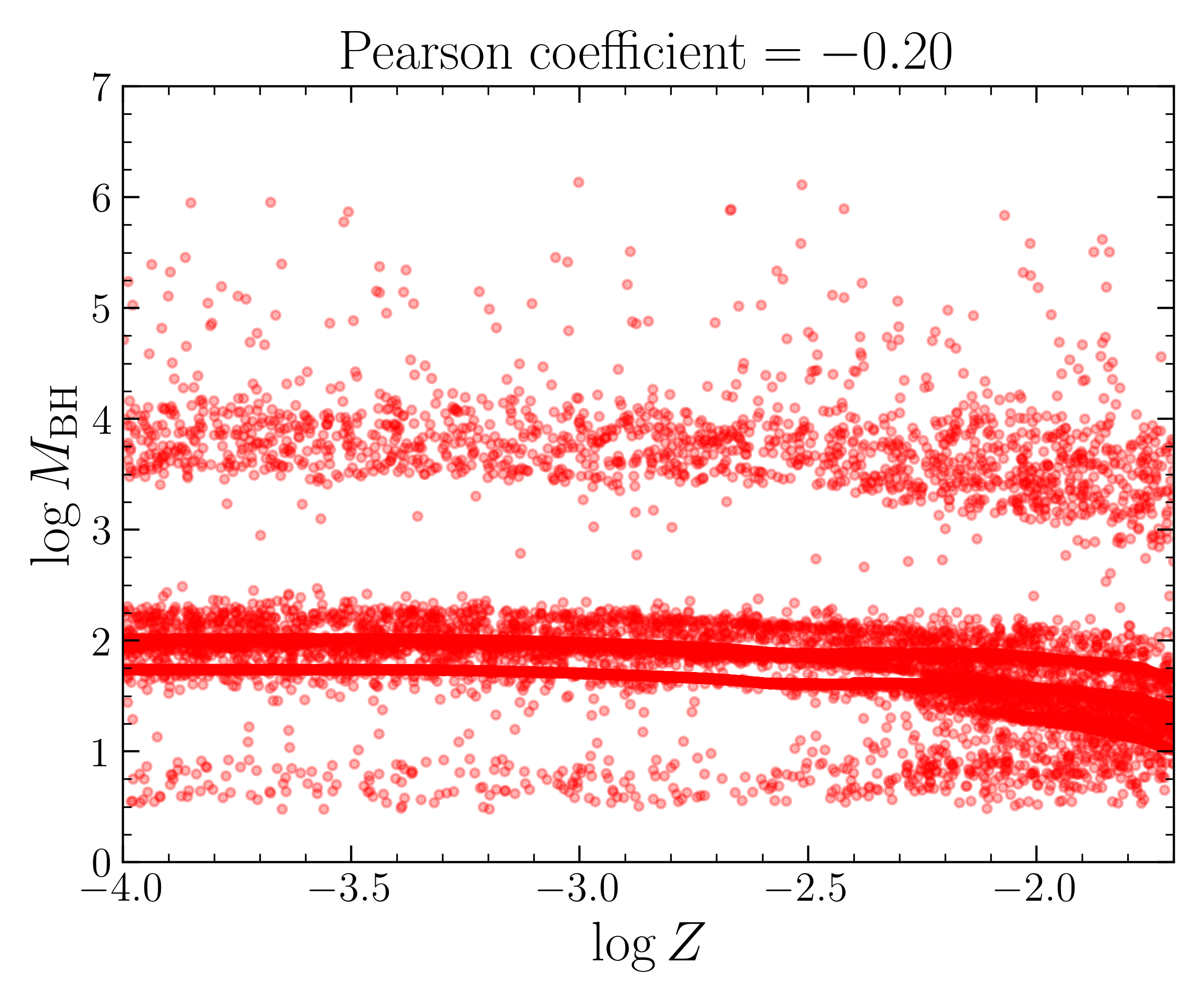}
    \includegraphics[width=0.24\linewidth]{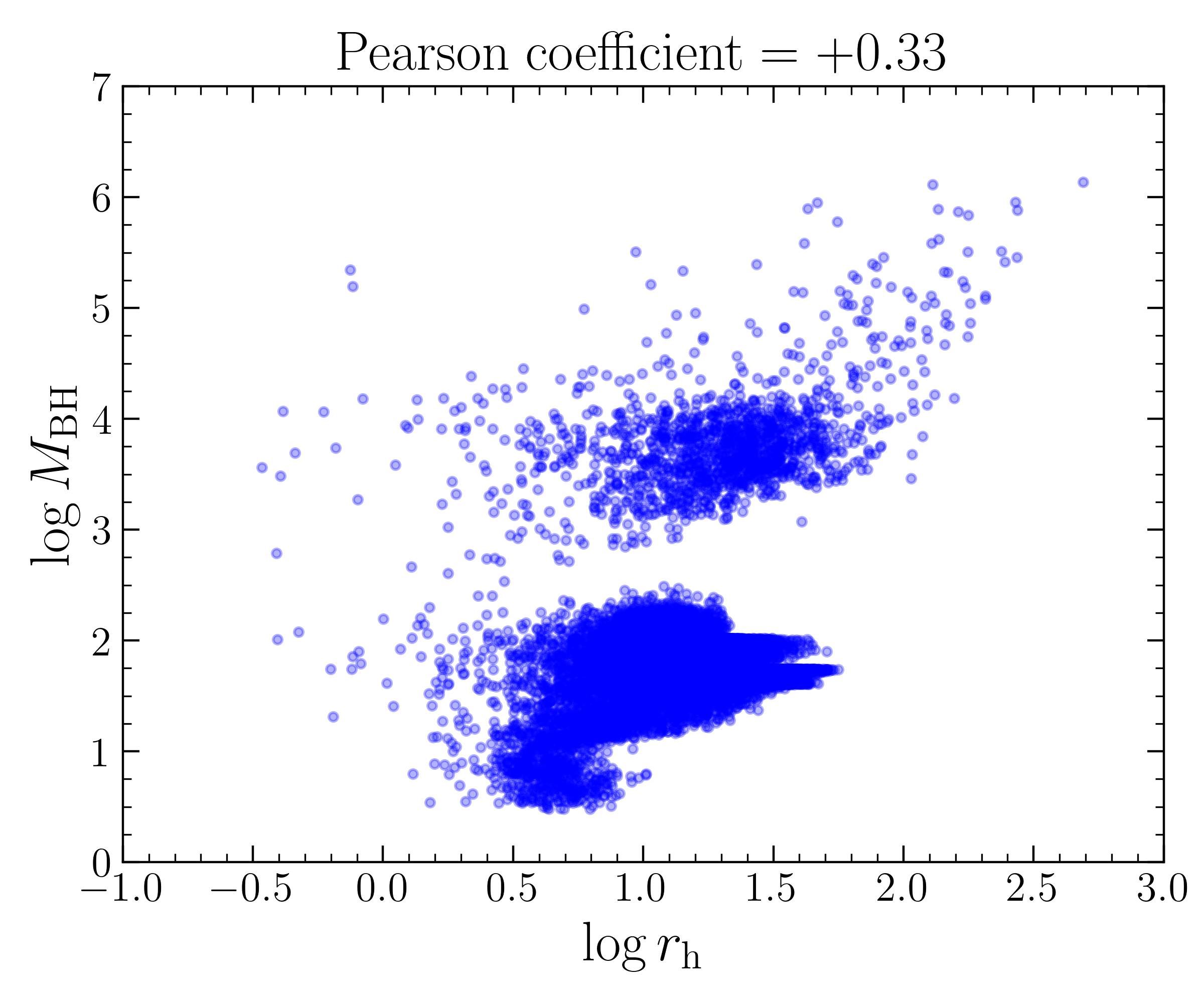}
    \includegraphics[width=0.24\linewidth]{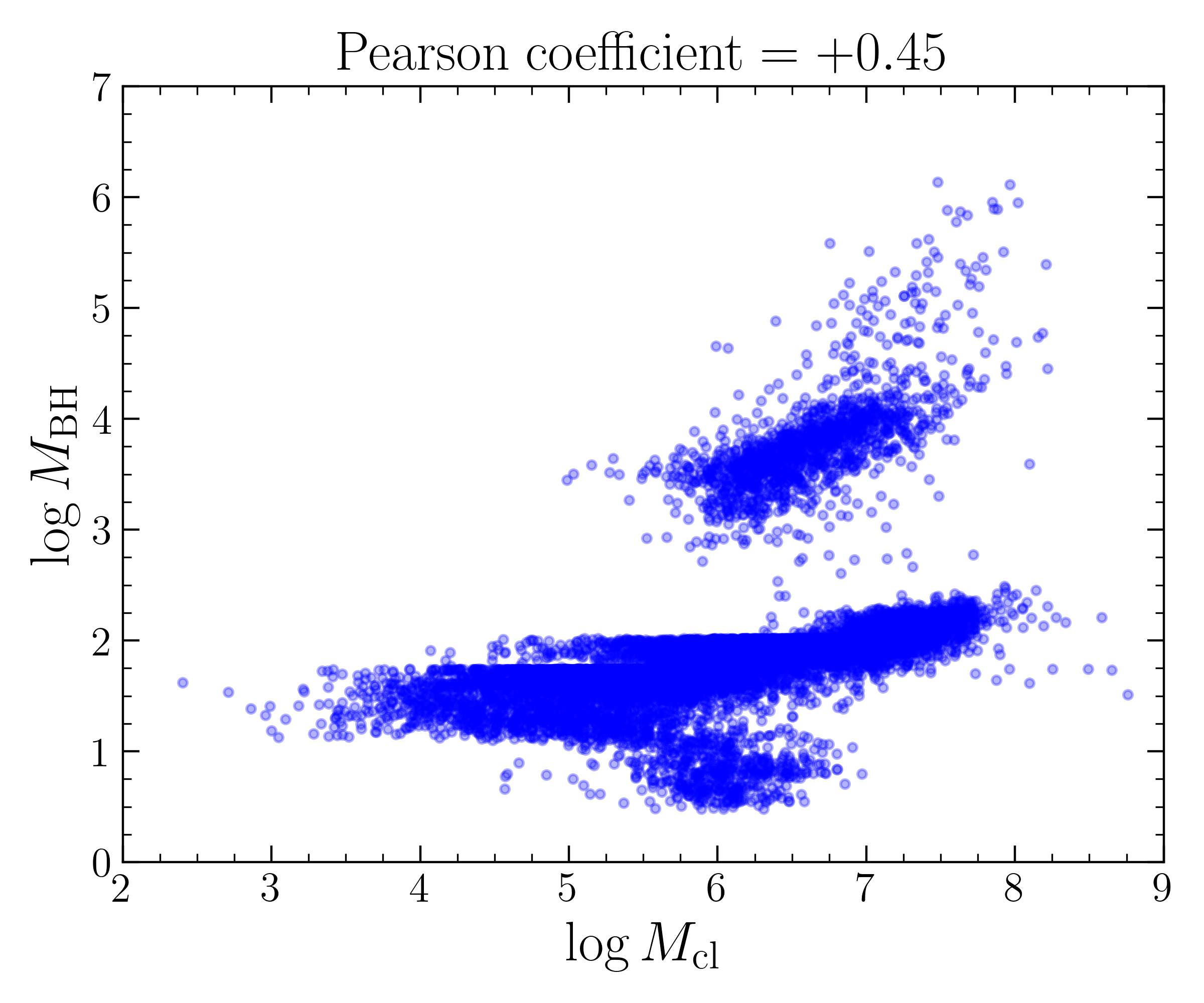}
    \includegraphics[width=0.24\linewidth]{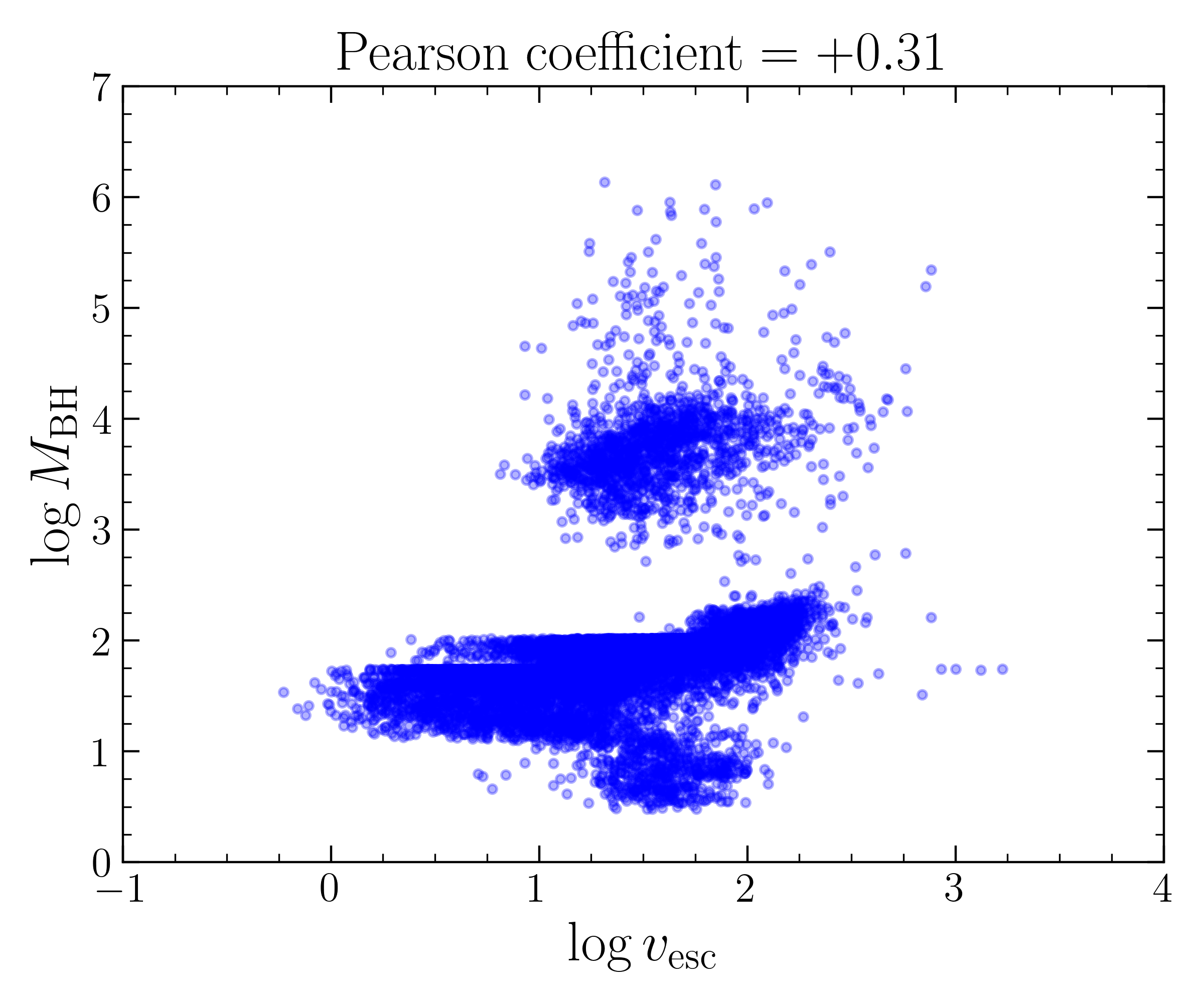}
    \includegraphics[width=0.24\linewidth]{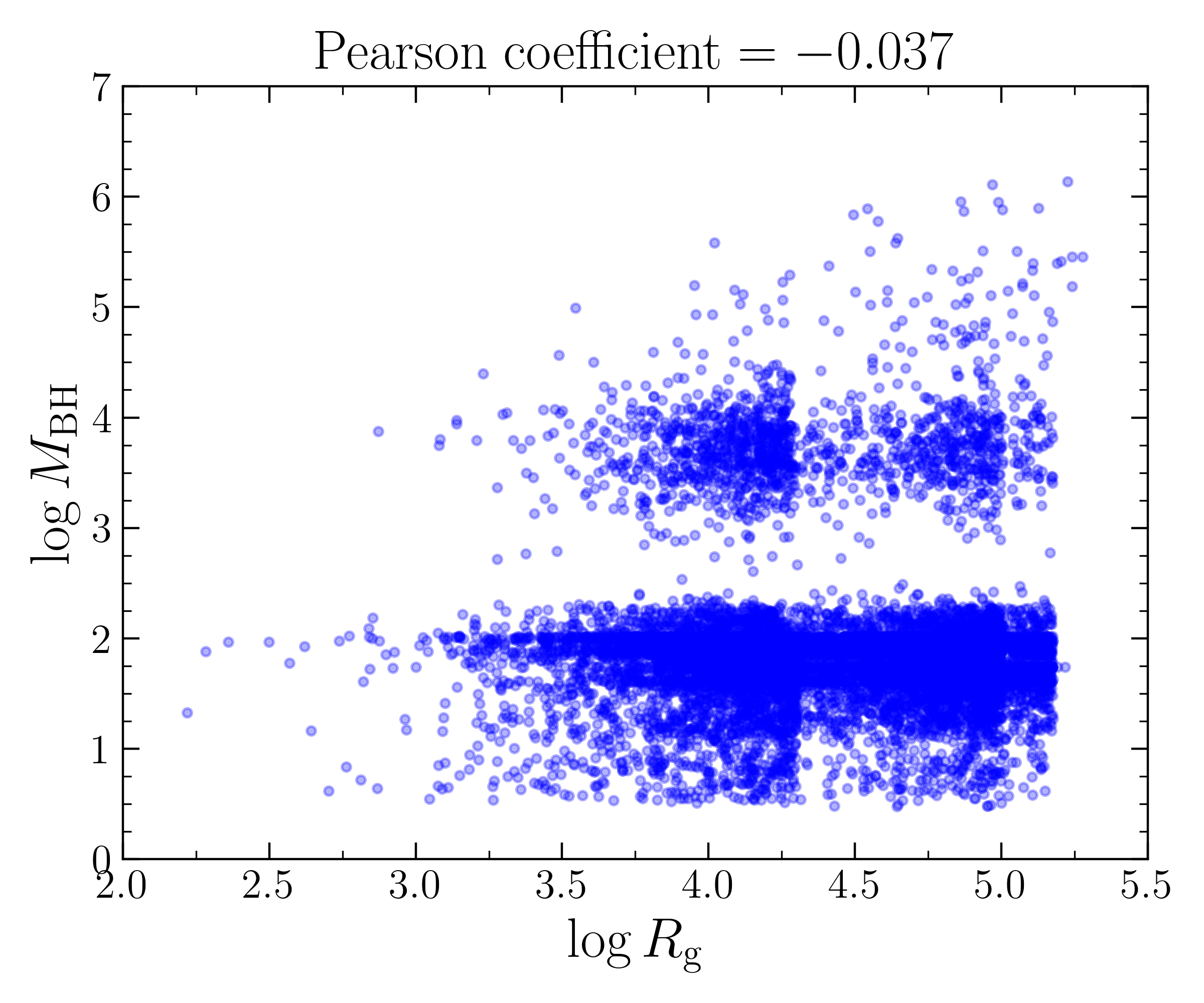}
    \caption[Pairwise correlations with final BH mass]{Logarithm of the heaviest BH mass ($M_\odot$) in the cluster at $z=0$ versus the logarithm of each cluster property considered in this work. We only show simulations that survived until $z=0$. The initial properties are labeled with subscript $0$, whereas the final properties are labeled without a subscript. The quantities shown from left to right in the top row are: initial half-mass radius (in $\rm pc$), initial cluster mass (in $M_\odot$), initial escape velocity (in $\rm km\,s^{-1}$), and absolute metallicity; and, from left to right in the bottom row: final half-mass radius (in $\rm pc$), final cluster mass (in $M_\odot$), final escape velocity (in $\rm km\,s^{-1}$), and final galactocentric radius (in $\rm pc$). We show the Pearson correlation coefficient at the top of each panel. The nonlinear dependence of $M_{\rm BH}$ on the cluster properties, together with the large scatter and multimodal relationships, makes the $M_{\rm BH}$ prediction problem challenging for traditional linear regression models.}
    \label{fig:pearson-coefficients}
\end{figure*}

\section{Machine learning models}

Our goal is to use the present-day properties of star clusters to infer the heaviest BH mass predicted by our {\sc Rapster}-based formation model. We therefore seek a mapping of the form
\begin{equation}
\{M_{\rm cl}, r_{\rm h}, R_{\rm g}, \ldots\} \to M_{\rm BH},
\end{equation}
where the inputs are the final, evolved properties of the clusters, as those can be probed in present-day observations. We also perform diagnostic tests on simulations in this section and use a combination of initial and final cluster properties to understand how much information is lost when only final observables are available.
In this section we define the machine learning models and evaluate their performance on simulated data. Applications to observed GCs, NSCs, and YMCs are deferred to Sec.~\ref{sec:Predicting-IMBH-masses}.

We show results from three classes of supervised regression models: random forest regressors (RFRs), neural networks (NNs), and symbolic regression (SR) later in this section. The motivation for training different algorithms is that each has its own strengths and weaknesses, and comparing their outputs helps us check the robustness of our predictions. In addition to these, we also explored normalizing flows for predicting the BH mass posterior. However, based on log-likelihood evaluations on the test set, we found their predictions to be comparatively worse than those from RFRs or NNs. This is likely due to the limited amount of training data in regions where the BH mass posterior becomes multimodal.

We convert all positive-valued parameters to log space to reduce the dynamic range of the data and make it easier to train our algorithms. We split the simulated dataset into training and testing sets with proportions 70\% and 30\%, respectively.
In the context of simulation-based inference, the distribution of simulated samples encodes prior information. Because our simulations were drawn from a broad sampling distribution rather than directly from the observed GC or NSC populations, we use importance weighting to train two population-specific models for GCs and NSCs. The simulation samples are reweighted separately for the GC and NSC cases before training the population-specific models.

\subsection{Random forest regressor (RFR)}
\label{sec:Random-forest-regressors}

\begin{figure*}
    \centering
    \includegraphics[width=0.49\linewidth]{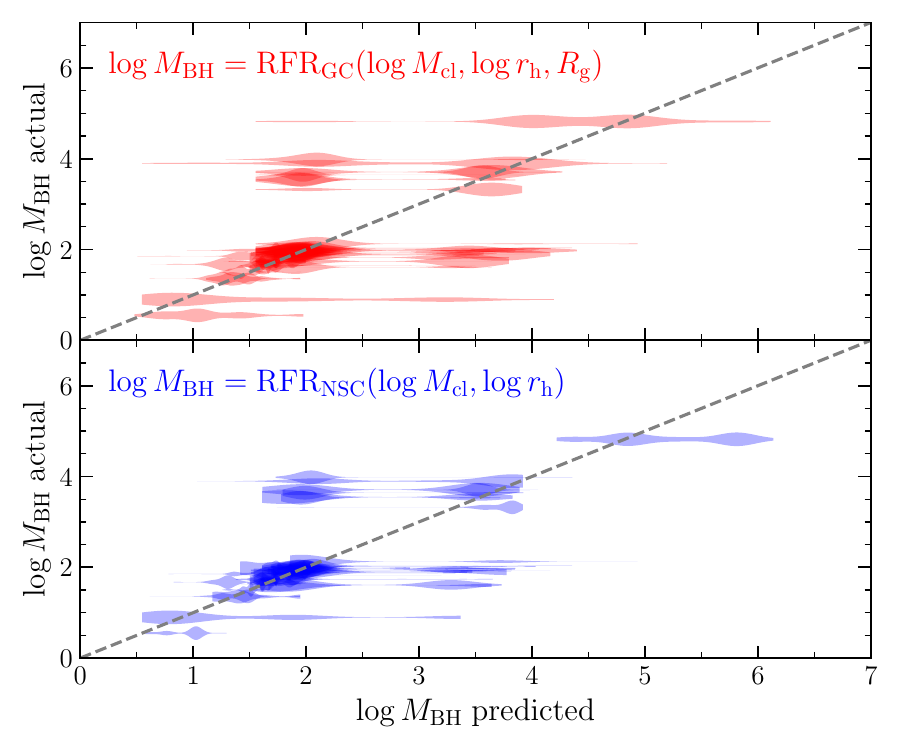}
    \includegraphics[width=0.49\linewidth]{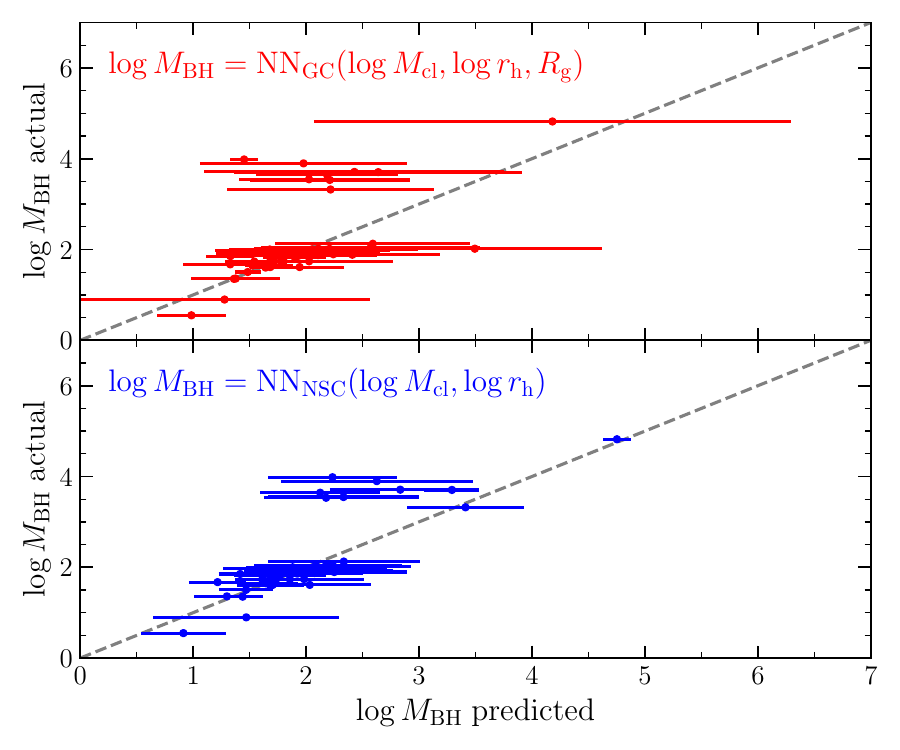}
    \caption{We use ML algorithms to predict the heaviest BH mass from the final ($z=0$) properties of star clusters, and we compare the true versus predicted BH masses using a random forest regressor (RFR, left column) and a deep neural network (NN, right column) from the simulation test set. Top and bottom rows show the results for the different models trained using the GC and NSC priors (see Fig.~\ref{fig:priors}). 
    % We show the first 50 points from the test set. 
    For the RFR models we show the distribution of predictions from the 100 individual decision trees, whereas the NN predicts a mean and standard deviation under a Gaussian likelihood. See the text for an explanation of the bimodality seen in the RFR prediction for some clusters.}
    \label{fig:pp-plots}
\end{figure*}

Random forests average the predictions of many decision trees trained on resampled data and randomized feature choices~\cite{Breiman2001RandomForests}. Related randomized-tree methods are widely used for nonlinear regression problems, because they can capture interactions among input variables without requiring a parametric functional form~\cite{Geurts2006ExtremelyRandomizedTrees}. They can also be more sample efficient than deep neural networks, requiring less training data to capture sharp features. We implement the RFRs using \texttt{scikit-learn}~\cite{Pedregosa2011ScikitLearn}.

In Fig.~\ref{fig:pearson-coefficients} we explore the pairwise correlations between the present-day (i.e., redshift $z=0$) BH mass and other cluster parameters. We restrict our interest to the subset of simulations that did not dissolve and survived until $z=0$. We find that metallicity and final galactocentric radius correlate only weakly with $M_{\rm BH}$. On the other hand, the initial escape velocity (or initial cluster mass) has the largest positive Pearson coefficient. This is expected for our hierarchical-merger channel, because the initial escape velocity controls the efficiency with which hierarchical-merger remnants are retained in the cluster following relativistic recoil (cf.~Fig.~\ref{fig:imbh-formation-probability}).
In fact, systems with $v_{\rm esc,0}\gtrsim500\,\rm km\, s^{-1}$ form an IMBH with a mass of at least $1000\,M_\odot$ within a few hundred $\rm Myr$ through repeated BH mergers in more than $90\%$ of our simulations. Furthermore, we observe that most clusters that have formed a $>1000\,M_\odot$ IMBH have expanded to a half-mass radius of $>10\,\rm pc$, and in some rare cases where a $\sim10^6\,M_\odot$ BH was formed, the radius expanded to $>300\,\rm pc$. This is also visible in the upper-left panel of Fig.~\ref{fig:priors}: only systems that formed a massive BH in their core reach cluster radii larger than $100\,\rm pc$.

We first train population-specific RFRs using only final cluster properties, matching the information available for observed clusters. Their performance on the simulation test set is shown in the left column of Fig.~\ref{fig:pp-plots}. For each test point, the violin shows the empirical distribution of predictions from the $100$ individual decision trees, each trained on a bootstrapped resample of the training set and a randomized subset of features. This non-parametric ensemble gives a useful measure of predictive uncertainty and can naturally expose multimodality in the predicted $M_{\rm BH}$, in contrast to the unimodal Gaussian summary returned by the NN. We compared this approach against a conditional normalizing flow trained on the same data and found that the RFR tree-based predictive distributions give a higher mean log probability on the held-out test set. We also verified that the RFR uncertainties are close to calibrated, in the sense that the empirical coverage of the predicted intervals tracks the nominal credible levels across the test set.

The bimodality visible for some clusters reflects a physical degeneracy in the mapping from final cluster properties to BH mass: clusters with similar present-day masses and radii may either host a relatively massive BH because they were initially much more compact and subsequently expanded due to energy injected by the growing BH, or host a lower-mass BH because they were already diffuse at birth. Since the final cluster properties alone do not fully distinguish these two evolutionary histories, the RFR can assign support to both outcomes. Knowledge of initial conditions can theoretically break this degeneracy, as discussed below and shown in Fig.~\ref{fig:rfr-all-model}.

The role of the simulation prior in this inference can be made explicit. At fixed final cluster properties $X_{\rm f}$, the posterior over the heaviest BH mass can be written schematically as a mixture over physically distinct evolutionary modes,
\begin{align}
    p(M_{\rm BH}\mid X_{\rm f}) &= \sum_k w_k(X_{\rm f})\,p_k(M_{\rm BH}\mid X_{\rm f}), \nonumber\\
    w_k(X_{\rm f}) &= \frac{I_k(X_{\rm f})}{\sum_j I_j(X_{\rm f})},\\
    I_k(X_{\rm f}) &= \int_{\mathcal{R}_k}\!dX_0\,p(X_{\rm f}\mid X_0)\,p(X_0).
\end{align}
Here the regions $\mathcal{R}_k$ in initial-condition space correspond to distinct evolutionary modes, such as a hierarchical-merger branch with a heavy retained remnant and a no-IMBH branch in which mergers were ejected or never assembled. When a single mode dominates, $w_k\to1$, and the prediction is insensitive to the choice of $p(X_0)$, the prior over initial cluster conditions. When two or more modes contribute appreciably, the mixture weights are explicit integrals of $p(X_0)$ over the corresponding regions of initial-condition space. Reweighting the simulation initial-condition prior, $p_{\rm simul}(X_{0})$, toward the GC or NSC target population, as described in Sec.~\ref{sec:cluster_suite} and Fig.~\ref{fig:priors}, leaves the locations of the modes largely unchanged but rescales their amplitudes. This is why the astrophysical prior matters most in the multimodal regime exposed by the RFR.

\subsection{Deep neural networks}
\label{sec:Deep-neural-networks}

We also train NNs to predict the heaviest BH mass from the final cluster properties. Unlike the RFR, the NN used here assumes the predictive distribution for the BH mass to be a Gaussian in log mass. The model predicts both a mean $\hat{y}^{(i)}$ and a standard deviation $\sigma^{(i)}$ for the target $y^{(i)}=\log M_{\rm BH}^{(i)}$. We train it by minimizing the weighted negative log likelihood of this Gaussian predictive distribution,
\begin{align}
    \mathcal{L}_{\rm NN} = \sum_{i} \frac{1}{2} w_i \left[ \frac{\big(y^{(i)} - \hat{y}^{(i)}\big)^2}{(\sigma^{(i)})^2 + \varepsilon} + \ln\left(2\pi \big[(\sigma^{(i)})^2+\varepsilon\big]\right) \right].
\end{align}
We use $\varepsilon = 0.001$ to prevent numerical instabilities from vanishing predicted variances. The weights for the GC model are defined as $w_i = p_{\rm GC}(X_i)/p_{\rm simul}(X_i)$, with an analogous expression for NSCs. The right column of Fig.~\ref{fig:pp-plots} shows the corresponding NN performance on the simulation test set. The RFR can expose multimodality through the spread of its individual-tree predictions, whereas the NN provides a compact Gaussian summary of the predictive distribution.

\begin{figure}
    \centering
    \includegraphics[width=\linewidth]{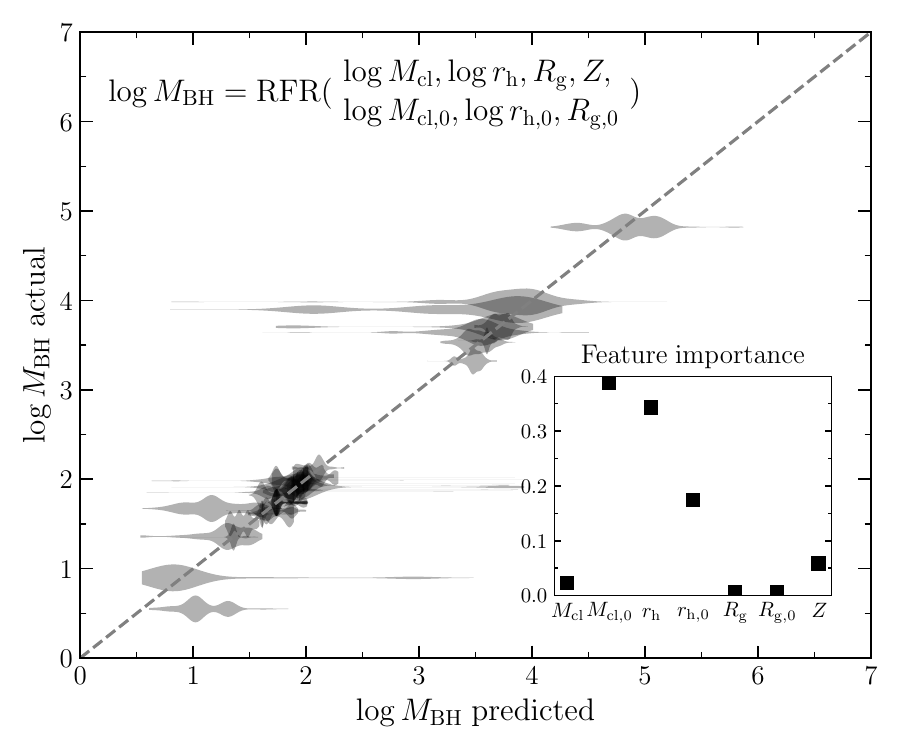}
    \caption{Same as the left panel of Fig.~\ref{fig:pp-plots}, but for the RFR model that was trained on both the final and initial cluster properties. Even though the initial conditions are not directly available for observed clusters, we include them in the training set to gain insight into the BH formation process in a cluster. The inset at the lower right shows the random-forest feature importances of the input parameters. Note that the multimodality seen in Fig.~\ref{fig:pp-plots} is no longer present. Knowing both the initial and final cluster properties constrains the effect of the BH on cluster evolution, and therefore the predictions of $M_{\rm BH}$ become more accurate.}
    \label{fig:rfr-all-model}
\end{figure}

As a diagnostic upper baseline, we also train an RFR to predict $\log M_{\rm BH}$ from all available initial and final cluster parameters ($\log M_{\rm cl}$, $\log M_{\rm cl,0}$, $\log r_{\rm h}$, $\log r_{\rm h,0}$, $R_{\rm g}$, $R_{\rm g,0}$, $\log Z$). This model uses information that would generally not be available for observed clusters, but it is useful for identifying how much predictive information is present in the simulations and which quantities are most informative. The resulting performance is shown in Fig.~\ref{fig:rfr-all-model}. Compared to the final-property RFR in Fig.~\ref{fig:pp-plots}, the multimodality is strongly reduced and the predictions are more accurate, because the initial conditions help break degeneracies in the subsequent cluster evolution.

\subsection{Symbolic regression}
\label{sec:Symbolic-regression}

Motivated by the ``M-$\sigma$'' relation that relates BH mass to the properties of the host stellar component, we also search for compact analytic approximations to the relation between the heaviest BH mass and cluster properties.
Symbolic regression has a long history as a tool for distilling analytic relations from data~\cite{SchLip09}, with recent physics-inspired and machine-learning-based developments substantially broadening its applicability~\cite{WuTeg18,UdrTeg20,CraSan20,KimLu19}.
We use the Python package \texttt{PySR}~\cite{pysr} to perform symbolic regression on our data.
Symbolic regression and related machine-learning approaches have previously been used to augment astrophysical scaling relations and reduce their scatter~\cite{Wadekar:2022cyw,WadThi22b}; more broadly, such techniques have proven useful for modeling nonlinear astrophysical and cosmological relations~\cite{astroexample2,astroexample3,Sha21,Lem22,VilAngGen20,DelWad22,ThiWad22}. In gravitational-wave astronomy, symbolic regression has been used for rapid inference and population-model emulation~\cite{Wong:2019uni,Cuoco:2020ogp,Chatterjee:2026tpy}, and also to obtain compact analytic waveform models and gravitational-wave-informed neutron-star relations~\cite{Tiglio:2019yec,Bejger:2025hbq,Islam:2026zob}.

From the RFR feature importances of the input parameters (shown in the inset of Fig.~\ref{fig:rfr-all-model}), we find that the initial and final galactocentric radii are the least informative parameters for predicting $M_{\rm BH}$. To obtain an analytic expression with low complexity, we restrict the symbolic-regression search to the most informative variables. Using $10\%$ feature importance as a heuristic threshold suggests a functional form $M_{\rm BH}=f(M_{\rm cl,0}, r_{\rm h}, r_{\rm h,0})$. While lower metallicity environments lead to the formation of heavier first-generation BHs from massive stars, we find only a weak correlation between metallicity ($Z$) and the final IMBH mass. To further reduce the input dimensionality, and motivated by Fig.~\ref{fig:pearson-coefficients}, we combine the initial mass and initial half-mass radius into the initial escape velocity, $v_{\rm esc,0}=2\sqrt{0.4GM_{\rm cl,0}r_{\rm h,0}^{-1}}$. Thus, we search for a lower-complexity relation of the form $M_{\rm BH}=g(v_{\rm esc,0}, r_{\rm h})$.

We train the symbolic-regression model with the following operators: $\exp(x)\equiv e^x$, ${\rm inv}(x)\equiv x^{-1}$, and ${\rm sqrt}(x)\equiv\sqrt{x}$. \texttt{PySR} returns a Pareto front of candidate expressions that trade off predictive loss against algebraic complexity. We select a low-complexity expression with good predictive performance, rather than the absolute lowest-loss expression, to avoid overfitting and preserve interpretability. The resulting fit for the final mass of the heaviest BH in the cluster is
\begin{align}
    M_{\rm BH}^{\rm SR} \simeq r_{\rm h}^{0.1968}\exp\left[\left(\frac{v_{\rm esc,0}}{6.186}\right)^{0.4343}\right]\,,
\end{align}
where $v_{\rm esc,0}$ is the initial escape velocity (in $\rm km\,s^{-1}$), and $r_{\rm h}$ is the final half-mass radius (in $\rm pc$). The positive dependence on the final radius reflects the fact that systems that form an IMBH generate more energy and expand by a larger factor. The exponential dependence on the initial escape velocity captures the sharp suppression of hierarchical growth when the escape velocity is too small to retain merger remnants (we find the transition to occur at $\sim300\,\rm km\,s^{-1}$, see also Fig.~\ref{fig:imbh-formation-probability}).

\begin{figure}
    \centering
    \includegraphics[width=\linewidth]{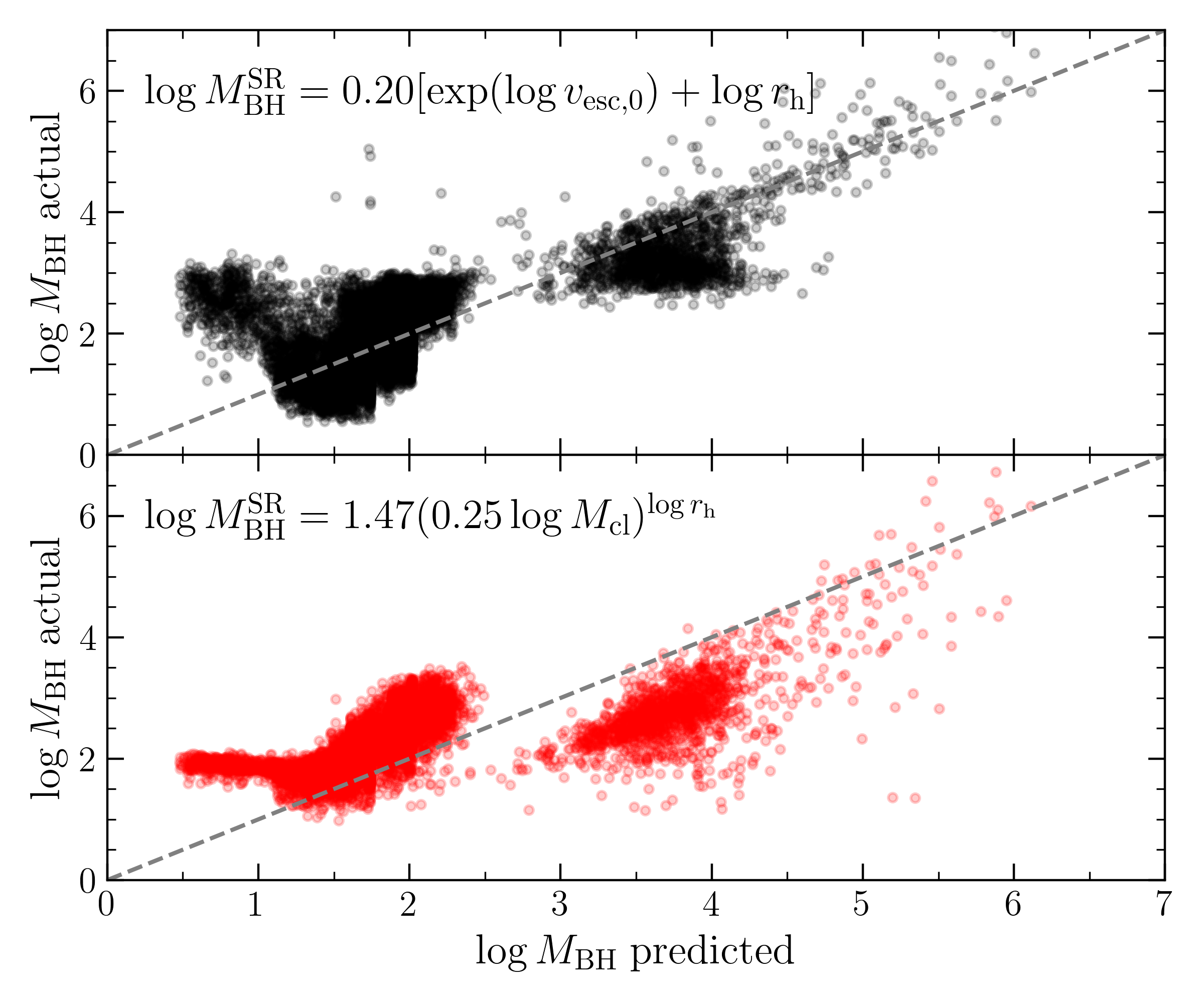}
    \caption{Same as in Fig.~\ref{fig:rfr-all-model} but for the predictions of the symbolic regression formulas, fitting the BH mass to $(v_{\rm esc,0}, r_{\rm h})$ (upper panel) and to $(M_{\rm cl},r_{\rm h})$ (lower panel). Although the prediction accuracy of symbolic regression is lower than that of the RFR, it provides interpretable equations that can give insight into physical processes in clusters. All logarithms are base 10.}
    \label{fig:symbolic-regression}
\end{figure}

For observed clusters, the initial conditions are generally unknown. We therefore also fit a symbolic-regression model using only the final cluster properties. As expected, this restriction increases the prediction loss. In terms of the final cluster mass and final radius, the simplest low-loss expression found by symbolic regression is
\begin{align}
    \log M_{\rm BH}^{\rm SR} = 1.468(0.2475 \log M_{\rm cl}) ^{\log r_{\rm h}}\,.
\end{align}
Despite its unusual form, this expression correctly captures the qualitative trend that heavier BH masses correspond to larger $M_{\rm cl}$ and $r_{\rm h}$. Since $0.2475\log M_{\rm cl} < 1$ for typical cluster masses, a larger $r_{\rm h}$ acting as the exponent yields a net positive correlation with both cluster mass and radius.
This equation has lower physical interpretability than the relation involving $v_{\rm esc,0}$, but it captures the qualitative trend that heavier and more expanded clusters tend to host heavier BHs: see Fig.~\ref{fig:symbolic-regression}.

These symbolic expressions are useful because they are compact and interpretable, but they provide point predictions and do not by themselves quantify predictive uncertainty. Comparing with RFR results, we see that RFRs trained on the same inputs produce more accurate predictions and uncertainty estimates, but at the cost of being less interpretable because they do not provide a simple analytic form.

\begin{figure*}
    \centering
    \includegraphics[width=\linewidth]{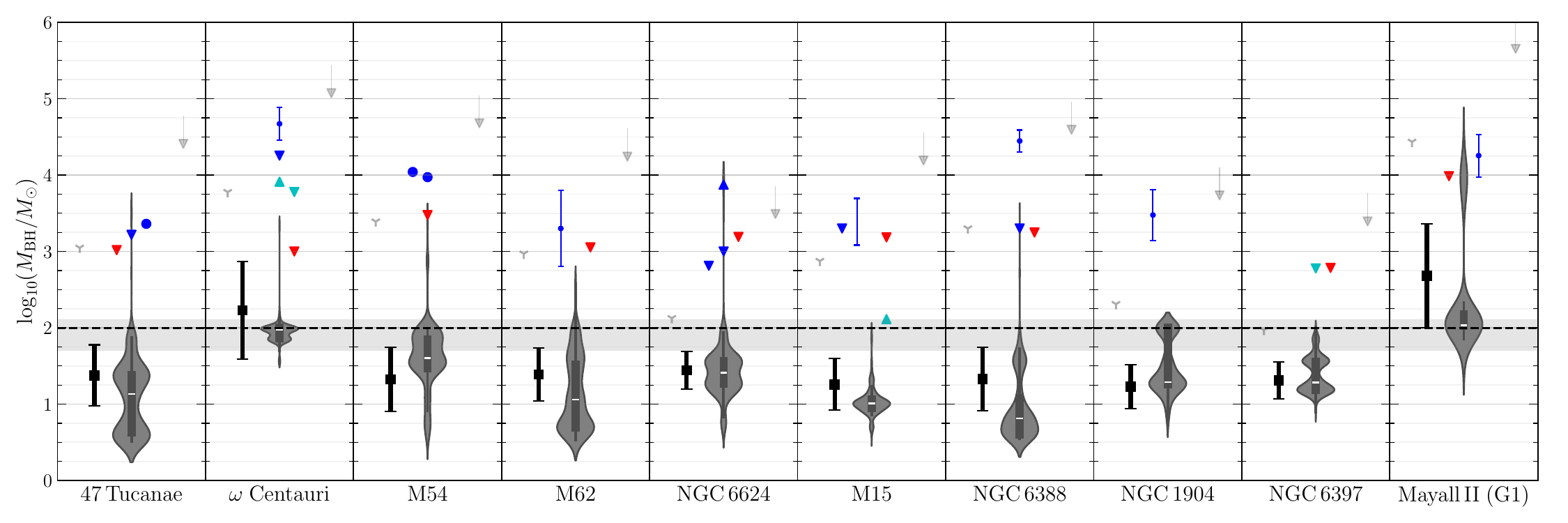}
    \caption{Predicted maximum BH mass for a subset of GCs. The black squares with error bars correspond to predictions from the NN, and the gray violin plots correspond to estimates from the trained RFR. The light gray downward triangles correspond to theoretical upper limits on the BH mass that can form from repeated BH mergers~\cite{Antonini:2018auk}, assuming a half-mass radius expansion factor of 10 and negligible cluster mass loss. The vertical light gray arrows show upper limits on the total BH mass that could be retained initially in the cluster, assuming a high (low) metallicity value corresponding to the arrow's lower (higher) end. The remaining data points are observational: triangles pointing downward (upward) indicate upper (lower) limits, while points with or without error bars correspond to measurements. Blue data are based on dynamical models from Table~3 of~\cite{2020ARA&A..58..257G}, and the red data points are limits from~\cite{Tremou:2018rvq}. Moreover, the cyan limit for $\omega$ Centauri is from~\cite{2024Natur.631..285H}, for M15 from~\cite{Huang:2024gpv}, and for NGC 6397 from~\cite{2016A&A...588A.149K}. Finally, the horizontal black dashed line arbitrarily defines the lower edge of the IMBH range, and the light gray band is the mass range corresponding to the upper mass gap $\sim[50,\,130]\,M_\odot$.}
    \label{fig:predictions_for_GCs}
\end{figure*}

\section{Predicting the heaviest BH mass of observed clusters}
\label{sec:Predicting-IMBH-masses}

In this section we apply the models trained in Sec.~\ref{sec:Random-forest-regressors}--\ref{sec:Symbolic-regression} to observed star clusters. We focus on predictions from the final-property RFR and NN models, because these use quantities available for observed systems, and compare them with analytic upper limits for hierarchical growth where applicable.

\subsection{Globular clusters}
\label{sec:Globular-clusters}

An upper limit on the BH mass assembled through repeated mergers was derived in Eq.~(21) of Ref.~\cite{Antonini:2018auk}. We rewrite it here in terms of the initial mass and half-mass radius as
\begin{align}
    m_{\rm BH}^{\rm max} &= m_{\rm BH,0} \nonumber\\&+ 70\,M_\odot \left(\frac{M_{\rm cl,0}}{10^5\,M_\odot}\right)^{51/42} \left(\frac{1\,\rm pc}{r_{\rm h,0}}\right)^{15/42}\left(\frac{m_2}{10\,M_\odot}\right)^{2/7},
    \label{eq:max-BH-mass-Antonini}
\end{align}
where $m_2$ is the typical secondary BH mass with which the primary BH is merging, and roughly equal to $\overline{m}_{\rm BH}$ in our simulations.
In applying this formula to the GC data, we assume that clusters have expanded by a factor of 10, as a ballpark estimate.
This estimate in Eq.~\eqref{eq:max-BH-mass-Antonini} overestimates the maximum BH mass because it does not account for ejections of BHs from the BH subsystem, so there are always enough BHs available in the cluster. The merger rate is reduced, and the mass of the runaway BH plateaus at this maximum as a consequence of cluster expansion. The ejection of BHs reduces the BH-BH merger rate, lowering the maximum BH mass value.

The ultimate maximum value of the BH mass that could theoretically be assembled via repeated BH mergers is assumed to be formed through the merger of all BHs in the cluster. 
This estimate relies on computing the fraction of the cluster's mass locked in BHs. We may compute this fraction given an initial mass function and a stellar evolution model.
Assuming a Kroupa initial mass function with high mass index $\alpha=-2.3$~\cite{Kroupa:2002ky}, we have computed this fraction to range between about 3\% and 7\% for metallicity $Z=0.01$ and $Z=0.0001$, respectively.
To compute these fractions, we calculated remnant masses given the stellar masses and the metallicity using the stellar evolution models {\tt SEVN}~\cite{Spera:2017fyx} and {\tt SSE}~\cite{Fryer:2011cx}.
These fractions do not depend on the cluster mass and assume that BHs receive no kicks at birth.
This upper limit is never obtained in practice, because it neglects cluster evolution and the ejection of BHs during cluster evolution and at BH formation. In particular, as a hard binary BH hardens to its merger, it ejects, on average, a few BHs from the cluster during energetic binary-single interactions in the core.
Moreover, the fraction of the cluster's mass in BH remnants is initially smaller for low cluster initial escape velocities, provided that natal kicks are strong and accounted for.

In Fig.~\ref{fig:predictions_for_GCs} we show our predictions for a subset of GCs with previous claims or constraints on a central IMBH. We also include the heaviest GC of the Andromeda galaxy, Mayall II (G1).

\subsubsection{47 Tucanae}

Claims for an IMBH in the GC 47 Tucanae have been reported based on pulsar acceleration measurements, suggesting a central mass in the range $\sim 1450$-$3800\,M_\odot$~\cite{2017Natur.542..203K}. However, subsequent observational constraints have significantly tightened the allowed mass range. Radio observations place upper limits on the IMBH mass of $M_{\rm IMBH} < 2060\,M_\odot$ in 47 Tucanae and NGC 6397, with more stringent constraints from~\cite{Tremou:2018rvq} yielding $M_{\rm IMBH} < 1040\,M_\odot$ for 47 Tucanae. Independent limits from deep, high-resolution X-ray imaging with the Chandra X-ray Observatory further restrict the mass to $M_{\rm IMBH} < 470\,M_\odot$, with detected X-ray sources consistent with millisecond pulsars, flaring binary stars, and accreting stellar remnants rather than a central IMBH~\cite{2005ApJ...625..796H}.

Within our framework, the RFR gives a model-based score $p(M_{\rm IMBH} > 1000\,M_\odot) \approx 0.4\%$, indicating that the presence of a high-mass IMBH formed through repeated mergers is strongly disfavored. Both the NN and RFR models instead suggest that the system is more likely devoid of BHs more massive than $100\,M_\odot$. Nevertheless, we find non-negligible support at the $\sim 10\%$ level for the presence of a BH with mass $> 60\,M_\odot$, implying that 47 Tucanae may have hosted merger events within the upper mass gap. A recent study~\cite{Chattopadhyay:2026siv} performed semianalytical cluster simulations using the \texttt{cBHBd} code, and also predicts the formation of a massive BH in 47 Tucanae.

\subsubsection{$\omega$~Centauri}

The heaviest GC in the Milky Way is $\omega$~Centauri, with a mass of about $3.9\times10^6\,M_\odot$~\cite{Baumgardt:2018pyl}.
It was suggested that $\omega$~Centauri hosts a $(4$--$5)\times10^4\,M_\odot$ IMBH based on a rise in the core velocity dispersion~\cite{Noyola:2008kt,2010ApJ...719L..60N}. 
However, anisotropic modeling limits the IMBH mass to $<1.8\times10^4\,M_\odot$ at $3\sigma$~\cite{2010ApJ...710.1063V}.
On the other hand, the existence of a $4\times10^4\,M_\odot$ IMBH is supported by N-body simulations based on properties of the cluster profile~\cite{2017MNRAS.464.2174B}.
Furthermore, the absence of a strong central radio source suggests an upper limit on the IMBH mass of $1040\,M_\odot$, assuming an extrapolation of the fundamental-plane relation into the IMBH regime~\cite{Tremou:2018rvq}.

The recent observation of a few fast-moving stars in the center of $\omega$~Centauri requires the existence of a central dark mass of at least $8200\,M_\odot$~\cite{2024Natur.631..285H}.
Nevertheless, dynamical modeling taking into account both stellar kinematics and line-of-sight accelerations of millisecond pulsars prefers an extended dark mass, placing a $3\sigma$ upper limit of $6\times10^3\,M_\odot$ on the IMBH mass~\cite{2025A&A...693A.104B}. This dark mass component may instead consist of stellar remnants, but it is more concentrated than stars.

The RFR model favors a heaviest retained BH near $100\,M_\odot$ in $\omega$~Centauri, within the upper-mass gap. The model-based probability that a $\sim10^3\,M_\odot$ IMBH formed through repeated mergers is $\sim0.3$\%.
Similarly, the NN predicts an IMBH mass in the hundreds of solar masses, but not larger than $10^3\,M_\odot$.
While an IMBH may be present in $\omega$~Centauri, at the time of writing, direct measurement of the accelerations of the fast-moving stars is required to prove the IMBH scenario unambiguously, in a way similar to how Sagittarius A$^*$ was discovered.
Moreover, our results suggest that such an IMBH could not have formed solely through repeated mergers, indicating that other growth channels would be required if the IMBH's mass is above $1000\,M_\odot$. 

\subsubsection{Messier 15}

The increase in the central velocity dispersion of Messier 15 (M15) indicates the presence of a central dark mass, which, if due to a point IMBH, may be $(3.9\pm2.2)\times10^3\,M_\odot$~\cite{2002AJ....124.3270G}.
However, the observed kinematic data of M15 can be reproduced even without the existence of an IMBH~\cite{Baumgardt:2002rc}.
Again, stellar remnants could contribute the majority of the dark mass, although the presence of a $500\,M_\odot$--$1000\,M_\odot$ IMBH cannot be excluded.
In addition, the non-detection of radio emission from the core of M15 limits the mass of any putative IMBH to $<1530\,M_\odot$ based on the fundamental-plane relation extrapolated to the IMBH regime~\cite{Tremou:2018rvq}.

Our RFR and NN predictions for M15 are consistent with these constraints, and the models prefer that the heaviest BH in the system be a few tens of solar masses or less, below the upper-mass gap. The RFR-based score for M15 hosting a $\sim100\,M_\odot$ BH is $\sim1\%$.
Moreover, Eq.~\eqref{eq:max-BH-mass-Antonini} constrains the mass of a putative IMBH formed via repeated mergers to be $\lesssim800\,M_\odot$.

During the close interaction of a stellar binary with an IMBH, a hyper-velocity star may be produced during binary separation through the Hills mechanism~\cite{1988Natur.331..687H}.
Such a fast-moving star was detected in the galactic halo, with an orbit that previously intersected M15's trajectory. The presence of an IMBH in this cluster was suggested to cause the ejection~\cite{Huang:2024gpv}.
Given the uncertainty in the binary-star parameters, such as semimajor axis, the data do not strongly constrain the IMBH mass, except that it should exceed $100\,M_\odot$ at 98\% confidence. If an IMBH of a few hundred solar masses does exist in the core of M15, then our results suggest that it is unlikely to have formed through repeated BH mergers.
Again, our results suggest alternative channels for the formation of M15's IMBH, if one is indeed hosted in its core.

\subsubsection{NGC 6397}

NGC 6397 is one of the nearest GCs to Earth, and its core can be studied in great detail.
This system has the most stringent constraints on BH mass.
Kinematic data suggest a low central velocity dispersion, limiting the mass of a putative central IMBH to $<600\,M_\odot$~\cite{2016A&A...588A.149K}.
The GC catalog of~\cite{Baumgardt:2018pyl} gives for NGC 6397 the following parameters: $M_{\rm cl}=8.24\times 10^4\,M_\odot$, $r_{\rm h}=3.16\,\rm pc$, and $R_{\rm g}=6.01\,\rm kpc$.
Using these data as inputs to our NN and RFR models, we predict that this system should not host a massive BH formed through repeated mergers, consistent with the above kinematical constraint.
The corresponding prediction from Eq.~\eqref{eq:max-BH-mass-Antonini} (cf.~Ref.~\cite{Antonini:2018auk}) for this system is about $94\,M_\odot$, assuming an expansion factor of 10. If the expansion factor had been 100, the BH could be up to $200\,M_\odot$, which is still less than the kinematically determined upper limit of $600\,M_\odot$. Finally, the non-detection of radio waves from the core of this cluster constrains the mass of any putative IMBH to less than $610\,M_\odot$, and is not in conflict with any of the above constraints.
We conclude that NGC 6397 is a GC that does not currently host an IMBH, and could not have formed one through hierarchical mergers. If formed via other channels, such as runaway stellar collisions, then the IMBH could not be heavier than $\sim600\,M_\odot$, or it was ejected in the past.

\subsubsection{Mayall II (G1)}

Based on kinematic data from the Hubble Space Telescope, Ref.~\cite{Gebhardt:2005cy} suggested that Mayall II (G1), which is the heaviest GC of M31 (the Andromeda galaxy), with a mass of about double that of $\omega$~Centauri, hosts an IMBH with mass $(1.8\pm0.5)\times10^4\,M_\odot$. In contrast, Ref.~\cite{Baumgardt:2003an} conducted N-body simulations and concluded that the existence of an IMBH in the core of Mayall II (G1) is not required. 
Instead, the observed increase in the velocity dispersion towards the center can be attributed to a subpopulation of stellar remnants.
Deep radio observations of its core did not detect radio emission, constraining the mass of any putative IMBH to $<9700\,M_\odot$~\cite{Miller-Jones:2012trw}.
The previously detected non-ultraluminous X-ray source in the core of Mayall II (G1) could be a low-mass X-ray binary.

Our RFR results indicate that Mayall II (G1) could contain an IMBH with mass $\sim10^4\,M_\odot$ with a model-based probability of $\approx20\%$, while the most likely outcomes are a heaviest BH of $\sim100\,M_\odot$ within the upper mass gap or a few hundred solar masses. The distribution predicted by the RFR is thus bimodal. The NN predicts a mass in the range $\sim10^2$--$10^3\,M_\odot$ and does not capture the bimodality of the distribution. This is expected, because the NN model used here represents the predictive distribution with a single Gaussian, whereas the RFR can expose multimodality through the spread of its individual-tree predictions.
Moreover, the initial total mass of BHs for this cluster should initially have been above $10^5\,M_\odot$, and the theoretical upper limit from Eq.~\eqref{eq:max-BH-mass-Antonini} is several $10^4\,M_\odot$. Therefore, in principle, there should have been enough resources for a $\sim10^4\,M_\odot$ IMBH to have formed via repeated mergers. Thus, Mayall II (G1) is one of the best candidate star clusters in the local Universe that could have hosted a hierarchical merger chain in the past, forming an IMBH in its core.

\subsubsection{Other systems}

Palomar 5 is a very sparse GC with a mass of $1.3\times10^4\,M_\odot$ and half-mass radius of $27.63\,\rm pc$.
The most likely heaviest BH mass predicted for this cluster is about $47\,M_\odot$; 
however, the RFR gives a model-based probability of 1\% that it hosts a $\sim3800\,M_\odot$ IMBH. 
The NN predicts a typical heaviest BH mass of about $150\,M_\odot$ for this cluster.

Nevertheless, our models predict a non-negligible population of BHs in the upper mass gap. This has consequences for GW observations, since LIGO is sensitive to BHs with masses around $100\,M_\odot$.

The detailed results of our analysis are reported in Table~\ref{tab:mwgc_full_results} in Appendix~\ref{app:full-table}.

\subsection{Nuclear star clusters}
\label{sec:Nuclear-star-clusters}

An important limitation of our work is the assumption of monolithic evolution. In particular, some of the most massive NSCs in our catalogs may have experienced different star formation histories. Our model assumes a single burst of star formation at one point in time, followed by subsequent mass loss and expansion. Alternatively, a large-mass NSC may assemble gradually over time, rather than in a single burst, which would result in a different expansion history. However, in the latter scenario, the mass of the assembled BH would be comparatively small.

Electromagnetic counterparts provide key evidence for low-level accretion onto massive BHs in nearby galaxies. In the compact elliptical galaxy M32, both radio and X-ray detections have been reported~\cite{Ho:2003xd,Yang:2015ada}, indicating weak but persistent nuclear activity associated with its central massive BH. These observations place M32 among the clearest examples of a low-luminosity galactic nucleus in the Local Group.

In contrast, the dwarf elliptical companion of M31, NGC 205, offers an important constraint on the high-mass end of the IMBH mass function. Stellar kinematical modeling yields an upper limit of $M_{\rm BH} \lesssim 4 \times 10^{4}\,M_\odot$~\cite{Valluri:2005up}, placing it among the lowest inferred SMBH masses in galactic nuclei. At the same time, these analyses highlight significant systematic uncertainties inherent in stellar dynamical modeling, including degeneracies in orbital structure and mass-to-light ratios, which can complicate robust BH mass determinations~\cite{Valluri:2002xs}.

Moving to bulgeless systems, the dwarf disk galaxy NGC~4395 provides compelling evidence that SMBHs are not exclusively associated with classical bulges. NGC~4395 hosts both an NSC and a Type 1 Seyfert nucleus, making it a prototypical example of a low-mass active galactic nucleus in a pure disk galaxy. Ref.~\cite{2003ApJ...588L..13F} measured NSC properties of $r_{\rm NSC} = 3.9\,\mathrm{pc}$ and $M_{\rm NSC} = 2 \times 10^{6}\,M_\odot$, emphasizing the coexistence of dense stellar systems and central BHs in such environments.

Multiple independent techniques have been used to constrain the BH mass in NGC~4395. Reverberation mapping yields $M_{\rm BH} = (3.6 \pm 1.1) \times 10^{5}\,M_\odot$~\cite{Peterson:2005yp}, while gas dynamical modeling provides a consistent estimate of $M_{\rm BH} = 4.0^{+8}_{-3} \times 10^{5}\,M_\odot$~\cite{2015ApJ...809..101D}. These measurements establish NGC~4395 as one of the best-constrained low-mass SMBH systems. More broadly, recent work has demonstrated that even lower-mass BHs can reside in bulgeless dwarf galaxies, such as the $M_{\rm BH} = 9.1^{+1.5}_{-1.6} \times 10^{3}\,M_\odot$ detection reported by Ref.~\cite{2019NatAs...3..755W}.

Applying our trained models to the nearby NSC sample, we find that NGC~5102 and NGC~5206 are the strongest candidates for hosting IMBHs assembled through repeated mergers in our framework.
We have confirmed that this result is robust under perturbations of the input parameters for NGC~5102 and NGC~5206.
By contrast, NGC~205 and NGC~4395 have low predicted masses despite their nuclear environments. This comparison emphasizes that NSC mass alone does not carry enough information: the final radius and the location of the system relative to the simulation prior also affect the inferred BH mass.

\begin{figure}
    \centering
    \includegraphics[width=\linewidth]{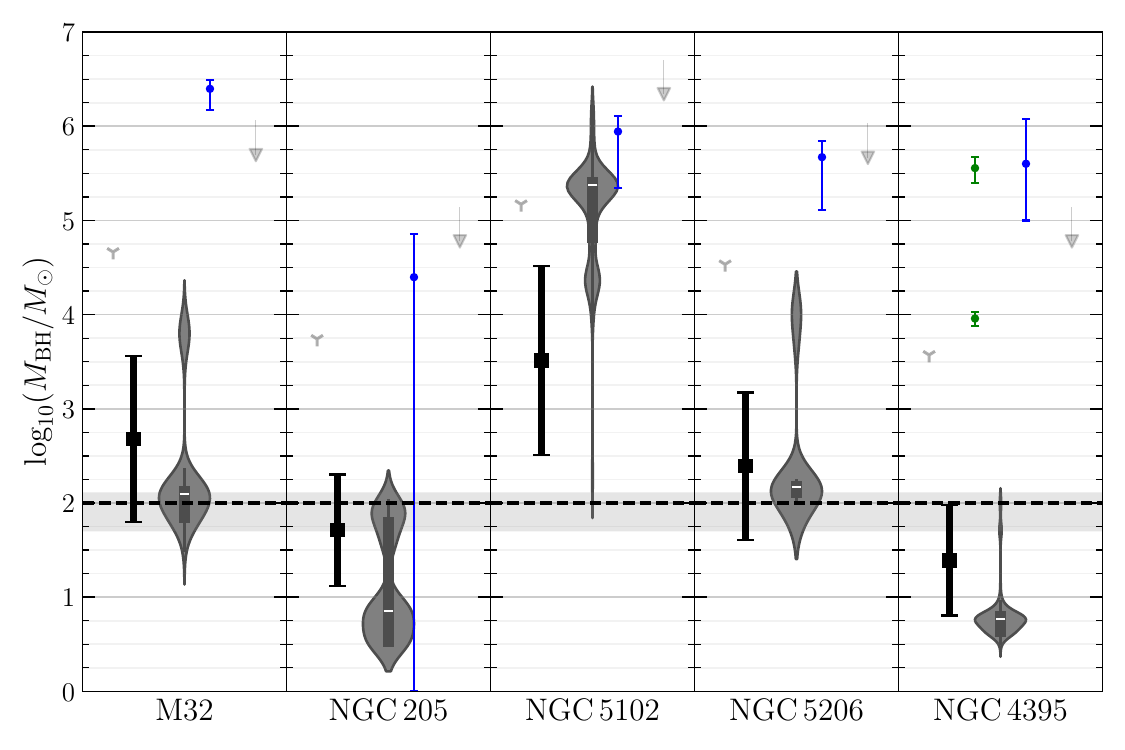}
    \caption{Same as Fig.~\ref{fig:predictions_for_GCs}, but for the subset of NSCs examined in this work. The measurement data for M32, NGC~205, NGC~5102, and NGC~5206 are taken from~\cite{2018ApJ...858..118N}, while observational data for NGC~4395 are taken from Table~2 of~\cite{2020ARA&A..58..257G}. Blue symbols correspond to dynamically inferred BH masses, while green symbols for NGC~4395 refer to BH masses estimated with reverberation mapping.}
    \label{fig:predictions_for_NSCs}
\end{figure}

\subsection{Young massive clusters}
\label{sec:Young-massive-clusters}

We also consider young massive clusters (YMCs) in the Milky Way, the Magellanic Clouds, and nearby galaxies, drawing on the compilation of cluster properties presented in~\cite{2010ARA&A..48..431P}. Our dataset includes systems in the Milky Way (Table 2 of that work), the Local Group -- specifically the LMC, SMC, M31, and NGC~6822 (Table 3) -- and galaxies beyond the Local Group (Table 4), most of which are starburst galaxies: ESO 338-IG, M51, M82, NGC~1140, NGC~1487, NGC~1569, NGC~1705, NGC~4038, NGC~4449, NGC~5236, and NGC~6946. For each cluster we adopt photometric masses ($\log M_{\rm phot}$) and effective radii ($r_{\rm eff}$) as inputs. 

\begin{figure*}
    \centering
    \includegraphics[width=\linewidth]{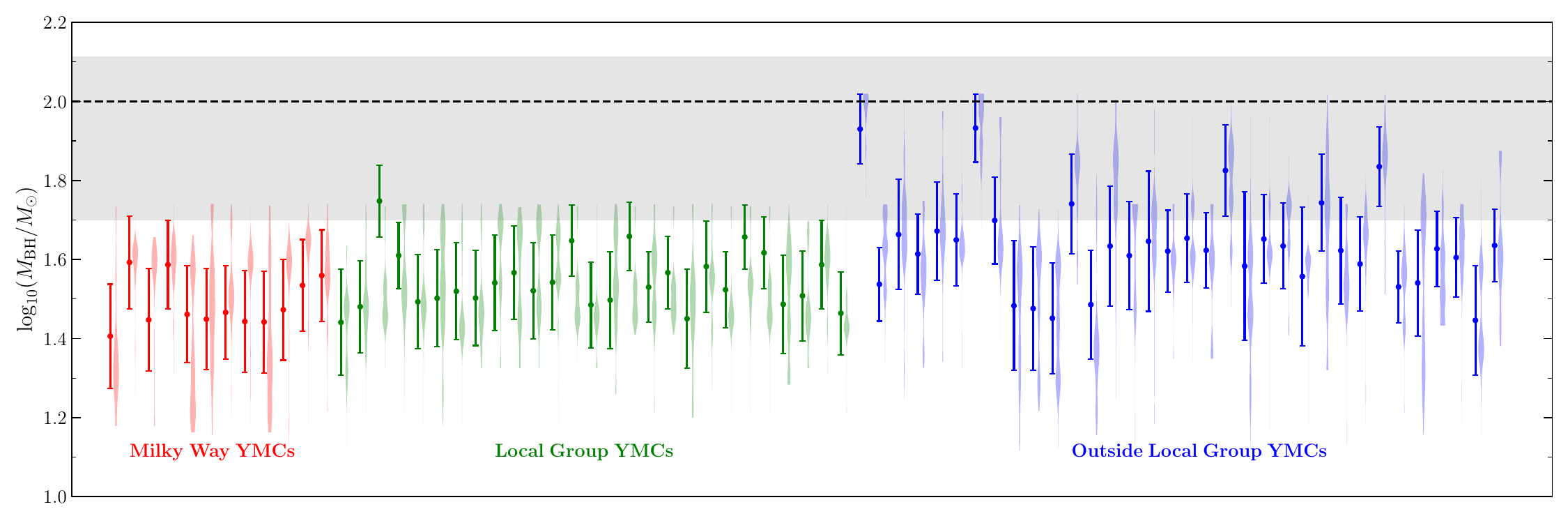}
    \caption{Predictions of the most massive BH mass formed through hierarchical mergers in nearby young massive clusters. We use the catalog in Tables 2, 3, and 4 from Ref.~\cite{2010ARA&A..48..431P} and consider only bound systems. The order of the systems shown here follows the order in which they appear in those tables. The violins show the RFR predictions, and the error bars show the NN predictions.}
    \label{fig:predictions-for-YMCs}
\end{figure*}

To ensure that we analyze only bound systems, we restrict the sample to clusters with ages larger than $t_{\rm dyn}$; objects failing this criterion are likely unbound associations and are excluded. Using these data, we assess whether massive BHs can form in such environments and generate predictions for each system, acknowledging that these predictions are not directly testable at present. Our results are summarized in Fig.~\ref{fig:predictions-for-YMCs}.

We emphasize that the dataset is subject to significant selection bias. In particular, flux-limited observations preferentially detect more massive and extended clusters at larger distances. As a result, our analysis probes the formation of IMBHs only in observed systems, not in the full underlying population. We are therefore more sensitive to heavier systems with larger radii, and those clusters are predicted to form heavier BHs, some of which lie confidently within the upper mass gap.

In the Milky Way, most YMCs are either insufficiently massive or not compact enough to support significant BH growth through repeated mergers. Consequently, none of the currently known Galactic YMCs are expected to produce IMBHs or BHs in the upper mass gap. A similar conclusion holds for YMCs in the Magellanic Clouds and elsewhere in the Local Group.

We quantify the likelihood of forming BHs with $M_{\rm BH} > 50\,M_\odot$ by applying the RFR model and computing the fraction of individual-tree predictions satisfying $\log M_{\rm BH} > 1.7$ for each cluster. These fractions should be interpreted as model-based scores rather than fully calibrated probabilities. These systems are of particular interest for gravitational-wave astronomy, given detections by the LVK network (e.g., GW190521 and GW231123) that include components within the upper mass gap.

Within the Milky Way, representative clusters such as Arches, Quintuplet, and Westerlund 1 exhibit relatively low model-based scores for forming BHs above $50\,M_\odot$: 6.6\%, 1.6\%, and 6.4\%, respectively. The well-studied R136 cluster in the Large Magellanic Cloud is likewise not expected to produce upper mass gap BHs. In contrast, several more distant systems, such as ESO 338-IG 23, M82~F, and NGC~4449~N-1, show scores exceeding 80\%. These objects are super star clusters in nearby starburst galaxies beyond the Local Group, and their high inferred scores largely reflect observational selection effects that favor detecting the most massive clusters at greater distances.

We stress that our input cluster masses likely represent lower limits, since the system's total dynamical mass could be higher if we used dynamical mass estimates or corrected for the mass-to-light ratio. Nevertheless, to ensure consistency across all systems, we use photometric mass estimates; thus, our predictions are conservative and limited by the observational resolution.

\section{Initial conditions for collisional runaway}

Repeated BH mergers require escape velocities of order several hundred $\rm km\, s^{-1}$, implying that the present-day properties of GCs, as we have seen in the previous sections, are generally inconsistent with the formation of IMBHs through hierarchical mergers. Nevertheless, observational evidence in specific systems, such as $\omega$~Centauri, suggests the presence of massive central BH candidates. This motivates consideration of alternative formation channels that operate efficiently during the earliest stages of cluster evolution. In particular, collisional runaway scenarios---driven by repeated stellar collisions within the first few Myr---offer a viable pathway for IMBH formation. In this section, we infer the initial conditions of GCs to assess whether their early environments are conducive to such processes.

Reconstructing GC initial conditions is inherently difficult due to strong degeneracies in their dynamical evolution. To address this, we employ a normalizing flow trained on cluster simulations to approximate the posterior distribution of initial cluster parameters conditioned on the observed final properties. These inferred initial conditions allow us to evaluate whether clusters could have satisfied the requirements for collisional runaway. Because the inferred runaway probabilities depend on posterior samples from the flow, we verify its calibration with the P--P plot shown in Fig.~\ref{fig:pp-normalizing-flow}. A key theoretical constraint arises from the condition derived by Miller \& Davies~\cite{Miller:2012ys}: if the initial one-dimensional velocity dispersion satisfies $\sigma_0 \gtrsim 40\,\rm km\, s^{-1}$, binary-star heating is insufficient to prevent core collapse, leading in practice to a collisional runaway. We use the following expression for relating the velocity dispersion to cluster properties:
\begin{equation}
\sigma_0 \simeq 13 \, \mathrm{km \, s^{-1}} \left( \frac{M_{\mathrm{cl}}}{10^5 \, M_\odot} \right)^{1/2} \left( \frac{r_{\mathrm{h}}}{1 \, \mathrm{pc}} \right)^{-1/2}
\end{equation}
However, this velocity dispersion condition alone is not sufficient. If the core-collapse timescale exceeds the lifetime of massive stars, a subsystem of stellar-mass BHs forms prior to collapse. These compact remnants can form hard binaries that dynamically stabilize the cluster and halt further contraction. As shown in Appendix~A of~\cite{Kritos:2022non}, binary BHs can provide sufficient heating to support clusters with total masses up to $10^8\,M_\odot$.

\begin{figure*}
    \centering
    \includegraphics[width=0.49\textwidth]{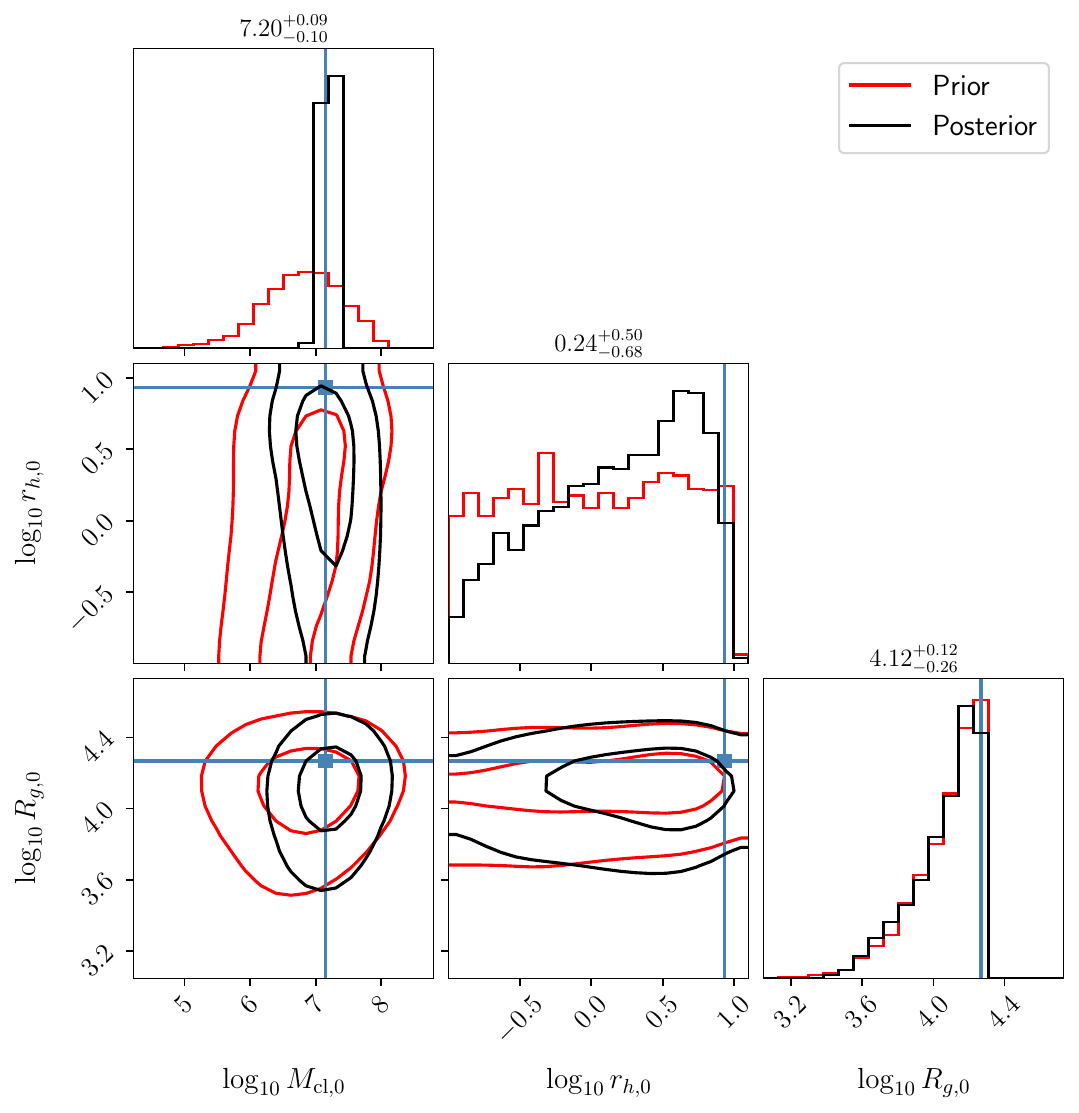}\includegraphics[width=0.49\textwidth]{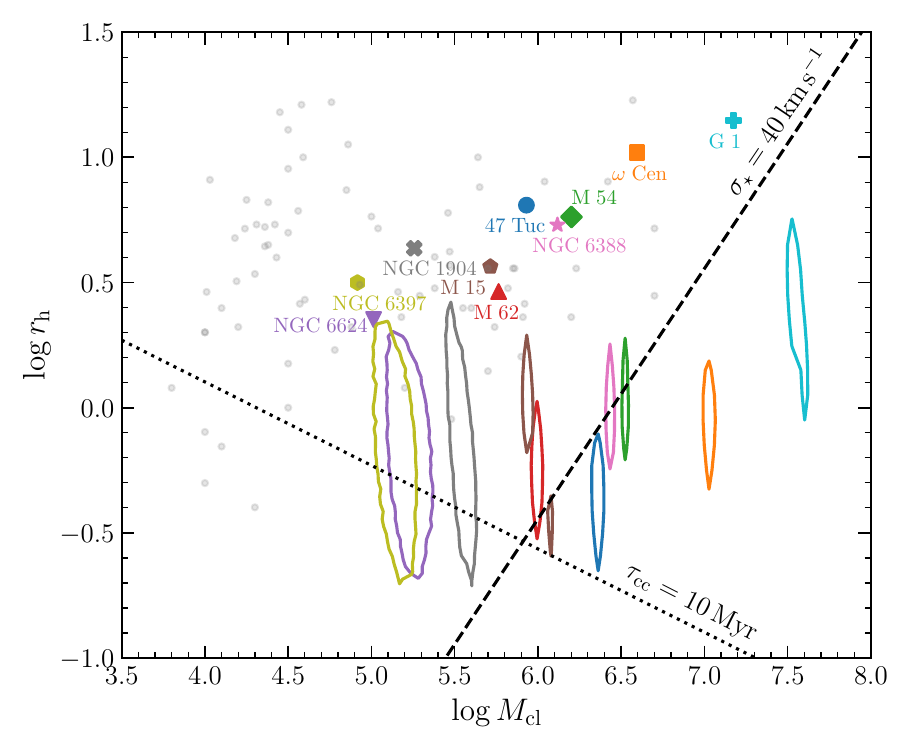}
    \caption{Initial cluster properties (denoted by a subscript $0$) predicted from their observed final properties using normalizing flows.
        Left: Simulation prior (in black) and posterior (in red) of initial properties for one of our simulated clusters. The prior and posterior are significantly different, showing the predictive power contained in the data.
    Right: Final observed properties of different clusters (colored points) are compared with the predicted initial conditions (contours) with the two-dimensional $1\sigma$ ($39.3\%$) contours.}
    \label{fig:initial-conditions}
\end{figure*}

The formation of the BH subsystem is gradual because of the mass dependence of stellar evolution timescales. Within $\sim 10\,\rm Myr$, essentially all stars capable of producing stellar-mass BH remnants will have done so. Consequently, in addition to the requirement $\sigma_0 \gtrsim 40\,\rm km\, s^{-1}$, we impose that the initial core-collapse timescale satisfies $t_{\rm cc} \simeq 0.2\,t_{\rm rh} < 10\,\rm Myr$. For reference, 
\begin{equation}
t_{\mathrm{cc}} \simeq 10 \, \mathrm{Myr} \left( \frac{M_{\mathrm{cl}}}{10^5 \, M_\odot} \right)^{\frac{1}{2}} \left( \frac{r_h}{0.5 \, \mathrm{pc}} \right)^{\frac{3}{2}} \frac{0.6 M_\odot}{\bar{m}} \frac{9}{\ln \Lambda} \frac{1}{\psi}
\end{equation}

where $\bar{m}$ is the average stellar mass in the cluster, which is approximately $0.6 \, M_\odot$ under the assumption of the Kroupa IMF in the range $[0.08, 150] \, M_\odot$, and $\Lambda \approx 0.02 N$ corresponds to the Coulomb logarithm. $\psi \equiv \frac{\langle m^{5/2} \rangle}{\bar{m}^{5/2}}$ is the dimensionless mass moment factor and accounts for a cluster with multiple mass components.

The $t_{\mathrm{cc}}$ criterion ensures that core collapse occurs before the cluster becomes BH-dominated. While more restrictive conditions have been proposed (e.g., Ref.~\cite{Rasio:2003sz} considers a $3\,\rm Myr$ threshold), our requirement reflects the need to form a complete BH subsystem before binary heating becomes effective.

Our reconstruction of initial conditions from the simulations, shown in the left panel of Fig.~\ref{fig:initial-conditions}, indicates that the initial cluster mass is more robustly constrained than the initial half-mass radius. Cluster mass evolves relatively deterministically due to the steady removal of stars in the high-velocity tail over successive half-mass relaxation times. In contrast, the evolution of the cluster radius is more stochastic, exhibiting significant variance across realizations with different seeds (cf.~Fig.~\ref{fig:cluster-evolution-realizations}). This increased variability broadens the posterior over initial radii. As a result, predicting the final IMBH mass from inferred initial conditions is not straightforward. Instead, we focus on estimating the probability that a given cluster experienced a collisional runaway phase by computing the fraction of posterior samples satisfying the two conditions for collisional runaway: $\sigma_0>40\,\rm km\,s^{-1}$ and $t_{\rm cc}<10\,\rm Myr$.

From our inferred initial conditions, we find that, on average, GCs have a $\sim 7\%$ probability of having undergone a collisional runaway early in their evolution. We show the collisional-runaway probabilities for all Milky Way GCs in the last column of Table~\ref{tab:mwgc_full_results}. This suggests that a non-negligible fraction of present-day GCs may host central IMBHs formed through this channel. However, current properties of YMCs do not favor such environments for runaway formation. Our analysis explicitly computes the probability of a runaway event, noting that IMBHs may also form over longer timescales through alternative channels.

In principle, the normalizing flow should learn physical constraints from the training data itself, but sharp boundaries in a finite training set can be learned only approximately. We therefore additionally enforce the following constraints on the normalizing-flow samples to ensure physically consistent evolutionary histories: $r_{\rm h} > r_{\rm h,0}$, $M_{\rm cl} < M_{\rm cl,0}$, and $R_{\rm g,0} > R_{\rm g}$. We emphasize that real clusters may have experienced more complex evolutionary pathways, whereas our framework assumes evolution within a static galactic potential. Under these assumptions, nearly all Milky Way GCs (except Mayall II/G1) exhibit a non-zero probability of having undergone a runaway phase in the past. In the case of Mayall II/G1, its large initial mass and velocity dispersion likely led to long relaxation timescales, allowing massive stars to evolve into BHs before significant collisional activity could occur. This favors hierarchical BH mergers as the dominant growth channel for Mayall II/G1. A similar argument applies to NSCs, where long relaxation times suppress early collisional runaways and instead promote IMBH formation through repeated BH mergers.

When a collisional runaway does occur, an order-of-magnitude estimate for the resulting IMBH mass can be obtained from~\cite{PortegiesZwart:2002iks}, yielding $M_{\rm IMBH} \sim 10^{-3} M_{\rm cl,0}$. This should be regarded as a lower bound, as the BH may continue to grow through accretion of stars and compact remnants over the cluster's lifetime. Finally, we note an important caveat: while we estimate the probability of a runaway occurrence, the formation of an IMBH would alter the cluster's subsequent dynamical evolution, potentially affecting the validity of backward inferences of its initial conditions.

\section{Discussion and Conclusions}
\label{sec:Conclusions}

We have trained machine-learning models on cluster simulation datasets generated with the {\sc Rapster} code. Training such models directly on N-body cluster simulations remains challenging because only limited simulation suites are available, and because the dense, massive stellar environments relevant for hierarchical mergers are computationally expensive to evolve. The rapid semi-analytic framework implemented in {\sc Rapster} self-consistently follows cluster evolution together with the assembly, evolution, and merger of BH binaries, enabling the large simulation set used in this work.

Predicting the heaviest BH mass from final cluster properties alone is intrinsically difficult because different evolutionary histories can lead to similar present-day observables. In particular, an expanded cluster may have been born compact and later expanded through energy injection from a growing IMBH, or it may have been diffuse from birth and never formed a massive retained remnant. This degeneracy explains the multimodal structure exposed by the RFR predictions. Including initial conditions largely removes this ambiguity in the simulations, but those quantities are generally unavailable for observed clusters.

Among the models considered here, the RFR gives the best overall performance on simulated test data. This is partly because ensembles of decision trees can represent the bimodal structure of the predicted BH mass distribution more naturally than the single-Gaussian NN used in this work. Symbolic-regression models provide compact analytic approximations and useful physical insight, but they have larger test losses than the RFR and NN models.

Our results suggest that present-day GCs are unlikely to form IMBHs through repeated BH mergers alone. Low escape velocities and strong recoil kicks prevent sustained hierarchical growth in most systems. If an IMBH is present in a GC, especially with mass well above $10^3\,M_\odot$, our results favor additional growth channels such as stellar collisions, tidal disruption events, or gas accretion during early cluster phases~\cite{Kritos:2024upo}. If $\omega$~Centauri is the stripped nucleus of a dwarf galaxy~\cite{Bekki:2003qw}, episodic gas inflows in its past host galaxy may have enabled additional BH growth through accretion~\cite{Alexander:2014noa}.

This interpretation is also consistent with recent machine-learning searches for IMBH host candidates in GCs. The analysis of~\cite{2024ApJ...965...89P} identified NGC~6569, Pal~6, NGC~6638, and NGC~6333 as leading IMBH candidates. It also identified NGC~6388, NGC~1904, and NGC~6397 as possible hosts, whereas our models assign these clusters low probabilities of hosting an IMBH assembled through repeated BH mergers. If such systems do host IMBHs, our results suggest that those BHs likely formed or grew through a different channel.

NSCs are more favorable environments for hierarchical growth because of their larger masses, higher densities, and higher escape velocities. We find that the probability of forming a BH more massive than $150\,M_\odot$ through repeated mergers exceeds $40\%$ when the initial escape velocity is greater than $400\,\rm km\,s^{-1}$. However, our NSC analysis assumes monolithic cluster evolution, whereas galactic nuclei can experience gas inflows, repeated star formation, and mergers with inspiraling GCs; such processes are observed in nearby dwarf galaxy nuclei~\cite{2025Natur.640..902P}. These caveats are especially important for interpreting the most massive NSCs in our sample.

The simulations also reveal a correlation between IMBH mass and final cluster radius: clusters that form more massive BHs tend to undergo greater expansion because of the energy injected during IMBH assembly. In extreme cases this energy input can dissolve the cluster and leave behind a free-floating IMBH. In our {\sc Rapster} simulation suite, about $5\%$ of all simulated clusters evaporated and left behind such a free IMBH, with about $1\%$ leaving behind a free IMBH more massive than $10^3\,M_\odot$.

Even when GCs and young massive clusters are inefficient at producing IMBHs above $\sim10^3\,M_\odot$ through repeated mergers, they can still form BHs in the upper mass gap, with masses up to $\sim150\,M_\odot$. This is relevant for gravitational-wave observations of high-mass events such as GW190521 and GW231123, whose components lie in the $\sim[60,\,120]\,M_\odot$ range. We note, however, that our assumption of initially non-spinning BHs enhances hierarchical-merger retention; allowing large first-generation BH spins would further suppress IMBH formation through this channel.

Future observational prospects for IMBH detection include white dwarf tidal disruption events~\cite{Maguire:2020lad}, IMBH-IMBH mergers detectable with next-generation gravitational-wave observatories such as the Einstein Telescope and Cosmic Explorer~\cite{Fairhurst:2023beb,Reali:2024hqf}, and high-resolution stellar monitoring in cluster centers with facilities such as the Habitable Worlds Observatory. The spin distribution of IMBHs may also provide clues to their formation pathways~\cite{Kritos:2024kpn}. Additional mechanisms that could enhance IMBH growth include collisions with massive stars~\cite{Baumgarte:2025udg} and hybrid scenarios involving simultaneous gas accretion and runaway mergers~\cite{Kritos:2024upo} (see also~\cite{Partmann:2024ees}). Constraints on intermediate mass-ratio inspiral rates from N-body studies are discussed in~\cite{2025arXiv250322109L}, and theoretical considerations such as the Miller-Davies instability criterion remain relevant~\cite{Miller:2012ys}.

Overall, repeated BH mergers can contribute to IMBH formation in dense nuclear environments, but they are unlikely to explain the full IMBH mass budget in all star clusters. In GCs, alternative formation or growth channels are likely required for any confirmed high-mass IMBH. Future gravitational-wave detections of IMBH binaries up to $\sim10^3\,M_\odot$ with next-generation detectors~\cite{Reali:2024hqf,Fairhurst:2023beb} may help constrain cluster properties and formation histories~\cite{Kritos:2025bby}, particularly if such binaries originate from cluster mergers. We use the recoil-kick prescriptions given in Ref.~\cite{Lousto:2010xk,Zlochower:2015wga}, but note that the recoil-kick prescriptions have recently been updated using new simulation data and normalizing-flow modeling techniques~\cite{Islam:2025drw,Islam:2026yxx}. Using the new recoil-kick prescriptions may modestly increase IMBH retention probabilities in merger remnants; we leave a systematic study of this effect to future work.

\acknowledgments

We thank Manuel Arca-Sedda, Debatri Chattopadhyay, Carl Rodriguez, Cole Miller, Mark Ho-Yeuk Cheung, Muryel Guolo, Francesco Iacovelli, Tousif Islam, Michela Mapelli, Lavinia Paiella, Johan Samsing, and Alessandro Alberto Trani for discussions.
K.K. is supported by the Onassis Foundation - Scholarship ID: F ZT 041-1/2023-2024.
K.K., D.W. and E.B. are supported by NSF Grants No.~AST-2307146, No.~PHY-2513337, No.~PHY-090003, and No.~PHY-20043, by NASA Grant No.~21-ATP21-0010, by John Templeton Foundation Grant No.~62840, by the Simons Foundation [MPS-SIP-00001698, E.B.], by the Simons Foundation International [SFI-MPS-BH-00012593-02], and by Italian Ministry of Foreign Affairs and International Cooperation Grant No.~PGR01167.
This work was carried out at the Advanced Research Computing at Hopkins (ARCH) core facility~\cite{archjhu}, which is supported by the NSF Grant No.~OAC-1920103.

\appendix

\begin{figure}
    \centering
    \includegraphics[width=\linewidth]{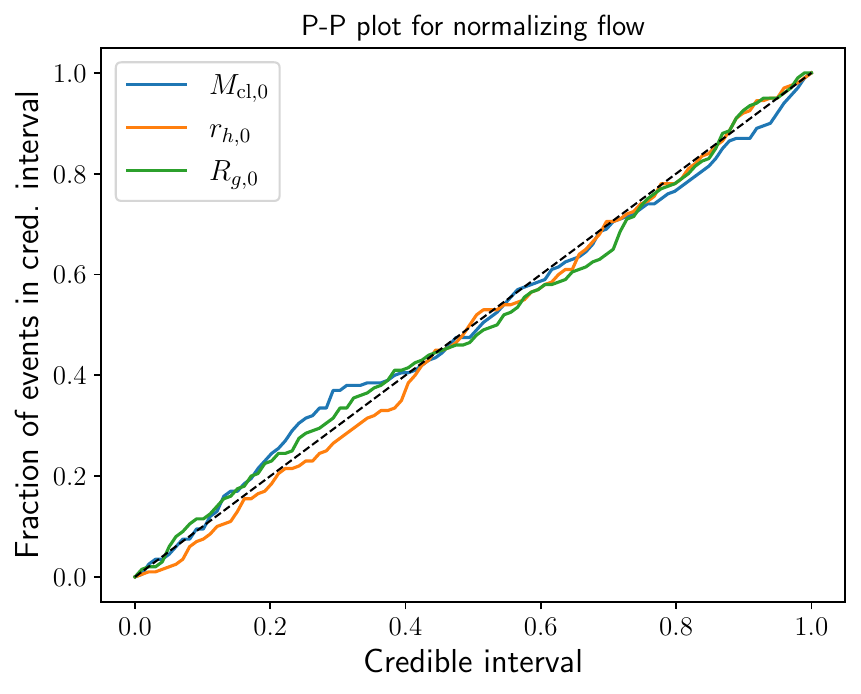}
    \caption{P--P plot for the normalizing-flow reconstruction of the initial cluster properties (used in Fig.~\ref{fig:initial-conditions}). The three curves correspond to $\log M_{\rm cl}$, $\log r_{\rm h}$, and $\log R_{\rm g}$. All three curves closely follow the diagonal, indicating that the normalizing-flow posteriors are well calibrated for the parameters relevant to our initial-condition inference.}
    \label{fig:pp-normalizing-flow}
\end{figure}

\section{Validity of the Breen \& Heggie theory}
\label{app:Validity-of-the-Breen-&-Heggie-theory}

The \texttt{Rapster} code relies on the theory developed and tested against N-body simulations by Breen \& Heggie in~\cite{Breen:2013vla}. In short, energy is produced in the core of the BH subsystem via tightening of hard BH binaries formed via three-body interactions until their merger or ejection from the cluster. The heat generated as kinetic energy flows through the system via two-body relaxation is shared among all BHs and then conducted into the bulk stellar population, thus supplying the energy required for the cluster to expand.
In the steady state, the rate of energy production balances the energy flow per unit time via the system's half-mass radius. 
According to Ref.~\cite{Breen:2013vla}, the theory can be trusted as long as the number of BHs in the subcluster is $\gtrsim40$. However, it is unclear whether the microphysics of the energy generation mechanism will remain the same if an IMBH forms in the system through repeated BH mergers. At the time of writing there is no tested theory or semi-analytic model for the evolution of two-component systems containing a massive BH in the center.
Therefore, here we derive an approximate condition for the heaviest allowed BH mass in the system without violating the microphysical picture of energy production in the center described above.

If the remnant of a previous BH binary merger is retained in the cluster, it will preferentially form another BH binary, which may merge again. This process is repeated, creating a heavy BH in the core if the gravitational-wave recoil received by the merger remnant does not exceed the escape velocity.
Intermediate mass ratio BH binaries composed of a $\sim10\,M_\odot$ BH and a heavy growing BH with mass $m_{\rm BH}^{\rm max}$ are assumed to form through three-body interactions and release energy in the core as they harden. 
However, if the mass of the growing BH exceeds some critical value, we argue that the energy production mechanism should change. The system will evolve to a state in which a massive BH sits roughly at the center of the cluster, undergoing small Brownian motion, and some BHs are bound to it within its influence radius. The latter radius is given by $r_{\rm infl}=Gm_{\rm BH}^{\rm max}/\langle v_{\rm BH}^2\rangle$, where $\langle v_{\rm BH}^2\rangle$ is the mean-square velocity dispersion in the BH subsystem.
In this configuration, BHs merge with the central IMBH when their velocity vector lies within the relativistic loss cone, and relaxation is dominated by changes in the angular momentum of the orbits and not their energy~\cite{Amaro-Seoane:2012lgq}. In other words, binaries do not harden through binary-single interactions as described above, and energy is produced at $\approx r_{\rm infl}$.
Assuming there is a single IMBH in the cluster, this IMBH will always be in equipartition with the other BHs in the subsystem, and thus the wandering radius will be $r_{\rm wand}=r_{\rm h,BH}\sqrt{\overline{m}_{\rm BH}/m_{\rm BH}^{\rm max}}$.
Based on the discussion above, the microphysical description of energy production through binary-single hardening in the core will be valid as long as $r_{\rm wand}\gtrsim r_{\rm infl}$, or
\begin{align}
    m_{\rm BH}^{\rm max}\lesssim0.54\overline{m}_{\rm BH}N_{\rm BH}^{2/3}\simeq 2500\,M_\odot \frac{\overline{m}_{\rm BH}}{10\,M_\odot}\left(\frac{N_{\rm BH}}{10^4}\right)^{2/3}\,.
\end{align}
In deriving this expression we have assumed the virial theorem for the BH subsystem, $\langle v_{\rm BH}^2\rangle=\kappa G\overline{m}_{\rm BH}N_{\rm BH}/r_{\rm h,BH}$, where $N_{\rm BH}$ is the number of stellar-mass BHs in the core and $\kappa\simeq0.4$~\cite{1971ApJ...164..399S}.
For example, a cluster with $10^7$ stars at solar metallicity will form about $10^4$ BHs with mass $\sim10\,M_\odot$, and the heaviest BH that can form satisfying the above conditions is $\simeq2500\,M_\odot$. This critical value scales with the mean BH mass and is thus larger at lower metallicities.
We stress that the condition $r_{\rm wand}\gtrsim r_{\rm infl}$ is itself derived from order-of-magnitude arguments, and the absence of direct N-body validation of {\sc Rapster} in the regime where an IMBH is already present in the cluster constitutes the dominant theoretical uncertainty in the estimate of this section.

\section{Table of cluster properties and BH mass predictions}
\label{app:full-table}

Table~\ref{tab:mwgc_full_results} lists the properties and BH mass predictions for all Milky Way GCs and nearby NSCs in our sample. For each cluster, we report the total mass $M_{\rm cl}$ and half-mass radius $r_{\rm h}$ from the catalog, alongside the BH mass estimates from the NN, RFR, and SR models, and the probabilities $p_{>100}$, $p_{>10^3}$, and $p_{\rm run}$ of hosting a BH formed from repeated mergers more massive than either $100\,M_\odot$ or $10^3\,M_\odot$, and of having undergone a runaway stellar collision sequence, respectively.

\newpage
\onecolumngrid
\begingroup
\small
\setlength{\tabcolsep}{8pt}
\begin{longtable}{llcccrrrr}
    \caption{Cluster properties and BH mass estimates. We use the GC catalog from~\cite{Baumgardt:2018pyl} and data for nearby NSCs from~\cite{2018ApJ...858..118N}.} \label{tab:mwgc_full_results} \\
    \toprule
    Cluster & $M_{\rm cl}$ ($M_{\odot}$) & $r_{\rm h}$ (pc) & $M_{\rm BH}^{\rm NN}$ ($M_\odot$) & $M_{\rm BH}^{\rm RFR}$ ($M_\odot$) & $M_{\rm BH}^{\rm SR}$ ($M_\odot$) & $p_{>100}$ & $p_{>10^3}$ & $p_{\rm run}$ \\
    \midrule
    \endfirsthead
    
    \multicolumn{9}{c}{{\bfseries \tablename\ \thetable{} -- continued from previous page}} \\
    \toprule
    Cluster & $M_{\rm cl}$ ($M_{\odot}$) & $r_{\rm h}$ (pc) & $M_{\rm BH}^{\rm NN}$ ($M_\odot$) & $M_{\rm BH}^{\rm RFR}$ ($M_\odot$) & $M_{\rm BH}^{\rm SR}$ ($M_\odot$) & $p_{>100}$ & $p_{>10^3}$ & $p_{\rm run}$ \\
    \midrule
    \endhead
    
    \midrule
    \multicolumn{8}{r}{{Continued on next page}} \\
    \bottomrule
    \endfoot
    \bottomrule
    \endlastfoot

NGC 104 & $8.53 \times 10^{5}$ & $6.44 \times 10^{0}$ & $2.15^{+3.12}_{-1.27} \times 10^{1}$ & $1.13^{+2.11}_{-0.74} \times 10^{1}$ & $1.00 \times 10^{2}$ & 0.004 & 0.004 & 0.116 \\
NGC 288 & $9.62 \times 10^{4}$ & $8.56 \times 10^{0}$ & $2.42^{+1.47}_{-0.91} \times 10^{1}$ & $2.21^{+1.21}_{-0.78} \times 10^{1}$ & $6.09 \times 10^{1}$ & 0.000 & 0.000 & 0.031 \\
NGC 362 & $2.52 \times 10^{5}$ & $3.36 \times 10^{0}$ & $1.91^{+1.87}_{-0.94} \times 10^{1}$ & $1.57^{+0.81}_{-0.53} \times 10^{1}$ & $5.13 \times 10^{1}$ & 0.000 & 0.000 & 0.112 \\
Whiting 1 & $1.37 \times 10^{3}$ & $1.65 \times 10^{1}$ & $6.32^{+1.05}_{-0.90} \times 10^{1}$ & $3.78^{+1.57}_{-1.11} \times 10^{1}$ & $1.20 \times 10^{1}$ & 0.000 & 0.000 & 0.006 \\
NGC 1261 & $1.72 \times 10^{5}$ & $4.86 \times 10^{0}$ & $1.67^{+1.55}_{-0.81} \times 10^{1}$ & $1.94^{+0.98}_{-0.65} \times 10^{1}$ & $5.67 \times 10^{1}$ & 0.000 & 0.000 & 0.075 \\
Pal 1 & $9.26 \times 10^{2}$ & $3.43 \times 10^{0}$ & $2.77^{+0.84}_{-0.65} \times 10^{1}$ & $1.99^{+1.15}_{-0.73} \times 10^{1}$ & $1.75 \times 10^{1}$ & 0.000 & 0.000 & 0.010 \\
AM 1 & $1.96 \times 10^{4}$ & $2.03 \times 10^{1}$ & $2.76^{+0.95}_{-0.71} \times 10^{1}$ & $2.81^{+0.95}_{-0.71} \times 10^{1}$ & $3.88 \times 10^{1}$ & 0.000 & 0.000 & 0.006 \\
Eridanus & $9.29 \times 10^{3}$ & $1.88 \times 10^{1}$ & $2.62^{+0.80}_{-0.61} \times 10^{1}$ & $2.81^{+0.84}_{-0.65} \times 10^{1}$ & $2.72 \times 10^{1}$ & 0.000 & 0.000 & 0.015 \\
Pal 2 & $2.20 \times 10^{5}$ & $7.64 \times 10^{0}$ & $1.98^{+1.69}_{-0.91} \times 10^{1}$ & $2.22^{+1.33}_{-0.83} \times 10^{1}$ & $7.55 \times 10^{1}$ & 0.000 & 0.000 & 0.103 \\
NGC 1851 & $2.83 \times 10^{5}$ & $3.16 \times 10^{0}$ & $1.99^{+2.02}_{-1.00} \times 10^{1}$ & $1.79^{+0.93}_{-0.61} \times 10^{1}$ & $5.07 \times 10^{1}$ & 0.000 & 0.000 & 0.151 \\
NGC 1904 & $1.81 \times 10^{5}$ & $4.33 \times 10^{0}$ & $1.70^{+1.59}_{-0.82} \times 10^{1}$ & $3.07^{+3.79}_{-1.70} \times 10^{1}$ & $5.44 \times 10^{1}$ & 0.000 & 0.000 & 0.076 \\
NGC 2298 & $4.98 \times 10^{4}$ & $3.20 \times 10^{0}$ & $1.98^{+1.32}_{-0.79} \times 10^{1}$ & $2.87^{+2.61}_{-1.37} \times 10^{1}$ & $3.83 \times 10^{1}$ & 0.000 & 0.000 & 0.013 \\
NGC 2419 & $7.83 \times 10^{5}$ & $2.64 \times 10^{1}$ & $7.01^{+4.98}_{-2.91} \times 10^{1}$ & $6.64^{+2.06}_{-1.57} \times 10^{1}$ & $3.23 \times 10^{2}$ & 0.032 & 0.000 & 0.075 \\
Pyxis & $3.22 \times 10^{4}$ & $2.35 \times 10^{1}$ & $5.28^{+1.75}_{-1.31} \times 10^{1}$ & $5.88^{+2.43}_{-1.72} \times 10^{1}$ & $5.07 \times 10^{1}$ & 0.000 & 0.000 & 0.014 \\
NGC 2808 & $7.91 \times 10^{5}$ & $3.75 \times 10^{0}$ & $1.78^{+2.29}_{-1.00} \times 10^{1}$ & $1.14^{+1.10}_{-0.56} \times 10^{1}$ & $6.66 \times 10^{1}$ & 0.012 & 0.000 & 0.149 \\
E 3 & $2.56 \times 10^{3}$ & $5.51 \times 10^{0}$ & $2.61^{+0.81}_{-0.62} \times 10^{1}$ & $1.80^{+0.62}_{-0.46} \times 10^{1}$ & $1.97 \times 10^{1}$ & 0.000 & 0.000 & 0.007 \\
Pal 3 & $1.85 \times 10^{4}$ & $2.76 \times 10^{1}$ & $3.36^{+0.91}_{-0.72} \times 10^{1}$ & $3.65^{+0.47}_{-0.42} \times 10^{1}$ & $3.87 \times 10^{1}$ & 0.000 & 0.000 & 0.008 \\
NGC 3201 & $1.93 \times 10^{5}$ & $8.68 \times 10^{0}$ & $2.58^{+2.18}_{-1.18} \times 10^{1}$ & $1.96^{+0.64}_{-0.48} \times 10^{1}$ & $7.73 \times 10^{1}$ & 0.000 & 0.000 & 0.129 \\
Pal 4 & $1.53 \times 10^{4}$ & $2.11 \times 10^{1}$ & $2.71^{+0.86}_{-0.65} \times 10^{1}$ & $2.90^{+0.88}_{-0.67} \times 10^{1}$ & $3.45 \times 10^{1}$ & 0.000 & 0.000 & 0.004 \\
Crater & $1.21 \times 10^{4}$ & $2.76 \times 10^{1}$ & $3.69^{+0.94}_{-0.75} \times 10^{1}$ & $3.00^{+0.78}_{-0.62} \times 10^{1}$ & $3.09 \times 10^{1}$ & 0.000 & 0.000 & 0.027 \\
NGC 4147 & $4.51 \times 10^{4}$ & $3.47 \times 10^{0}$ & $1.97^{+1.29}_{-0.78} \times 10^{1}$ & $2.32^{+2.18}_{-1.13} \times 10^{1}$ & $3.84 \times 10^{1}$ & 0.000 & 0.000 & 0.010 \\
NGC 4372 & $1.89 \times 10^{5}$ & $8.34 \times 10^{0}$ & $2.55^{+2.17}_{-1.17} \times 10^{1}$ & $2.42^{+1.23}_{-0.82} \times 10^{1}$ & $7.53 \times 10^{1}$ & 0.000 & 0.000 & 0.134 \\
Rup 106 & $3.40 \times 10^{4}$ & $1.11 \times 10^{1}$ & $2.94^{+1.33}_{-0.92} \times 10^{1}$ & $3.21^{+1.63}_{-1.08} \times 10^{1}$ & $4.52 \times 10^{1}$ & 0.000 & 0.000 & 0.012 \\
NGC 4590 & $1.28 \times 10^{5}$ & $7.27 \times 10^{0}$ & $2.19^{+1.58}_{-0.92} \times 10^{1}$ & $1.99^{+1.94}_{-0.98} \times 10^{1}$ & $6.25 \times 10^{1}$ & 0.000 & 0.000 & 0.049 \\
BH 140 & $6.12 \times 10^{4}$ & $9.73 \times 10^{0}$ & $2.97^{+1.50}_{-1.00} \times 10^{1}$ & $4.97^{+2.91}_{-1.83} \times 10^{1}$ & $5.43 \times 10^{1}$ & 0.000 & 0.000 & 0.021 \\
NGC 4833 & $1.86 \times 10^{5}$ & $4.58 \times 10^{0}$ & $1.78^{+1.66}_{-0.86} \times 10^{1}$ & $1.76^{+1.29}_{-0.74} \times 10^{1}$ & $5.61 \times 10^{1}$ & 0.000 & 0.000 & 0.069 \\
NGC 5024 & $5.02 \times 10^{5}$ & $1.00 \times 10^{1}$ & $3.25^{+4.07}_{-1.81} \times 10^{1}$ & $3.73^{+1.33}_{-0.98} \times 10^{1}$ & $1.18 \times 10^{2}$ & 0.000 & 0.000 & 0.174 \\
NGC 5053 & $6.28 \times 10^{4}$ & $1.70 \times 10^{1}$ & $4.76^{+2.20}_{-1.50} \times 10^{1}$ & $4.47^{+1.02}_{-0.83} \times 10^{1}$ & $6.50 \times 10^{1}$ & 0.000 & 0.000 & 0.040 \\
NGC 5139 & $3.94 \times 10^{6}$ & $1.04 \times 10^{1}$ & $1.70^{+5.66}_{-1.31} \times 10^{2}$ & $8.70^{+3.02}_{-2.24} \times 10^{1}$ & $2.61 \times 10^{2}$ & 0.296 & 0.004 & 0.006 \\
NGC 5272 & $4.09 \times 10^{5}$ & $5.47 \times 10^{0}$ & $1.71^{+1.94}_{-0.91} \times 10^{1}$ & $1.89^{+1.63}_{-0.88} \times 10^{1}$ & $7.42 \times 10^{1}$ & 0.002 & 0.000 & 0.209 \\
NGC 5286 & $4.24 \times 10^{5}$ & $3.58 \times 10^{0}$ & $1.83^{+2.04}_{-0.97} \times 10^{1}$ & $1.82^{+1.96}_{-0.94} \times 10^{1}$ & $5.80 \times 10^{1}$ & 0.000 & 0.000 & 0.219 \\
AM 4 & $7.46 \times 10^{2}$ & $1.13 \times 10^{1}$ & $5.10^{+0.88}_{-0.75} \times 10^{1}$ & $3.04^{+1.20}_{-0.86} \times 10^{1}$ & $1.06 \times 10^{1}$ & 0.000 & 0.000 & 0.006 \\
NGC 5466 & $5.61 \times 10^{4}$ & $1.38 \times 10^{1}$ & $4.00^{+1.91}_{-1.29} \times 10^{1}$ & $3.86^{+0.83}_{-0.68} \times 10^{1}$ & $5.81 \times 10^{1}$ & 0.000 & 0.000 & 0.021 \\
NGC 5634 & $2.47 \times 10^{5}$ & $7.72 \times 10^{0}$ & $2.11^{+1.89}_{-1.00} \times 10^{1}$ & $3.04^{+2.45}_{-1.36} \times 10^{1}$ & $7.88 \times 10^{1}$ & 0.000 & 0.000 & 0.145 \\
NGC 5694 & $2.69 \times 10^{5}$ & $4.35 \times 10^{0}$ & $1.73^{+1.83}_{-0.89} \times 10^{1}$ & $2.36^{+3.11}_{-1.34} \times 10^{1}$ & $5.92 \times 10^{1}$ & 0.002 & 0.002 & 0.159 \\
IC 4499 & $1.50 \times 10^{5}$ & $1.50 \times 10^{1}$ & $4.60^{+3.26}_{-1.91} \times 10^{1}$ & $4.07^{+0.88}_{-0.72} \times 10^{1}$ & $9.19 \times 10^{1}$ & 0.000 & 0.000 & 0.129 \\
NGC 5824 & $7.46 \times 10^{5}$ & $6.31 \times 10^{0}$ & $1.75^{+2.26}_{-0.99} \times 10^{1}$ & $1.41^{+1.40}_{-0.70} \times 10^{1}$ & $9.54 \times 10^{1}$ & 0.000 & 0.000 & 0.133 \\
Pal 5 & $1.34 \times 10^{4}$ & $2.76 \times 10^{1}$ & $1.46^{+0.39}_{-0.31} \times 10^{2}$ & $4.75^{+3.70}_{-2.08} \times 10^{1}$ & $3.26 \times 10^{1}$ & 0.022 & 0.014 & 0.009 \\
NGC 5897 & $1.67 \times 10^{5}$ & $1.09 \times 10^{1}$ & $3.41^{+2.59}_{-1.47} \times 10^{1}$ & $3.61^{+0.48}_{-0.43} \times 10^{1}$ & $8.23 \times 10^{1}$ & 0.000 & 0.000 & 0.143 \\
NGC 5904 & $3.92 \times 10^{5}$ & $5.60 \times 10^{0}$ & $1.88^{+2.16}_{-1.01} \times 10^{1}$ & $1.39^{+1.09}_{-0.61} \times 10^{1}$ & $7.45 \times 10^{1}$ & 0.000 & 0.000 & 0.205 \\
NGC 5927 & $2.93 \times 10^{5}$ & $5.92 \times 10^{0}$ & $2.11^{+2.26}_{-1.09} \times 10^{1}$ & $1.74^{+1.10}_{-0.67} \times 10^{1}$ & $7.14 \times 10^{1}$ & 0.000 & 0.000 & 0.185 \\
NGC 5946 & $1.13 \times 10^{5}$ & $2.83 \times 10^{0}$ & $2.18^{+1.72}_{-0.96} \times 10^{1}$ & $2.09^{+1.87}_{-0.99} \times 10^{1}$ & $4.20 \times 10^{1}$ & 0.004 & 0.002 & 0.018 \\
NGC 5986 & $2.99 \times 10^{5}$ & $4.35 \times 10^{0}$ & $1.78^{+1.86}_{-0.91} \times 10^{1}$ & $1.81^{+2.56}_{-1.06} \times 10^{1}$ & $6.05 \times 10^{1}$ & 0.000 & 0.000 & 0.159 \\
FSR 1716 & $5.66 \times 10^{4}$ & $5.26 \times 10^{0}$ & $2.03^{+1.27}_{-0.78} \times 10^{1}$ & $2.05^{+0.79}_{-0.57} \times 10^{1}$ & $4.47 \times 10^{1}$ & 0.002 & 0.000 & 0.014 \\
Pal 14 & $1.91 \times 10^{4}$ & $3.77 \times 10^{1}$ & $7.98^{+1.80}_{-1.47} \times 10^{1}$ & $4.96^{+0.75}_{-0.65} \times 10^{1}$ & $4.05 \times 10^{1}$ & 0.000 & 0.000 & 0.011 \\
Lynga 7 & $6.75 \times 10^{4}$ & $4.97 \times 10^{0}$ & $1.98^{+1.36}_{-0.81} \times 10^{1}$ & $2.05^{+1.03}_{-0.68} \times 10^{1}$ & $4.59 \times 10^{1}$ & 0.000 & 0.000 & 0.015 \\
NGC 6093 & $3.21 \times 10^{5}$ & $2.82 \times 10^{0}$ & $2.35^{+2.47}_{-1.20} \times 10^{1}$ & $3.54^{+4.59}_{-2.00} \times 10^{1}$ & $4.87 \times 10^{1}$ & 0.006 & 0.004 & 0.165 \\
RLGC 1 & $2.85 \times 10^{5}$ & $1.16 \times 10^{1}$ & $3.30^{+3.08}_{-1.59} \times 10^{1}$ & $3.96^{+1.10}_{-0.86} \times 10^{1}$ & $1.05 \times 10^{2}$ & 0.000 & 0.000 & 0.181 \\
NGC 6101 & $1.74 \times 10^{5}$ & $1.39 \times 10^{1}$ & $4.67^{+3.69}_{-2.06} \times 10^{1}$ & $4.10^{+2.09}_{-1.38} \times 10^{1}$ & $9.44 \times 10^{1}$ & 0.008 & 0.008 & 0.161 \\
NGC 6121 & $9.04 \times 10^{4}$ & $3.95 \times 10^{0}$ & $1.86^{+1.40}_{-0.80} \times 10^{1}$ & $2.11^{+1.20}_{-0.77} \times 10^{1}$ & $4.55 \times 10^{1}$ & 0.000 & 0.000 & 0.014 \\
NGC 6139 & $3.52 \times 10^{5}$ & $2.49 \times 10^{0}$ & $2.73^{+2.99}_{-1.43} \times 10^{1}$ & $3.32^{+4.20}_{-1.85} \times 10^{1}$ & $4.61 \times 10^{1}$ & 0.018 & 0.012 & 0.172 \\
NGC 6144 & $8.46 \times 10^{4}$ & $4.97 \times 10^{0}$ & $1.98^{+1.43}_{-0.83} \times 10^{1}$ & $2.50^{+1.96}_{-1.10} \times 10^{1}$ & $4.84 \times 10^{1}$ & 0.000 & 0.000 & 0.025 \\
Ter 3 & $3.25 \times 10^{4}$ & $6.24 \times 10^{0}$ & $2.30^{+1.15}_{-0.76} \times 10^{1}$ & $2.43^{+2.14}_{-1.14} \times 10^{1}$ & $4.00 \times 10^{1}$ & 0.000 & 0.000 & 0.012 \\
NGC 6171 & $6.12 \times 10^{4}$ & $3.72 \times 10^{0}$ & $1.93^{+1.31}_{-0.78} \times 10^{1}$ & $2.77^{+2.24}_{-1.24} \times 10^{1}$ & $4.14 \times 10^{1}$ & 0.000 & 0.000 & 0.012 \\
ESO 452-SC11 & $7.32 \times 10^{3}$ & $3.26 \times 10^{0}$ & $2.13^{+0.93}_{-0.65} \times 10^{1}$ & $1.94^{+1.26}_{-0.76} \times 10^{1}$ & $2.72 \times 10^{1}$ & 0.000 & 0.000 & 0.010 \\
NGC 6205 & $4.84 \times 10^{5}$ & $5.21 \times 10^{0}$ & $1.69^{+2.02}_{-0.92} \times 10^{1}$ & $1.46^{+1.76}_{-0.80} \times 10^{1}$ & $7.49 \times 10^{1}$ & 0.000 & 0.000 & 0.198 \\
NGC 6218 & $1.06 \times 10^{5}$ & $4.08 \times 10^{0}$ & $1.87^{+1.47}_{-0.82} \times 10^{1}$ & $1.80^{+1.45}_{-0.80} \times 10^{1}$ & $4.75 \times 10^{1}$ & 0.000 & 0.000 & 0.021 \\
NGC 6229 & $2.47 \times 10^{5}$ & $4.80 \times 10^{0}$ & $1.66^{+1.73}_{-0.85} \times 10^{1}$ & $1.62^{+1.25}_{-0.71} \times 10^{1}$ & $6.12 \times 10^{1}$ & 0.000 & 0.000 & 0.143 \\
FSR 1735 & $9.98 \times 10^{4}$ & $2.69 \times 10^{0}$ & $2.30^{+1.77}_{-1.00} \times 10^{1}$ & $2.18^{+1.56}_{-0.91} \times 10^{1}$ & $4.06 \times 10^{1}$ & 0.006 & 0.000 & 0.016 \\
NGC 6235 & $9.60 \times 10^{4}$ & $5.07 \times 10^{0}$ & $1.94^{+1.45}_{-0.83} \times 10^{1}$ & $1.87^{+1.18}_{-0.72} \times 10^{1}$ & $5.03 \times 10^{1}$ & 0.000 & 0.000 & 0.024 \\
NGC 6254 & $1.89 \times 10^{5}$ & $4.75 \times 10^{0}$ & $1.81^{+1.70}_{-0.88} \times 10^{1}$ & $2.00^{+1.66}_{-0.91} \times 10^{1}$ & $5.73 \times 10^{1}$ & 0.002 & 0.000 & 0.085 \\
NGC 6256 & $1.11 \times 10^{5}$ & $4.21 \times 10^{0}$ & $1.89^{+1.51}_{-0.84} \times 10^{1}$ & $1.84^{+1.81}_{-0.91} \times 10^{1}$ & $4.85 \times 10^{1}$ & 0.000 & 0.000 & 0.039 \\
Pal 15 & $5.25 \times 10^{4}$ & $2.70 \times 10^{1}$ & $6.61^{+2.35}_{-1.73} \times 10^{1}$ & $5.59^{+1.71}_{-1.31} \times 10^{1}$ & $6.81 \times 10^{1}$ & 0.098 & 0.000 & 0.035 \\
NGC 6266 & $5.81 \times 10^{5}$ & $2.91 \times 10^{0}$ & $2.45^{+3.01}_{-1.35} \times 10^{1}$ & $1.32^{+2.57}_{-0.87} \times 10^{1}$ & $5.38 \times 10^{1}$ & 0.024 & 0.000 & 0.224 \\
NGC 6273 & $7.19 \times 10^{5}$ & $4.70 \times 10^{0}$ & $1.80^{+2.37}_{-1.02} \times 10^{1}$ & $8.41^{+18.03}_{-5.73} \times 10^{0}$ & $7.65 \times 10^{1}$ & 0.030 & 0.000 & 0.149 \\
NGC 6284 & $1.74 \times 10^{5}$ & $4.04 \times 10^{0}$ & $1.82^{+1.64}_{-0.86} \times 10^{1}$ & $1.82^{+1.25}_{-0.74} \times 10^{1}$ & $5.23 \times 10^{1}$ & 0.000 & 0.000 & 0.053 \\
Gran 3 & $3.45 \times 10^{4}$ & $1.01 \times 10^{1}$ & $3.25^{+1.44}_{-1.00} \times 10^{1}$ & $6.63^{+15.50}_{-4.64} \times 10^{1}$ & $4.46 \times 10^{1}$ & 0.102 & 0.080 & 0.017 \\
Patchick 126 & $5.39 \times 10^{3}$ & $1.53 \times 10^{0}$ & $3.18^{+1.41}_{-0.98} \times 10^{1}$ & $2.33^{+1.17}_{-0.78} \times 10^{1}$ & $2.79 \times 10^{1}$ & 0.004 & 0.000 & 0.011 \\
NGC 6287 & $1.30 \times 10^{5}$ & $2.81 \times 10^{0}$ & $2.28^{+1.85}_{-1.02} \times 10^{1}$ & $2.23^{+1.77}_{-0.98} \times 10^{1}$ & $4.28 \times 10^{1}$ & 0.006 & 0.000 & 0.027 \\
NGC 6293 & $1.42 \times 10^{5}$ & $3.18 \times 10^{0}$ & $2.08^{+1.73}_{-0.95} \times 10^{1}$ & $1.99^{+1.04}_{-0.68} \times 10^{1}$ & $4.55 \times 10^{1}$ & 0.000 & 0.000 & 0.040 \\
Gran 2 & $4.20 \times 10^{4}$ & $1.23 \times 10^{1}$ & $4.03^{+1.85}_{-1.27} \times 10^{1}$ & $5.31^{+2.23}_{-1.57} \times 10^{1}$ & $5.01 \times 10^{1}$ & 0.002 & 0.002 & 0.021 \\
NGC 6304 & $1.03 \times 10^{5}$ & $3.05 \times 10^{0}$ & $2.13^{+1.65}_{-0.93} \times 10^{1}$ & $1.74^{+0.92}_{-0.60} \times 10^{1}$ & $4.26 \times 10^{1}$ & 0.000 & 0.000 & 0.021 \\
NGC 6316 & $3.47 \times 10^{5}$ & $6.08 \times 10^{0}$ & $2.10^{+2.38}_{-1.12} \times 10^{1}$ & $1.96^{+1.88}_{-0.96} \times 10^{1}$ & $7.58 \times 10^{1}$ & 0.000 & 0.000 & 0.209 \\
NGC 6325 & $6.19 \times 10^{4}$ & $2.38 \times 10^{0}$ & $2.53^{+1.75}_{-1.03} \times 10^{1}$ & $3.45^{+4.12}_{-1.88} \times 10^{1}$ & $3.67 \times 10^{1}$ & 0.016 & 0.014 & 0.012 \\
NGC 6333 & $3.08 \times 10^{5}$ & $3.81 \times 10^{0}$ & $1.86^{+1.93}_{-0.95} \times 10^{1}$ & $2.45^{+2.45}_{-1.23} \times 10^{1}$ & $5.67 \times 10^{1}$ & 0.000 & 0.000 & 0.192 \\
NGC 6341 & $2.73 \times 10^{5}$ & $3.59 \times 10^{0}$ & $1.84^{+1.84}_{-0.92} \times 10^{1}$ & $1.80^{+1.25}_{-0.74} \times 10^{1}$ & $5.38 \times 10^{1}$ & 0.000 & 0.000 & 0.140 \\
NGC 6342 & $3.77 \times 10^{4}$ & $1.86 \times 10^{0}$ & $3.02^{+1.89}_{-1.16} \times 10^{1}$ & $2.62^{+2.08}_{-1.16} \times 10^{1}$ & $3.29 \times 10^{1}$ & 0.004 & 0.004 & 0.012 \\
NGC 6352 & $6.02 \times 10^{4}$ & $4.80 \times 10^{0}$ & $2.00^{+1.33}_{-0.80} \times 10^{1}$ & $1.91^{+1.48}_{-0.83} \times 10^{1}$ & $4.42 \times 10^{1}$ & 0.000 & 0.000 & 0.011 \\
NGC 6355 & $9.91 \times 10^{4}$ & $3.34 \times 10^{0}$ & $2.02^{+1.56}_{-0.88} \times 10^{1}$ & $1.74^{+1.14}_{-0.69} \times 10^{1}$ & $4.37 \times 10^{1}$ & 0.000 & 0.000 & 0.034 \\
NGC 6356 & $5.65 \times 10^{5}$ & $7.73 \times 10^{0}$ & $2.56^{+3.47}_{-1.47} \times 10^{1}$ & $2.99^{+2.10}_{-1.23} \times 10^{1}$ & $1.02 \times 10^{2}$ & 0.006 & 0.000 & 0.161 \\
IC 1257 & $1.77 \times 10^{4}$ & $5.53 \times 10^{0}$ & $2.17^{+1.06}_{-0.71} \times 10^{1}$ & $2.28^{+1.10}_{-0.74} \times 10^{1}$ & $3.34 \times 10^{1}$ & 0.000 & 0.000 & 0.011 \\
Ter 2 & $8.05 \times 10^{4}$ & $4.31 \times 10^{0}$ & $1.95^{+1.43}_{-0.83} \times 10^{1}$ & $3.01^{+2.69}_{-1.42} \times 10^{1}$ & $4.57 \times 10^{1}$ & 0.002 & 0.000 & 0.050 \\
Ter 4 & $1.81 \times 10^{5}$ & $5.67 \times 10^{0}$ & $2.20^{+2.01}_{-1.05} \times 10^{1}$ & $4.08^{+5.87}_{-2.41} \times 10^{1}$ & $6.16 \times 10^{1}$ & 0.002 & 0.000 & 0.195 \\
HP 1 & $1.37 \times 10^{5}$ & $4.19 \times 10^{0}$ & $1.90^{+1.60}_{-0.87} \times 10^{1}$ & $2.20^{+1.77}_{-0.98} \times 10^{1}$ & $5.06 \times 10^{1}$ & 0.000 & 0.000 & 0.080 \\
FSR 1758 & $4.91 \times 10^{5}$ & $1.62 \times 10^{1}$ & $1.33^{+2.05}_{-0.81} \times 10^{2}$ & $1.66^{+6.32}_{-1.32} \times 10^{2}$ & $1.67 \times 10^{2}$ & 0.258 & 0.258 & 0.095 \\
NGC 6362 & $1.17 \times 10^{5}$ & $7.24 \times 10^{0}$ & $2.39^{+1.69}_{-0.99} \times 10^{1}$ & $2.33^{+2.54}_{-1.22} \times 10^{1}$ & $6.07 \times 10^{1}$ & 0.032 & 0.000 & 0.054 \\
NGC 6366 & $4.66 \times 10^{4}$ & $6.42 \times 10^{0}$ & $2.20^{+1.18}_{-0.77} \times 10^{1}$ & $1.83^{+1.12}_{-0.70} \times 10^{1}$ & $4.46 \times 10^{1}$ & 0.000 & 0.000 & 0.009 \\
Liller 1 & $1.01 \times 10^{6}$ & $2.42 \times 10^{0}$ & $3.56^{+5.33}_{-2.13} \times 10^{1}$ & $2.57^{+8.77}_{-1.99} \times 10^{1}$ & $5.11 \times 10^{1}$ & 0.032 & 0.018 & 0.165 \\
NGC 6380 & $3.41 \times 10^{5}$ & $4.42 \times 10^{0}$ & $1.78^{+1.92}_{-0.92} \times 10^{1}$ & $1.60^{+1.48}_{-0.77} \times 10^{1}$ & $6.28 \times 10^{1}$ & 0.000 & 0.000 & 0.209 \\
Ter 1 & $1.99 \times 10^{5}$ & $1.86 \times 10^{0}$ & $3.58^{+3.45}_{-1.76} \times 10^{1}$ & $2.57^{+2.45}_{-1.25} \times 10^{1}$ & $3.79 \times 10^{1}$ & 0.014 & 0.006 & 0.051 \\
Ton 2 & $4.31 \times 10^{4}$ & $4.26 \times 10^{0}$ & $2.01^{+1.25}_{-0.77} \times 10^{1}$ & $2.24^{+2.03}_{-1.06} \times 10^{1}$ & $3.98 \times 10^{1}$ & 0.000 & 0.000 & 0.013 \\
NGC 6388 & $1.31 \times 10^{6}$ & $5.37 \times 10^{0}$ & $2.14^{+3.45}_{-1.32} \times 10^{1}$ & $8.71^{+13.68}_{-5.32} \times 10^{0}$ & $9.70 \times 10^{1}$ & 0.006 & 0.002 & 0.087 \\
NGC 6397 & $8.24 \times 10^{4}$ & $3.16 \times 10^{0}$ & $2.04^{+1.52}_{-0.87} \times 10^{1}$ & $2.44^{+1.69}_{-1.00} \times 10^{1}$ & $4.16 \times 10^{1}$ & 0.000 & 0.000 & 0.016 \\
NGC 6401 & $1.21 \times 10^{5}$ & $3.17 \times 10^{0}$ & $2.10^{+1.68}_{-0.93} \times 10^{1}$ & $1.75^{+0.82}_{-0.56} \times 10^{1}$ & $4.43 \times 10^{1}$ & 0.000 & 0.000 & 0.042 \\
NGC 6402 & $5.95 \times 10^{5}$ & $4.92 \times 10^{0}$ & $1.72^{+2.17}_{-0.96} \times 10^{1}$ & $9.69^{+6.43}_{-3.87} \times 10^{0}$ & $7.57 \times 10^{1}$ & 0.000 & 0.000 & 0.178 \\
Pal 6 & $8.56 \times 10^{4}$ & $2.91 \times 10^{0}$ & $2.21^{+1.65}_{-0.94} \times 10^{1}$ & $2.89^{+2.65}_{-1.38} \times 10^{1}$ & $4.07 \times 10^{1}$ & 0.022 & 0.002 & 0.019 \\
NGC 6426 & $7.31 \times 10^{4}$ & $7.88 \times 10^{0}$ & $2.17^{+1.22}_{-0.78} \times 10^{1}$ & $2.18^{+1.21}_{-0.78} \times 10^{1}$ & $5.41 \times 10^{1}$ & 0.000 & 0.000 & 0.012 \\
Djor 1 & $8.43 \times 10^{4}$ & $5.62 \times 10^{0}$ & $2.11^{+1.46}_{-0.86} \times 10^{1}$ & $3.12^{+4.06}_{-1.76} \times 10^{1}$ & $5.04 \times 10^{1}$ & 0.002 & 0.000 & 0.044 \\
Ter 5 & $1.09 \times 10^{6}$ & $4.03 \times 10^{0}$ & $1.99^{+2.84}_{-1.17} \times 10^{1}$ & $8.85^{+6.57}_{-3.77} \times 10^{0}$ & $7.44 \times 10^{1}$ & 0.002 & 0.000 & 0.116 \\
NGC 6440 & $5.69 \times 10^{5}$ & $2.80 \times 10^{0}$ & $2.58^{+3.18}_{-1.43} \times 10^{1}$ & $1.06^{+2.17}_{-0.71} \times 10^{1}$ & $5.24 \times 10^{1}$ & 0.014 & 0.004 & 0.231 \\
Gran 5 & $2.29 \times 10^{4}$ & $2.20 \times 10^{0}$ & $2.54^{+1.41}_{-0.91} \times 10^{1}$ & $2.94^{+4.47}_{-1.77} \times 10^{1}$ & $3.21 \times 10^{1}$ & 0.018 & 0.016 & 0.010 \\
NGC 6441 & $1.39 \times 10^{6}$ & $4.30 \times 10^{0}$ & $1.96^{+2.99}_{-1.19} \times 10^{1}$ & $9.31^{+12.31}_{-5.30} \times 10^{0}$ & $8.20 \times 10^{1}$ & 0.000 & 0.000 & 0.088 \\
Ter 6 & $1.00 \times 10^{5}$ & $1.91 \times 10^{0}$ & $3.30^{+2.61}_{-1.46} \times 10^{1}$ & $2.71^{+3.88}_{-1.60} \times 10^{1}$ & $3.62 \times 10^{1}$ & 0.018 & 0.014 & 0.019 \\
NGC 6453 & $1.68 \times 10^{5}$ & $3.87 \times 10^{0}$ & $1.90^{+1.68}_{-0.89} \times 10^{1}$ & $1.96^{+1.22}_{-0.75} \times 10^{1}$ & $5.09 \times 10^{1}$ & 0.000 & 0.000 & 0.069 \\
UKS 1 & $7.99 \times 10^{4}$ & $4.07 \times 10^{0}$ & $1.86^{+1.36}_{-0.79} \times 10^{1}$ & $2.49^{+2.00}_{-1.11} \times 10^{1}$ & $4.48 \times 10^{1}$ & 0.000 & 0.000 & 0.013 \\
VVV-CL001 & $1.54 \times 10^{5}$ & $3.23 \times 10^{0}$ & $2.08^{+1.77}_{-0.95} \times 10^{1}$ & $1.96^{+1.47}_{-0.84} \times 10^{1}$ & $4.64 \times 10^{1}$ & 0.000 & 0.000 & 0.080 \\
Gran 1 & $2.61 \times 10^{4}$ & $7.79 \times 10^{0}$ & $2.67^{+1.17}_{-0.82} \times 10^{1}$ & $4.68^{+3.62}_{-2.04} \times 10^{1}$ & $3.88 \times 10^{1}$ & 0.102 & 0.000 & 0.038 \\
NGC 6496 & $7.44 \times 10^{4}$ & $7.01 \times 10^{0}$ & $2.44^{+1.48}_{-0.92} \times 10^{1}$ & $2.89^{+2.45}_{-1.33} \times 10^{1}$ & $5.24 \times 10^{1}$ & 0.008 & 0.000 & 0.041 \\
Ter 9 & $1.37 \times 10^{5}$ & $2.58 \times 10^{0}$ & $2.44^{+2.01}_{-1.10} \times 10^{1}$ & $2.47^{+1.84}_{-1.05} \times 10^{1}$ & $4.17 \times 10^{1}$ & 0.002 & 0.000 & 0.024 \\
Djor 2 & $1.34 \times 10^{5}$ & $6.48 \times 10^{0}$ & $2.39^{+1.88}_{-1.05} \times 10^{1}$ & $3.37^{+5.33}_{-2.06} \times 10^{1}$ & $6.03 \times 10^{1}$ & 0.004 & 0.000 & 0.181 \\
NGC 6517 & $2.16 \times 10^{5}$ & $2.88 \times 10^{0}$ & $2.23^{+2.08}_{-1.08} \times 10^{1}$ & $2.32^{+2.23}_{-1.14} \times 10^{1}$ & $4.65 \times 10^{1}$ & 0.012 & 0.000 & 0.079 \\
Ter 10 & $3.19 \times 10^{5}$ & $4.67 \times 10^{0}$ & $1.81^{+1.94}_{-0.94} \times 10^{1}$ & $2.36^{+2.67}_{-1.25} \times 10^{1}$ & $6.38 \times 10^{1}$ & 0.002 & 0.000 & 0.200 \\
NGC 6522 & $2.14 \times 10^{5}$ & $3.15 \times 10^{0}$ & $2.12^{+1.97}_{-1.02} \times 10^{1}$ & $1.94^{+1.75}_{-0.92} \times 10^{1}$ & $4.84 \times 10^{1}$ & 0.002 & 0.002 & 0.108 \\
NGC 6535 & $1.98 \times 10^{4}$ & $3.43 \times 10^{0}$ & $2.03^{+1.08}_{-0.70} \times 10^{1}$ & $2.09^{+1.21}_{-0.77} \times 10^{1}$ & $3.29 \times 10^{1}$ & 0.000 & 0.000 & 0.009 \\
NGC 6528 & $9.44 \times 10^{4}$ & $5.01 \times 10^{0}$ & $2.01^{+1.49}_{-0.85} \times 10^{1}$ & $2.87^{+3.97}_{-1.67} \times 10^{1}$ & $4.99 \times 10^{1}$ & 0.006 & 0.000 & 0.092 \\
NGC 6539 & $2.15 \times 10^{5}$ & $5.16 \times 10^{0}$ & $1.92^{+1.88}_{-0.95} \times 10^{1}$ & $2.07^{+1.00}_{-0.67} \times 10^{1}$ & $6.14 \times 10^{1}$ & 0.000 & 0.000 & 0.135 \\
NGC 6540 & $5.62 \times 10^{4}$ & $4.49 \times 10^{0}$ & $1.99^{+1.33}_{-0.80} \times 10^{1}$ & $2.02^{+2.62}_{-1.14} \times 10^{1}$ & $4.28 \times 10^{1}$ & 0.000 & 0.000 & 0.013 \\
VVV-CL160 & $5.28 \times 10^{4}$ & $4.50 \times 10^{0}$ & $2.00^{+1.31}_{-0.79} \times 10^{1}$ & $2.32^{+3.09}_{-1.33} \times 10^{1}$ & $4.22 \times 10^{1}$ & 0.000 & 0.000 & 0.013 \\
NGC 6541 & $2.57 \times 10^{5}$ & $3.81 \times 10^{0}$ & $1.87^{+1.84}_{-0.93} \times 10^{1}$ & $1.98^{+1.16}_{-0.73} \times 10^{1}$ & $5.48 \times 10^{1}$ & 0.000 & 0.000 & 0.141 \\
NGC 6544 & $8.15 \times 10^{4}$ & $2.24 \times 10^{0}$ & $2.58^{+1.91}_{-1.10} \times 10^{1}$ & $2.84^{+2.25}_{-1.26} \times 10^{1}$ & $3.73 \times 10^{1}$ & 0.002 & 0.002 & 0.016 \\
2MASS-GC01 & $4.11 \times 10^{4}$ & $5.98 \times 10^{0}$ & $2.12^{+1.13}_{-0.74} \times 10^{1}$ & $2.03^{+0.75}_{-0.55} \times 10^{1}$ & $4.24 \times 10^{1}$ & 0.000 & 0.000 & 0.009 \\
ESO 280-SC06 & $4.01 \times 10^{4}$ & $9.69 \times 10^{0}$ & $2.72^{+1.28}_{-0.87} \times 10^{1}$ & $2.27^{+0.96}_{-0.67} \times 10^{1}$ & $4.67 \times 10^{1}$ & 0.000 & 0.000 & 0.012 \\
NGC 6553 & $2.29 \times 10^{5}$ & $2.88 \times 10^{0}$ & $2.25^{+2.14}_{-1.10} \times 10^{1}$ & $2.37^{+2.25}_{-1.15} \times 10^{1}$ & $4.69 \times 10^{1}$ & 0.012 & 0.000 & 0.087 \\
2MASS-GC02 & $1.56 \times 10^{4}$ & $2.85 \times 10^{0}$ & $2.16^{+1.12}_{-0.74} \times 10^{1}$ & $2.38^{+2.25}_{-1.16} \times 10^{1}$ & $3.11 \times 10^{1}$ & 0.012 & 0.002 & 0.010 \\
NGC 6558 & $3.13 \times 10^{4}$ & $1.96 \times 10^{0}$ & $2.87^{+1.71}_{-1.07} \times 10^{1}$ & $2.69^{+3.58}_{-1.54} \times 10^{1}$ & $3.27 \times 10^{1}$ & 0.014 & 0.014 & 0.011 \\
IC 1276 & $7.43 \times 10^{4}$ & $4.98 \times 10^{0}$ & $1.97^{+1.38}_{-0.81} \times 10^{1}$ & $2.18^{+1.38}_{-0.85} \times 10^{1}$ & $4.70 \times 10^{1}$ & 0.000 & 0.000 & 0.016 \\
Ter 12 & $3.76 \times 10^{4}$ & $2.71 \times 10^{0}$ & $2.23^{+1.36}_{-0.84} \times 10^{1}$ & $2.09^{+1.83}_{-0.97} \times 10^{1}$ & $3.54 \times 10^{1}$ & 0.004 & 0.000 & 0.013 \\
NGC 6569 & $2.29 \times 10^{5}$ & $3.46 \times 10^{0}$ & $1.94^{+1.84}_{-0.95} \times 10^{1}$ & $2.17^{+1.48}_{-0.88} \times 10^{1}$ & $5.12 \times 10^{1}$ & 0.000 & 0.000 & 0.103 \\
BH 261 & $2.38 \times 10^{4}$ & $4.45 \times 10^{0}$ & $2.06^{+1.08}_{-0.71} \times 10^{1}$ & $1.75^{+1.73}_{-0.87} \times 10^{1}$ & $3.51 \times 10^{1}$ & 0.000 & 0.000 & 0.010 \\
NGC 6584 & $1.15 \times 10^{5}$ & $5.37 \times 10^{0}$ & $1.90^{+1.49}_{-0.83} \times 10^{1}$ & $1.85^{+0.77}_{-0.54} \times 10^{1}$ & $5.37 \times 10^{1}$ & 0.000 & 0.000 & 0.023 \\
NGC 6624 & $1.03 \times 10^{5}$ & $2.25 \times 10^{0}$ & $2.78^{+2.16}_{-1.22} \times 10^{1}$ & $2.77^{+4.24}_{-1.68} \times 10^{1}$ & $3.83 \times 10^{1}$ & 0.022 & 0.016 & 0.020 \\
NGC 6626 & $2.70 \times 10^{5}$ & $2.33 \times 10^{0}$ & $2.88^{+2.95}_{-1.46} \times 10^{1}$ & $2.70^{+2.78}_{-1.37} \times 10^{1}$ & $4.33 \times 10^{1}$ & 0.014 & 0.006 & 0.106 \\
NGC 6637 & $1.38 \times 10^{5}$ & $3.19 \times 10^{0}$ & $2.07^{+1.72}_{-0.94} \times 10^{1}$ & $2.12^{+1.18}_{-0.76} \times 10^{1}$ & $4.54 \times 10^{1}$ & 0.000 & 0.000 & 0.040 \\
NGC 6638 & $1.24 \times 10^{5}$ & $2.28 \times 10^{0}$ & $2.77^{+2.26}_{-1.24} \times 10^{1}$ & $2.63^{+2.16}_{-1.19} \times 10^{1}$ & $3.93 \times 10^{1}$ & 0.008 & 0.004 & 0.024 \\
NGC 6642 & $3.95 \times 10^{4}$ & $1.73 \times 10^{0}$ & $3.27^{+2.08}_{-1.27} \times 10^{1}$ & $2.77^{+1.54}_{-0.99} \times 10^{1}$ & $3.26 \times 10^{1}$ & 0.002 & 0.000 & 0.008 \\
NGC 6652 & $4.09 \times 10^{4}$ & $1.85 \times 10^{0}$ & $3.04^{+1.94}_{-1.18} \times 10^{1}$ & $2.79^{+3.40}_{-1.53} \times 10^{1}$ & $3.32 \times 10^{1}$ & 0.020 & 0.018 & 0.013 \\
NGC 6656 & $4.70 \times 10^{5}$ & $5.21 \times 10^{0}$ & $1.78^{+2.13}_{-0.97} \times 10^{1}$ & $1.57^{+1.40}_{-0.74} \times 10^{1}$ & $7.44 \times 10^{1}$ & 0.002 & 0.000 & 0.202 \\
Pal 8 & $7.13 \times 10^{4}$ & $6.14 \times 10^{0}$ & $2.17^{+1.36}_{-0.83} \times 10^{1}$ & $2.72^{+1.92}_{-1.12} \times 10^{1}$ & $4.96 \times 10^{1}$ & 0.002 & 0.000 & 0.018 \\
NGC 6681 & $1.05 \times 10^{5}$ & $3.04 \times 10^{0}$ & $2.13^{+1.66}_{-0.93} \times 10^{1}$ & $1.71^{+0.97}_{-0.62} \times 10^{1}$ & $4.27 \times 10^{1}$ & 0.000 & 0.000 & 0.022 \\
RLGC 2 & $2.68 \times 10^{5}$ & $5.00 \times 10^{0}$ & $1.71^{+1.76}_{-0.87} \times 10^{1}$ & $1.50^{+0.73}_{-0.49} \times 10^{1}$ & $6.37 \times 10^{1}$ & 0.000 & 0.000 & 0.143 \\
NGC 6712 & $9.50 \times 10^{4}$ & $3.50 \times 10^{0}$ & $1.96^{+1.50}_{-0.85} \times 10^{1}$ & $2.19^{+1.36}_{-0.84} \times 10^{1}$ & $4.41 \times 10^{1}$ & 0.000 & 0.000 & 0.017 \\
NGC 6715 & $1.59 \times 10^{6}$ & $5.78 \times 10^{0}$ & $2.11^{+3.47}_{-1.31} \times 10^{1}$ & $4.06^{+6.19}_{-2.45} \times 10^{1}$ & $1.08 \times 10^{2}$ & 0.022 & 0.004 & 0.064 \\
NGC 6717 & $2.63 \times 10^{4}$ & $3.26 \times 10^{0}$ & $2.03^{+1.13}_{-0.73} \times 10^{1}$ & $1.85^{+1.23}_{-0.74} \times 10^{1}$ & $3.44 \times 10^{1}$ & 0.000 & 0.000 & 0.009 \\
NGC 6723 & $1.97 \times 10^{5}$ & $5.07 \times 10^{0}$ & $1.91^{+1.81}_{-0.93} \times 10^{1}$ & $2.11^{+1.62}_{-0.92} \times 10^{1}$ & $5.97 \times 10^{1}$ & 0.000 & 0.000 & 0.129 \\
NGC 6749 & $2.05 \times 10^{5}$ & $6.10 \times 10^{0}$ & $2.20^{+2.06}_{-1.06} \times 10^{1}$ & $1.89^{+1.43}_{-0.81} \times 10^{1}$ & $6.60 \times 10^{1}$ & 0.000 & 0.000 & 0.128 \\
NGC 6752 & $2.61 \times 10^{5}$ & $4.75 \times 10^{0}$ & $1.77^{+1.81}_{-0.89} \times 10^{1}$ & $1.64^{+1.29}_{-0.72} \times 10^{1}$ & $6.16 \times 10^{1}$ & 0.000 & 0.000 & 0.149 \\
NGC 6760 & $2.86 \times 10^{5}$ & $5.74 \times 10^{0}$ & $2.05^{+2.18}_{-1.06} \times 10^{1}$ & $1.84^{+1.00}_{-0.65} \times 10^{1}$ & $6.97 \times 10^{1}$ & 0.000 & 0.000 & 0.182 \\
NGC 6779 & $1.66 \times 10^{5}$ & $4.42 \times 10^{0}$ & $1.76^{+1.59}_{-0.83} \times 10^{1}$ & $1.56^{+1.23}_{-0.69} \times 10^{1}$ & $5.39 \times 10^{1}$ & 0.004 & 0.000 & 0.048 \\
Ter 7 & $2.23 \times 10^{4}$ & $1.22 \times 10^{1}$ & $3.31^{+1.33}_{-0.95} \times 10^{1}$ & $3.81^{+1.27}_{-0.95} \times 10^{1}$ & $3.88 \times 10^{1}$ & 0.000 & 0.000 & 0.006 \\
Pal 10 & $1.25 \times 10^{5}$ & $6.91 \times 10^{0}$ & $2.22^{+1.63}_{-0.94} \times 10^{1}$ & $1.94^{+1.09}_{-0.70} \times 10^{1}$ & $6.08 \times 10^{1}$ & 0.000 & 0.000 & 0.046 \\
Arp 2 & $3.87 \times 10^{4}$ & $1.93 \times 10^{1}$ & $5.39^{+2.09}_{-1.51} \times 10^{1}$ & $4.13^{+0.75}_{-0.63} \times 10^{1}$ & $5.35 \times 10^{1}$ & 0.000 & 0.000 & 0.014 \\
NGC 6809 & $1.97 \times 10^{5}$ & $6.78 \times 10^{0}$ & $2.30^{+2.08}_{-1.09} \times 10^{1}$ & $2.42^{+1.59}_{-0.96} \times 10^{1}$ & $6.88 \times 10^{1}$ & 0.008 & 0.000 & 0.144 \\
Ter 8 & $7.59 \times 10^{4}$ & $2.12 \times 10^{1}$ & $6.41^{+3.07}_{-2.07} \times 10^{1}$ & $5.13^{+6.84}_{-2.93} \times 10^{1}$ & $7.68 \times 10^{1}$ & 0.036 & 0.036 & 0.074 \\
Pal 11 & $1.42 \times 10^{4}$ & $9.75 \times 10^{0}$ & $3.16^{+1.21}_{-0.87} \times 10^{1}$ & $4.07^{+1.64}_{-1.17} \times 10^{1}$ & $3.22 \times 10^{1}$ & 0.000 & 0.000 & 0.007 \\
Sagittarius II & $3.05 \times 10^{4}$ & $3.94 \times 10^{1}$ & $9.91^{+2.59}_{-2.06} \times 10^{1}$ & $5.37^{+1.11}_{-0.92} \times 10^{1}$ & $5.41 \times 10^{1}$ & 0.000 & 0.000 & 0.016 \\
NGC 6838 & $3.78 \times 10^{4}$ & $4.81 \times 10^{0}$ & $2.00^{+1.15}_{-0.73} \times 10^{1}$ & $1.96^{+1.12}_{-0.71} \times 10^{1}$ & $3.96 \times 10^{1}$ & 0.000 & 0.000 & 0.009 \\
NGC 6864 & $4.59 \times 10^{5}$ & $2.94 \times 10^{0}$ & $2.10^{+2.38}_{-1.12} \times 10^{1}$ & $2.54^{+3.47}_{-1.47} \times 10^{1}$ & $5.24 \times 10^{1}$ & 0.004 & 0.000 & 0.217 \\
NGC 6934 & $1.50 \times 10^{5}$ & $4.71 \times 10^{0}$ & $1.74^{+1.55}_{-0.82} \times 10^{1}$ & $2.11^{+1.77}_{-0.96} \times 10^{1}$ & $5.42 \times 10^{1}$ & 0.000 & 0.000 & 0.042 \\
NGC 6981 & $8.12 \times 10^{4}$ & $5.79 \times 10^{0}$ & $1.94^{+1.33}_{-0.79} \times 10^{1}$ & $2.32^{+1.08}_{-0.74} \times 10^{1}$ & $5.04 \times 10^{1}$ & 0.000 & 0.000 & 0.015 \\
NGC 7006 & $1.32 \times 10^{5}$ & $6.59 \times 10^{0}$ & $1.95^{+1.57}_{-0.87} \times 10^{1}$ & $2.28^{+1.17}_{-0.77} \times 10^{1}$ & $6.05 \times 10^{1}$ & 0.000 & 0.000 & 0.038 \\
Laevens 3 & $2.68 \times 10^{3}$ & $9.85 \times 10^{0}$ & $2.53^{+0.85}_{-0.64} \times 10^{1}$ & $2.89^{+1.01}_{-0.75} \times 10^{1}$ & $1.76 \times 10^{1}$ & 0.000 & 0.000 & 0.009 \\
NGC 7078 & $5.18 \times 10^{5}$ & $3.66 \times 10^{0}$ & $1.81^{+2.13}_{-0.98} \times 10^{1}$ & $1.07^{+0.54}_{-0.36} \times 10^{1}$ & $6.08 \times 10^{1}$ & 0.000 & 0.000 & 0.203 \\
NGC 7089 & $6.24 \times 10^{5}$ & $4.76 \times 10^{0}$ & $1.59^{+1.97}_{-0.88} \times 10^{1}$ & $8.45^{+7.68}_{-4.02} \times 10^{0}$ & $7.48 \times 10^{1}$ & 0.000 & 0.000 & 0.179 \\
NGC 7099 & $1.21 \times 10^{5}$ & $4.27 \times 10^{0}$ & $1.82^{+1.50}_{-0.82} \times 10^{1}$ & $2.55^{+2.31}_{-1.21} \times 10^{1}$ & $4.97 \times 10^{1}$ & 0.000 & 0.000 & 0.018 \\
Pal 12 & $6.19 \times 10^{3}$ & $1.05 \times 10^{1}$ & $3.34^{+1.06}_{-0.81} \times 10^{1}$ & $3.32^{+1.91}_{-1.21} \times 10^{1}$ & $2.37 \times 10^{1}$ & 0.000 & 0.000 & 0.007 \\
Pal 13 & $2.78 \times 10^{3}$ & $1.57 \times 10^{1}$ & $5.84^{+1.26}_{-1.03} \times 10^{1}$ & $2.98^{+1.42}_{-0.96} \times 10^{1}$ & $1.63 \times 10^{1}$ & 0.000 & 0.000 & 0.008 \\
NGC 7492 & $1.97 \times 10^{4}$ & $1.06 \times 10^{1}$ & $2.72^{+1.14}_{-0.81} \times 10^{1}$ & $2.89^{+1.48}_{-0.98} \times 10^{1}$ & $3.65 \times 10^{1}$ & 0.000 & 0.000 & 0.006 \\
$\rm Mayall\ II\ (G1)$ & $1.50 \times 10^{7}$ & $1.40 \times 10^{1}$ & $4.80^{+18.10}_{-3.79} \times 10^{2}$ & $2.43^{+11.18}_{-1.99} \times 10^{2}$ & $6.83 \times 10^{2}$ & 0.880 & 0.182 & 0.000 \\
$\rm M\ 32$ & $1.65 \times 10^{7}$ & $4.40 \times 10^{0}$ & $4.78^{+31.61}_{-4.16} \times 10^{2}$ & $1.90^{+6.40}_{-1.46} \times 10^{2}$ & $1.35 \times 10^{2}$ & 0.692 & 0.146 & - \\
$\rm NGC\ 205$ & $2.00 \times 10^{6}$ & $1.30 \times 10^{0}$ & $5.16^{+14.93}_{-3.83} \times 10^{1}$ & $1.29^{+3.40}_{-0.94} \times 10^{1}$ & $3.50 \times 10^{1}$ & 0.016 & 0.000 & - \\
$\rm NGC\ 5102$ & $7.30 \times 10^{7}$ & $2.63 \times 10^{1}$ & $3.28^{+29.90}_{-2.96} \times 10^{3}$ & $1.31^{+3.18}_{-0.93} \times 10^{5}$ & $5.99 \times 10^{3}$ & 1.000 & 0.992 & - \\
$\rm NGC\ 5206$ & $1.54 \times 10^{7}$ & $8.10 \times 10^{0}$ & $2.46^{+12.46}_{-2.06} \times 10^{2}$ & $2.51^{+10.15}_{-2.01} \times 10^{2}$ & $3.00 \times 10^{2}$ & 0.854 & 0.166 & - \\
$\rm NGC\ 4395$ & $2.00 \times 10^{6}$ & $3.90 \times 10^{0}$ & $2.46^{+6.99}_{-1.82} \times 10^{1}$ & $6.22^{+4.84}_{-2.72} \times 10^{0}$ & $8.10 \times 10^{1}$ & 0.012 & 0.000 & - \\

\end{longtable}
\endgroup
\twocolumngrid

\bibliography{IMBH_ML}

\end{document}